%% file: Thesis-arx.tex
\documentclass[twoside, 12pt, BCOR=20mm]{scrbook}

\usepackage{hyperref}
\usepackage{amssymb,amstext,amsmath,amsthm}
\usepackage{graphicx}
\usepackage{latexsym}
\usepackage{psfrag}
\usepackage{amsfonts}
\usepackage{scrpage2}

\usepackage{amsmath,amsxtra,amsthm,amssymb,xr}
\usepackage[all]{xy}
\usepackage[]{mathrsfs}
\usepackage{verbatim}
\usepackage{color}  

\KOMAoptions{DIV=last}

\automark[section]{chapter}

\setheadwidth{textwithmarginpar}
\setfootwidth{textwithmarginpar}

\setheadsepline{.2mm}

\KOMAoptions{DIV=last}

\include{Macros}

\titlehead{School of Mathematics \hfill University of Nottingham}
\title{State Sums and Geometry}

\author{Frank Hellmann}
\date{}
\publishers{Thesis submitted to the University of Nottingham for the degree of Doctor of Philosophy\\October 2010}

\begin{document}

\maketitle 

\section*{\centering Abstract}
This thesis studies the construction of state sums from representation theory, and shows how to use geometric asymptotics to understand the leading order behaviour of the state sum weights in geometric terms.

In Chapter 1, I review the axiomatic definition of topological quantum field theories due to Atiyah. I review the definition of topological quantum field theories through state sums on triangulated manifolds. I describe the construction of state sum invariants of 3-manifolds from a graphical calculus and show how to evaluate the invariants as boundary amplitudes. As a particular example I define the Ponzano-Regge state sum through the graphical calculus of SU(2) representation theory.

I review various asymptotic geometricity results in Chapter 2. I define coherent boundary manifolds for state sums based on $\SU(2)$ representations. The geometry of the representations of $\Spin(4)$ and $\SL(2,\C)$ is given in terms of bivectors and spinors. I give a list of conditions necessary and sufficient for a set of bivectors to be the bivectors of a 4-simplex in $\R^4$ and $\R^{3,1}$.

In Chapter 3, I then derive the asymptotic geometry of the $\SU(2)$ based Ponzano-Regge invariant in three dimensions for arbitrary triangulations of a 3-ball using the formulation of the amplitude in terms of boundary amplitudes of Chapter 1 and the coherent boundary manifolds of Chapter 2.

I derive the asymptotic geometry of the 4-simplex weight of the 4-dimensional $\SU(2)\, BF$ Ooguri state sum is derived in Chapter 4, using the coherent boundary manifolds of Chapter 2. As a corollary I give the asymptotic behaviour of various spin foam models for Euclidean 4-dimensional lattice gravity recently proposed.

In Chapter 5 I derive the asymptotic geometry of the recently proposed spin foam model for Lorentzian 4-dimensional lattice gravity. The techniques differ substantially from those in Chapter 3 and 4 in that I make heavy use of spinor representations of $\SL(2,\C)$.

\vfill
\section*{Zueignung}
The results presented in chapters 2-5 were developed in collaboration with John Barrett, Richard Dowdall, Winston Fairbairn, Henrique Gomes and Roberto Pereira. They were published in \cite{Barrett2009a,Barrett2010,Barrett2010a}. The presentation herein is my own, and new except where indicated. The direct derivation of geometricity for the Ooguri model is new. The derivation of the geometricity results in Chapter 4 and Chapter 5 presented here is either new or significantly reorganised and simplified with respect to the papers.
\newpage

\section*{\centering Acknowledgements}

My thanks go out to my parents Ruth and Alfons, my two sisters Anja and Verena and especially to Karen who have supported me throughout my PhD in too many ways to count. I am deeply indebted to all my wonderful friends, in Nottingham and elsewhere. You have made my thesis years not just worthwhile but as enjoyable as any I have had. I am thankful to my friends and colleagues in the community; the discussions over the past years have been an invaluable part of my research and my life. I would like to specifically thank Carlo Rovelli, Jurek Lewandowski and Bianca Dittrich for hosting me during my PhD and for the many inspiring disussions that have resulted from these visits. I would like to thank my collaborators. It has been a joy to work with them over the past years. Finally, special thanks go to John Barrett who has been an outstanding advisor, and whom it has been a great pleasure and inspiration to work with and learn from.

\paragraph{}{\it Thank you all.}

\vfill
\begin{flushright}
 
{\it Ever tried. Ever failed. No matter.\\ Try again. Fail again. Fail better.}\\ -- Samuel Beckett

\end{flushright}

\tableofcontents

\chapter{State Sums}\label{chap-statesum}

We will begin with reviewing topological quantum field theories (TQFTs) as defined by the Atiyah-Segal axioms, and describe how a theory of quantum gravity can fit into this framework. We will then show how to define state sums on triangulated manifolds. We will give a set of algebraic relations that are sufficient to ensure that a state sum defines a TQFT and translate them into a diagrammatic calculus. We show that the representation theory of $\SU(2)$ can be used to define a diagrammatic calculus that satisfies these relations, except a finiteness condition, and thus defines a TQFT up to regularisation.

\section{Topological Quantum Field Theories}

TQFT is at the intersection of various developments in theoretical physics and mathematics over the last decades. Standard quantum field theory (QFT) can currently not be rigorously defined for the cases of interest. Because of this, Atiyah, following Segal's axiomatisation of conformal field theories in \cite{Segalf}, suggested a set of axioms to capture the essential structures of TQFT that make them of interest to mathematicians \cite{Atiyah1989a}. These axioms in turn have spurred the development of a large set of theories that satisfy them, as well as attempts at extending the axiomatisations in natural ways \cite{Baez1995a}.

\subsection{Atiyahs Axioms}

We will start by reviewing Atiyah's axioms of TQFT. This section is not self contained and we refer to the original (very readable) paper for further technical details \cite{Atiyah1989a}. Atiyah's axioms make TQFT an extension or ``categorification" of algebraic topology. Algebraic topology understood most generally is the study of functors from some category of topological spaces to an algebraic category. A TQFT, on the other hand, is defined as a functor from a category in which we interpret manifolds as morphisms between their boundaries to an algebraic category. In our case we will specify the algebraic category to be $\Vect$, the category with objects given by vector spaces and morphisms given by linear maps. Topological manifolds of dimension $n$ with $(n-1)$-dimensional boundary are made into a category by taking as objects the $(n-1)$-dimensional topological spaces and as morphisms the $n$-dimensional manifolds with boundary equal to the spaces in question. We call this category $\nCob^{top}$:

\begin{defi}[n-Cobordisms]
The set of objects of $\nCob^{top}$ is given by the oriented topological $(n-1)$-dimensional closed manifolds which we denote $\bo$. The morphisms between two objects are given by the $n$-dimensional oriented manifolds $\cbo$ with boundary the disjoint union of the source and target objects: $\cbo \in Morph(\bo_1, \bo_2)$ if and only if $\partial \cbo$, equipped with the boundary orientation, is $\bo_1 \cup \bo_2^*\ $, where $\bo^*$ is the manifold with opposite orientation. $\nCob$ is a monoidal category with the tensor product $\tensor$ given by the disjoint union of spaces.
\end{defi}

If the target of cobordism is the source of another they can be composed by identifying the spaces $\bo$ and $\bo^*$. The opposite orientations on the identified parts of the boundary ensure that the orientations on the cobordisms glue correctly, thus we obtain an oriented manifold with source that of the first cobordism and target that of the second.

Note that the components of the boundary ${\bo_{1/2}}$ are allowed to be empty, and the empty set is the unit of the tensor product. The use of topological spaces is not essential, we could just as well use smooth spaces or spaces equipped with a PL structure, and define a ``T"QFT on $\nCob^{diff}$ or $\nCob^{PL}$. Indeed, Atiyah's original definition was given in terms of smooth manifolds. A TQFT is then defined in the following way:

\begin{defi}[TQFT]\label{defi-TQFT}
A TQFT is a monoidal functor $\ZZ$ from a cobordism category $\nCob$ to $\Vect$ subject to the condition that $\ZZ(\bo^*) = \ZZ(\bo)^*$, where $\ZZ(\bo)^*$ is the dual vector space. It is required to be non-trivial in the sense of Atiyah (\cite{Atiyah1989a}).
\end{defi}

Functoriality in particular entails the following properties:

\begin{itemize}
\item $\ZZ(\cbo \cup \cbo') = \ZZ(\cbo) \tensor \ZZ(\cbo')$
\item $\ZZ(\cbo \cup_{\bo} \cbo') = \braket{\ZZ(\cbo)}{ \ZZ(\cbo')}_{\ZZ(\bo)}$ where $\bo \in \partial \cbo$ and $\bo^* \in \partial \cbo'$. $\cup_\bo$ refers to gluing the manifolds by identifying the (components of the) boundary on $\cbo$ and $\cbo'$ and $\braket{}{} _{\ZZ(\bo)} $ is contraction of the indices living in ${\ZZ(\bo)}$ and ${\ZZ(\bo)}^*$ 
\item $\ZZ(\bo \times I) = P \in \ZZ(\bo)\otimes\ZZ(\bo)^* = End(\ZZ(\bo))$ with $P^2 = P$. The non-triviality assumption implies that $P = \id$.
\end{itemize}

$\ZZ(\cbo)$ is usually called the partition function. Using an inner product for composition of morphisms on the $\Vect$ side a similar construction for the unoriented case can also be given.

\begin{figure}[htbp]
\begin{center}
\includegraphics[scale=0.5]{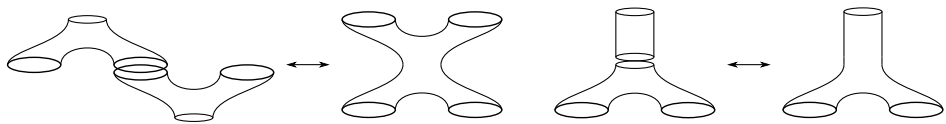}
\caption{Gluing identities}
\label{fig-cbo-gluing}
\end{center}
\end{figure}

These axioms formally capture the essence of the path-integral for quantum gravity. Though gravity is not a topological theory its background independent character implies that the path integral depends only on the smooth structure on the manifold. Heuristically we can think of $\ZZ_{gravity} (\bo_i) = L^2(\eta_i)$, where $\eta_i$ is a 3d metric on the $i$ component on the boundary, and $$\braket{\eta_1, \eta_2, \dots , \eta_n}{\ZZ_{gravity}(\cbo)} = \int_{\partial_i \eta = \eta_i} [D\eta]\exp\left(S_{EH}(\cbo, \eta)\right).$$

\subsection{Boundary Spaces}\label{sec-TQFT-boundaries}

The composition identity of $\ZZ$ across boundary spaces implies a kind of locality in the sense that the amplitude of a complicated cobordism can be calculated by chopping it up into smaller pieces. In the case of 2-dimensional cobordisms the partition function for all manifolds can be obtained simply from knowing the pair of trousers diagram (see diagram \ref{fig-cbo-gluing}) and the disc. In three dimensions the Heegaard splitting decomposes any manifold into socalled handlebodies and it is sufficient to know the partition function of these handlebodies.  This is therefore a remarkably powerful axiom. In the case of the TQFTs which we might hope have relevance to gravity we need as a further condition that we can glue parts of boundaries to produce spaces with new boundary topology. In particular the path integral shown above can formally be composed not just on entire boundaries but on subspaces of boundaries as well.

As an example consider a 3-ball. By identifying two disks on its surface we obtain a solid torus. In fact handlebodies in 3d are exactly the spaces obtained by sequences of such identifications. Thus if we have the ability to glue parts of boundary spaces the 3d theory is defined in terms of only the amplitude of the 3-ball. Cutting up the boundary manifolds along boundaries is an extension to TQFT already suggested in \cite{Atiyah1989a}. It leads to the notion of extended TQFTs considered in \cite{Baez1995a} which has not been fully axiomatized. The theories we will define in this chapter will achieve this gluing in a specific way. We will not consider the general picture further and instead refer the reader to \cite{Baez1995a}.

\begin{figure}[htbp]
\begin{center}
\includegraphics[scale=1]{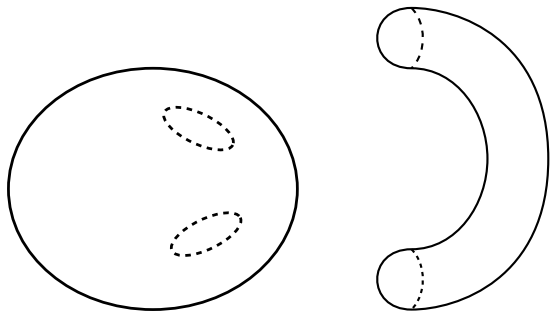}
\caption{Handle meets body}
\label{fig-handlebody-gluing}
\end{center}
\end{figure}

\section{Triangulations, State Sums and Coloured Networks}

The specific TQFTs we will describe are defined in terms of extra structure on the manifold, in particular triangulations. In a further step they are then shown to not depend on these extra structures. A triangulation on a manifold induces a triangulation on its boundary and the vector space will depend on this boundary triangulation. We will describe triangulations and how to relate two triangulations of the same manifold as well as their boundary spaces. We will then define state sums on triangulations and a set of conditions under which they become  independent of the triangulation and define TQFTs. We will then give a diagrammatic way of presenting triangulations and state sums on them. We will express the triangulation independence in terms of a diagrammatic calculus. Finally, we will show how to use $\SU(2)$ to define such a diagrammatic calculus explicitly. This will formally define a TQFT, the Ponzano-Regge model.

\subsection{Triangulations}

A triangulation $\TT$ of an $n$-dimensional manifold $\cbo$ is given by a set of maps from the abstract $n$-simplex $\sigma$ to the manifold such that the union of the images of the maps is $\cbo$ and the intersection of the image of two maps defines a map from a lower dimensional simplex to $\cbo$\footnote{Note that this differs from a $\Delta$-complex in that two $k$-simplices can only be glued on a single $k-1$-simplex.}. A triangulation defines a $PL$-structure on the manifold \cite{nla.cat-vn1972092}. The images of these maps and their intersections define a set of submanifolds of $\cbo$ called the $k$-simplices, $\sigma^k$, of the triangulation. We call the set of $k$-simplices $\TT_{k}$. The sub-simplices of a simplex $\sigma^k$ are the simplices completely contained in it.

Triangulations are convenient for us as we have at our disposal a powerful theorem due to Alexander and Newman (described for example in \cite{nla.cat-vn1972092}) and simplified by Pachner \cite{107898} (see also \cite{Barrett1996}):

\begin{figure}[htbp]
\begin{center}
\includegraphics[scale=1]{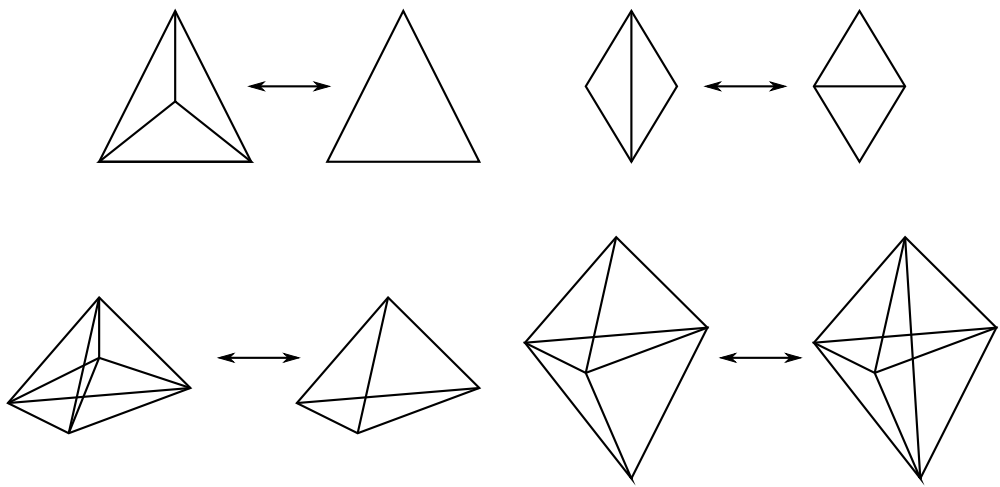}
\caption{Pachner moves in 2 and 3 dimensions.}
\label{fig-PAmoves}
\end{center}
\end{figure}

\begin{theo}[Pachner Moves]\label{Theo-PAmoves}
Any two finite triangulations $\TT$ of an $n$-manifold $\cbo$ defining the same PL structure can be related by a finite number of moves called Pachner moves. These moves are obtained by taking an $(n+1)$-simplex and splitting its surface, which is an $n$-sphere triangulated with $n+2$ $n$-simplices into two $n$-balls. These will be triangulated with $l$ and $n+2-l$ $n$-simplices ($0<l<n+2$). The $l \rightarrow n+2-l$ Pachner move is given by replacing a region of the triangulation $\TT$ isomorphic to the $l$ triangulated $n$-ball with the $n+2-l$ triangulated $n$-ball.
\end{theo}

The $l \rightarrow n+2-l$ Pachner move is the inverse of the $n+2-l \rightarrow l$ Pachner move. For an example of the Pachner moves in 2 and 3 dimensions see Figure \ref{fig-PAmoves}

\subsection{Colourings of Triangulations and State Sums}

In the theories we will consider we will colour the simplices of the triangulation and assign amplitudes to each colouring. That is, we will have functions $c^{k}$ from the simplices $\TT_{k}$ to finite sets of colours $C^{k}$ and amplitudes $f^{k}$ associated to the $k$-simplices that will depend on the colourings of the simplices contained in them as well as, in principle, on the orientations of the simplices though this will not be the case in the theories considered here. We will write $c: \TT \rightarrow C$ for a complete set of colourings of all types of simplices and $c|_{\sigma^k}: \sigma^k \rightarrow C$ for a complete colouring of a $k$-simplex $\sigma^k$ and its sub-simplices\footnote{In the examples we will consider it will actually be convenient to build up the colourings from the lowest dimensional simplices up and let the higher dimensional colouring sets depend on the lower dimensional colours. This structure is naturally captured by thinking of the colourings in a categorical way which we will not pursue here.}. The state sum will then be defined by summing over these colourings. It will then naturally define an element in the linear space $Span (c: \TT_\bo \rightarrow C)$, that is, the span of colourings of the triangulation $\TT_\bo$ of the boundary $\bo = \partial\cbo$. We will then define a state sum $\ZZ(\TT, \cbo)$ depending on the manifolds as well as their triangulations.

\begin{defi}[State sums]\label{defi-statesum}
An $n$-dimensional state sum is defined by a set of colours $C^{k}$ associated to $k$-simplices for $k<n$, and a set of finite amplitudes $f^{k}: \{c|_{\sigma^k}\} \rightarrow \C$ from the colourings of a $k$-simplex and its sub-simplices to the complex numbers. A state sum gives a partition function $\ZZ$ on triangulated manifolds by:
\begin{itemize}
\item $\ZZ(\TT, \bo) = Span (c: \TT_\bo \rightarrow C)$
\item $\ZZ(\TT, \cbo) = \displaystyle \sum_{c} \prod_k \prod_{\sigma^k \in \TT_{k}} f^k(c|_{\sigma^k})$
\end{itemize}
An inner product, in the sense of a bilinear form, on $\ZZ(\TT, \bo)$ is given by the $f^k$-weighted sum over colourings:

\begin{itemize}
 \item $\braket{a}{b} = \displaystyle \sum_{c} \prod_{k=0}^{n-1} \prod_{\sigma^k \in \TT_{k}} f^k(c|_{\sigma^k}) a_c b_c$ for $\ket{a} = \displaystyle \sum_{c} a_c \ket{c}$.
\end{itemize}

\end{defi}

Note that any colouring on the $n$-simplices could be absorbed into the $f^n$ weight. With this inner product distinct colourings are orthogonal. The inner product on a particular colouring is simply the state sum prescription $\displaystyle \sum_{c} \prod_{k=0}^{n-1} \prod_{\sigma^k \in \TT_{k}} f^k(c|_{\sigma^k})$ on the boundary $\bo$. Using this inner product there is a natural gluing of state sums. If $\TT_{\bo'} \in \TT_{\partial\cbo}$ and $\TT_{\bo''} \in \TT_{\partial\cbo'}$ are two identically glued sets of $(n-1)$-simplices we have that
$$\braket{\ZZ(\TT, \cbo)}{\ZZ(\TT', \cbo')} = \ZZ(\TT\cup\TT', \cbo \cup \cbo').$$
Here $\TT\cup\TT'$ and $\cbo \cup \cbo'$ are the triangulation and manifold obtained by identifying the simplices in $\TT_{\bo'}$ and $\TT_{\bo''}$ and the inner product is given by the $f^k$-weighted sum over the colourings of those simplices that are now no longer on the boundary. Thus the inner product simply matches the gluing and supplies the missing internal weights where simplices are identified. Using this inner product we can identify dual spaces with those components of the boundary we have designated as targets, the gluing is then just index contraction again.

As the set of colourings and the amplitudes are finite there is no question of convergence in the sum, at least for finite triangulations. If two components of the boundaries of two manifolds have the same triangulation this immediately defines a gluing in the TQFT sense. Note that this gluing condition is stronger than the one in the TQFT Definition \ref{defi-TQFT} as we have distinguished localized regions. The boundary vector spaces split in a natural way, that is they are local in the sense of Section \ref{sec-TQFT-boundaries}, and we can glue them locally. 
 
This data does not define a TQFT for two reasons. Due to the dependence of the boundary vector space on the triangulation of the boundary of the manifold it is not possible to glue the partition functions defined on manifolds with the same boundary but different triangulations. Furthermore the partition function $\ZZ(\TT, \cbo)$ itself depends on the triangulation of $\cbo$. Both issues can be overcome if the partition functions are compatible with the Pachner moves in a natural way. Call the triangulation of $n$-balls in terms of $l$ $n$-simplices arising in the Pachner moves $\TT^{(l)}$. We then can require the following equivalence of partition functions:

\begin{equation}\label{eq-PAmoveInv}
\ZZ(\TT^{(l)}, B^n) = \ZZ(\TT^{(n + 2 -l)}, B^n) \, \mbox{for all} \, 0<l<n+2.
\end{equation}

\begin{figure}[htbp]
\begin{center}
\includegraphics[scale=1]{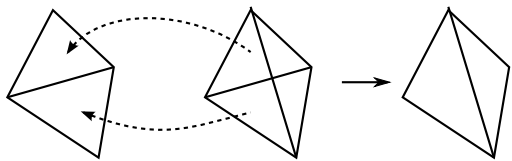}
\caption{The 2-dimensional 2-2 move from gluing a tetrahedron.}
\label{fig-22movefromtet}
\end{center}
\end{figure}

This is well defined as the boundary state spaces on both sides are the same. From Theorem \ref{Theo-PAmoves} and the gluing in Definition \ref{defi-statesum} it now immediately follows that $\ZZ(\TT, \cbo)$ does not depend on the interior triangulation. On the other hand observe that any two triangulations of a particular boundary manifold $\bo$ can be related by a sequence of $(n-1)$-dimensional Pachner moves. Remember from the definition that these are generated by replacing half of an $n$-simplex with its other half. Thus these moves can be generated on the boundary triangulation by gluing the partition function of a full $n$-simplex $\ZZ(\sigma^n, B^n)$ onto the surface on a set of $(n-1)$-simplices (see e.g. Figure \ref{fig-22movefromtet}). In principle such a move might produce a degenerate triangulation; this can always be avoided by choosing an appropriate interior triangulation of $\sigma^n$. A sequence of such gluings gives rise to a manifold of topology $\bo\times I$.

On the other hand if we have a triangulated manifold of the form $\bo\times I$ with the same triangulation on both parts of the boundary the invariance of the partition function under triangulation changes on the interior implies that its partition function is a projector:

$$\braket{\ZZ(\TT, \bo\times I)}{\ZZ(\TT, \bo\times I)} = \ZZ(\TT', \bo\times I) = \ZZ(\TT, \bo\times I).$$

The inner product on the left is the state sum inner product on $\ZZ(\TT, \bo)$. We now have a natural isomorphism between the images of this projector on every possible boundary triangulation given by constructing a triangulation of $\bo\times I$ taking the one to the other. By the argument above such a triangulation always exists due to Theorem \ref{Theo-PAmoves}.

Identifying these spaces under this isomorphism associates a boundary vector space to every boundary irrespective of its triangulation, and using this isomorphism any two triangulations can be glued. In this way a state sum satisfying equation \eqref{eq-PAmoveInv} defines a TQFT.

\subsection{The Dual Networks in 3d and their Evaluation}\label{sec-DualNets}

In order to get a better handle on the kind of amplitudes that satisfy \eqref{eq-PAmoveInv} we will introduce a notation of dual networks for the 3d case. In particular we will obtain a representation of the triangulation and its colouring in terms of 2-dimensional diagrams\footnote{The construction we will consider is not the only way to obtain a state sum invariant by decomposing 3-manifolds and using links and networks to present them, see for example \cite{Foxon:1994nq} for a construction similar to ours and \cite{Roberts93skeintheory} for a construction that clarifies the relationship to other 3-manifold invariants, as well as the references therein.}.

These diagrams are obtained by mapping the 2-dimensional Poincar\'e dual of the surface of every tetrahedron to the plane in an orientation preserving way. This means triangles ($\sigma^2$) get mapped to vertices in the diagram, edges ($\sigma^1$) to links, and vertices of the triangulation ($\sigma^0$) to triangles. If the triangles of two tetrahedra are glued in the manifold we denote this by a dotted line connecting the vertices dual to them in such a way that going clockwise around the one vertex and anticlockwise around the other vertex we will encounter the links associated to the same edges in the same order. This is always possible due to the requirement that the manifold be oriented and the map to the plane preserve the inherited orientation. Therefore the order in which the links go around a vertex will necessarily be opposite for glued triangles.

In this diagram many links will be dual to the same edge. As the ordering at the ends of a dotted line is choosen such that links dual to the same edge are encountered in the same order, the information which links are associated to the same edge is implicit in the unlabeled diagram already. In fact, there is a closed loop of alternating links and dotted lines associated to every edge. However it will be convenient for us to mark the presence of an edge by picking a fiducial link among those dual to it and placing a dot on it. Quantities associated to the edges can then be associated to the dot without having to keep track of many links.

\begin{figure}[htbp]
\begin{center}
\includegraphics[scale=1]{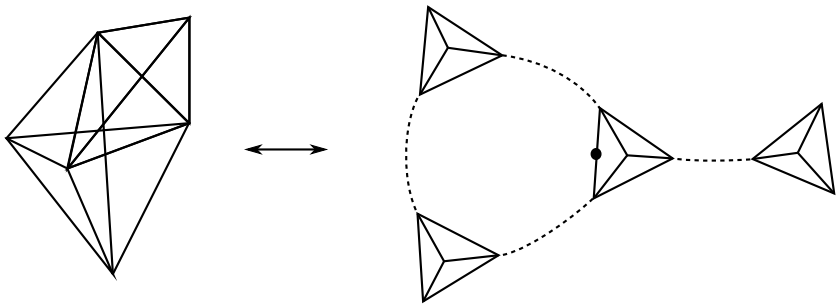}
\caption{A dual network.}
\label{fig-dualnets}
\end{center}
\end{figure}

An example of such a dual network is given in Figure \ref{fig-dualnets}.

We now colour this diagram in the natural way. The colours on triangles go to vertices of the diagram and edges go to links. In the theories we will consider vertices of the triangulation are not coloured so we ignore them here. We can then further treat each individual coloured diagram as the numerical value of the amplitudes $f$ evaluated at the given colouring. These amplitudes will then also be associated to the elements of the diagram, the triangle amplitude to dotted lines, the edge amplitude to the dot on a link, and the tetrahedral amplitude to the tetrahedral net. Of course this means that two diagrams corresponding to the same triangulation must correspond to the same number. The amplitudes cannot depend on the additional choices made in going to the 2-dimensional diagrams. Much of this and the next section will deal with making this condition precise.

The partition function is then given by simply summing the diagrams interpreted as amplitudes over all their colourings. Note that the diagram does not contain information on the vertices of the triangulation in any straightforward manner. This needs to be added by hand. In the cases we want to consider the vertices are not coloured and their weight is a constant. The gluing of partition functions is immediate.

\begin{figure}[htbp]
\begin{center}
\includegraphics[scale=1]{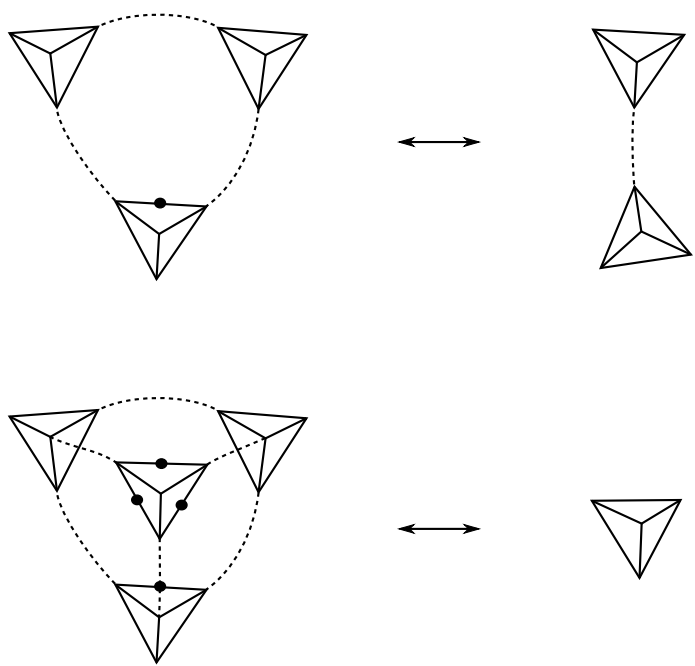}
\caption{The dual Pachner moves.}
\label{fig-dualmoves}
\end{center}
\end{figure}

The Pachner moves in terms of these dual nets are presented in Figure \ref{fig-dualmoves}.

\begin{figure}[htbp]
\begin{center}
\includegraphics[scale=.8]{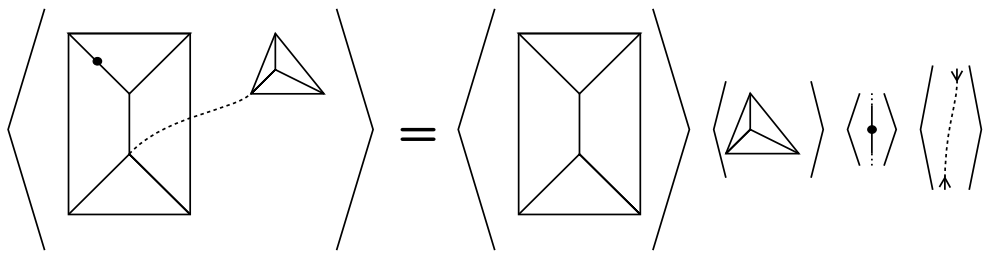}
\caption{The local evaluation of a network.}
\label{fig-evaluation}
\end{center}
\end{figure}

A particular way to define amplitudes satisfying the Pachner identities \eqref{eq-PAmoveInv} is given by taking the diagrammatic representation of the triangulation and the partition function seriously. That is, we define various graphical manipulations on the diagrams that are sufficient to establish the equality of the diagrams in Figure \ref{fig-dualmoves}. If we then have a consistent way to associate amplitudes with not just the diagrams that arise in Figure \ref{fig-dualnets}, but also those obtained through the graphical manipulation, these graphical manipulations can be translated back directly into equations. We will encapsulate a diagram in triangular brackets $\la\cdot\ra $ to denote its evaluation. That is $\la\cdot\ra : \text{coloured diagrams} \rightarrow \C$. As we want to identify the evaluation of the diagram with the amplitudes in the state sum, and the amplitudes are associated to particular parts of the diagrams, we require the evaluation to be local in the sense that each disconnected network of links, each dotted line and each dot on a link can be evaluated individually and multiplied together, e.g. see Figure \ref{fig-evaluation}.  We will usually suppress the colourings in the evaluation, in which case a sum over colourings is implied.

\begin{figure}[htbp]
\begin{center}
\includegraphics[scale=.8]{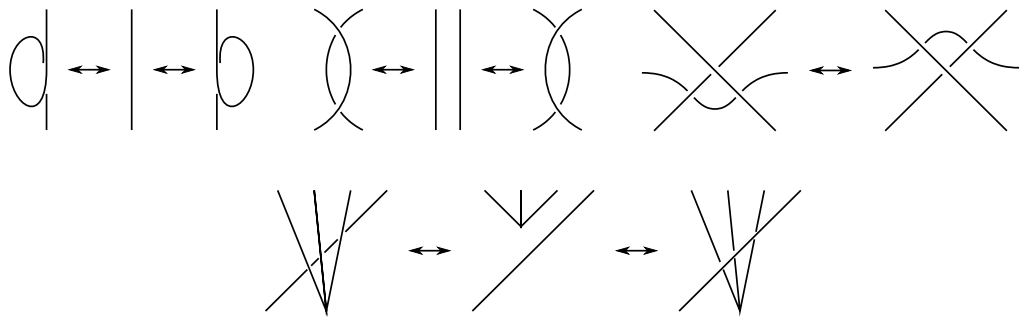}
\caption{The Reidemeister moves I, II, III in the first line and the vertex move. We have omitted the third Reidemeister move with the opposite crossing.}
\label{fig-Reidemeistered}
\end{center}
\end{figure}

The condition that the amplitudes should not depend on the particular diagrams chosen to represent the triangulation can be expressed as its invariance under deformations relating all different diagrams arising from the same triangulation. In particular we want to identify different ways of projecting the tetrahedra onto the plane. This means we need invariance under diffeomorphisms in the plane as well as the ability to move lines ``past the point at infinity''. Furthermore we want to allow crossing lines, dotted lines and vertices under and above each other. This can be done by allowing the Reidemeister moves extended by operations including the vertices of the diagrams of Figure \ref{fig-Reidemeistered}. Taken together the Reidemeister moves and the diffeomorphisms on the sphere are called 3-dimensional ambient isotopy. The diffeomorphisms themself are the ambient isotopy of the sphere, invariance under these would be sufficient to construct state sums and led to the definition of spherical categories \cite{Barrett1999,Barrett1996}. The condition of invariance under 3-dimensional ambient isotopy can be translated into specific equalities on the map $\la\cdot\ra $ which we will study in greater detail in the next section. As an immediate consequence, this implies that $\la\cdot\ra $ defines a knot invariant (e.g. \cite{Kauffmang}).

\begin{figure}[htbp]
\begin{center}
\psfrag{1}{$1$}
\psfrag{2}{$2$}
\psfrag{d}{\large$\delta_{12}$}
\psfrag{-1}{$-1$}
\includegraphics[scale=1]{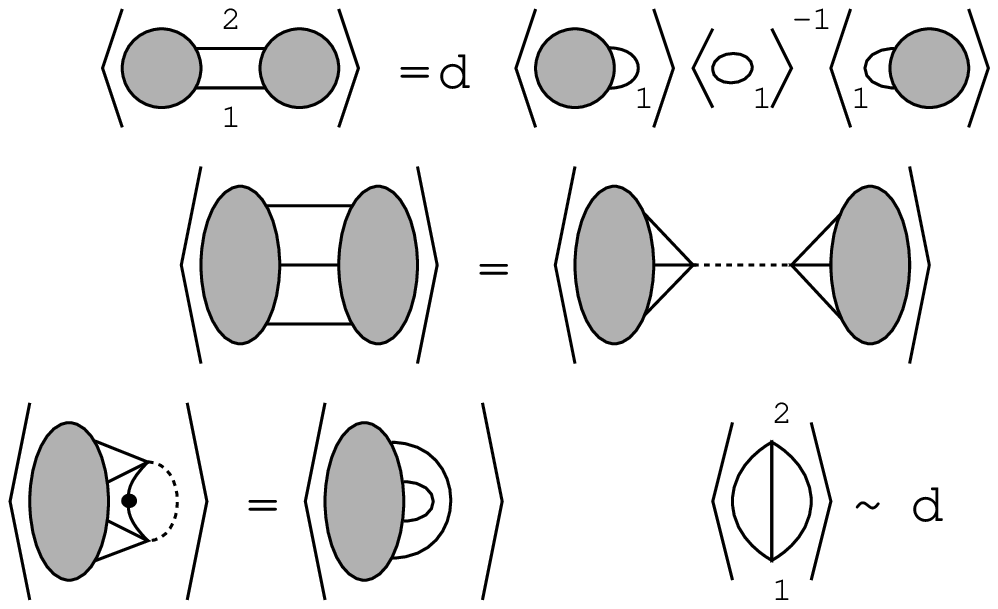}
\caption{Invariances in the top two lines, semi-simplicity and orthogonality in the last line. The grey circles are arbitrary diagrams. 1 and 2 are colourings of the links in the first equation and of the vertices in the last one.}
\label{fig-InvSemiSimple}
\end{center}
\end{figure}

Using ambient isotopy the dual Pachner moves in Figure \ref{fig-dualmoves} can be recovered from the first three identities given in Figure \ref{fig-InvSemiSimple} that act locally in the diagrams evaluated. We call them two- and three-valent invariance and semi-simplicity for reasons that will become apparent. The fourth identity in Figure \ref{fig-InvSemiSimple} is called orthogonality. Invariance applies when two or three links or a dotted line connect two separate networks of links, semi-simplicity applies always. As per the convention above we have marked the presence of an edge factor for the closed loop by placing a thick dot along one of its links. Note further that we are implicitly summing over labels in the diagrams. In particular for semi-simplicity we are summing over the edge colouring of the closed loop and the triangle colouring at the ends of the dotted line. In the case of three-valent invariance we sum over the triangle colouring. There is no sum in two-valent invariance. The colourings summed over are local in the part of the diagram we are changing. They do not appear anywhere else in the state sum and are removed by the moves. The fourth equality in Figure \ref{fig-InvSemiSimple}, orthogonality, does not involve a sum but simply relates two labels on different vertices of the network.

\begin{figure}[htbp]
\begin{center}
\includegraphics[scale=.89]{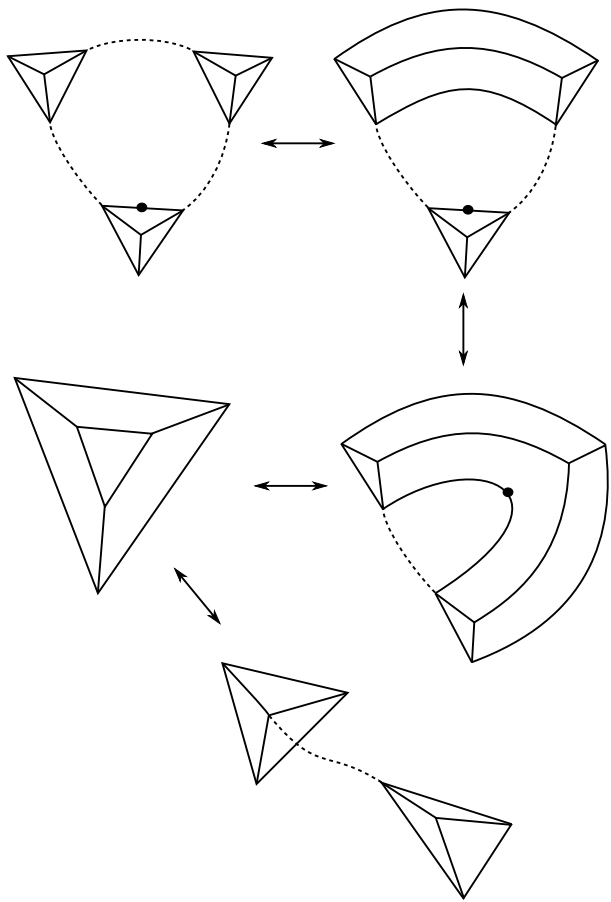}
\caption{2-3 Pachner move from invariance and semi-simplicity.}
\label{fig-PAprovediagrams2-3}
\end{center}
\end{figure}

\begin{figure}[htbp]
\begin{center}
\includegraphics[scale=.89]{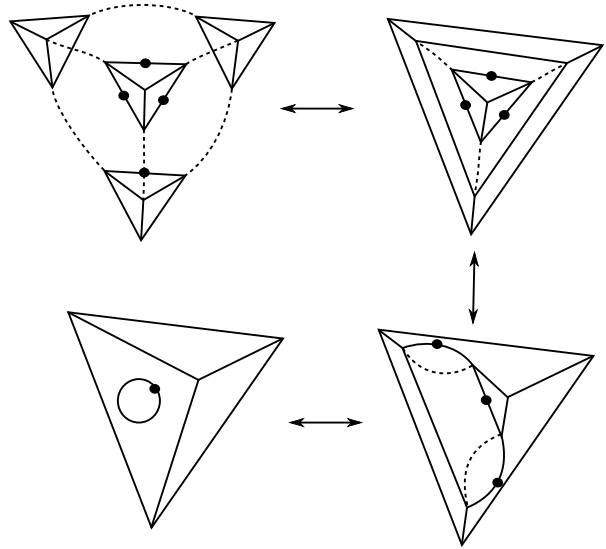}
\caption{4-1 Pachner move from invariance and semi-simplicity.}
\label{fig-PAprovediagrams1-4}
\end{center}
\end{figure}

The sequence of manipulations that allows one to recover the Pachner moves from invariance and semi-simplicity is given in Figure \ref{fig-PAprovediagrams2-3} for the 2-3 (3-2) move and in Figure \ref{fig-PAprovediagrams1-4} for the 1-4 (4-1) move. As mentioned it does not require orthogonality.

Note that the 1-4 move is only implemented up to a factor given by the completely disconnected diagram. This is a reflection of the fact that we did not include the vertices $\sigma^0$ of the triangulation in the diagrams and the 1-4 move creates or destroys a $\sigma^0$. Conversely we can take this calculation to determine the constant weight of a vertex in terms of link diagrams. Thus if the state sum is normalized by the factor $\prod_{\sigma^0} f^0 = (f^0)^{|\TT_0|}$ it is invariant under the 1-4 move.

\begin{figure}[htbp]
\begin{center}
\newcommand{\svg}{{\small $\,j$}}
\def\svgwidth{\columnwidth}
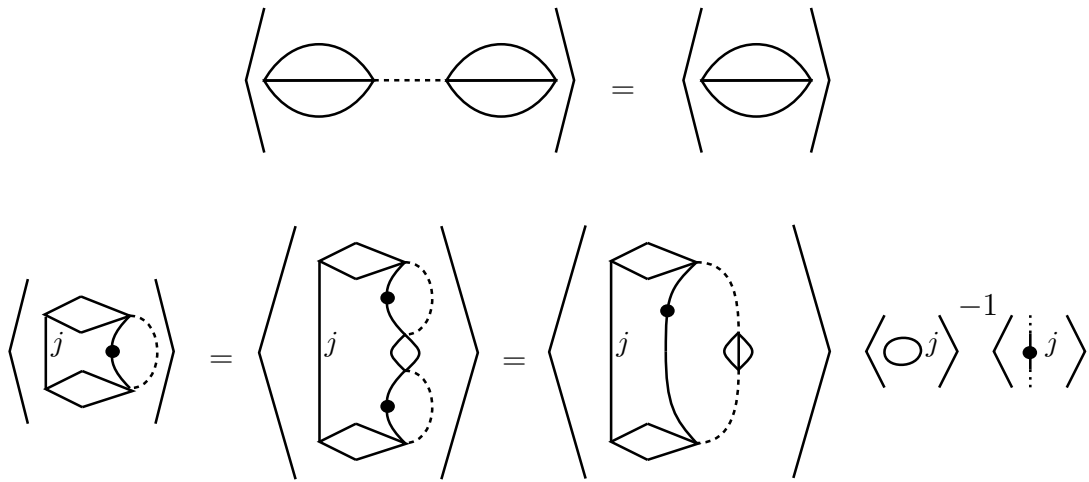
\caption{Lower dimensional amplitudes from link networks.}
\label{fig-lowerampsdiag}
\end{center}
\end{figure}

In fact if orthogonality holds, not only the factors $f^0$ can be expressed as networks of links through these moves, but also $f^1$ and $f^2$. To see this consider the equality of diagram evaluations in Figure \ref{fig-lowerampsdiag}. On the left hand side of the first equality we have two dotted lines and one network called the $\Theta$ network, on the right hand side we have one $\Theta$ network. As by definition we defined a dotted line to evaluate to the weight $f^2$, we conclude that $(f^2)^{-1}$ must equal the value assigned to the $\Theta$ network. Using this we can compare sides on the other equality to conclude that the thick dot representing the weight $f^1$ has to equal the value associated to the circle. Using this we can then conclude that $(f^0)^{-1}$ is the sum over all colourings of the evaluation of a circle squared. We could thus remove all dotted lines and thick dots from the diagram replacing them purely with networks of links and placing the triangle labels on the vertices. Denoting the tetrahedral network as $\tet$, the theta network as $\thet$ and the circle network as $\ci$ we can thus inversely define the partition function in terms of network evaluations:

\begin{defi}[State Sum of a Network Evaluation]\label{defi-graphicalstatesum}
Given a bracket evaluation $\la\cdot\ra$ on coloured networks invariant under ambient isotopy of the sphere we have an associated state sum defined by: 
\begin{equation}
\ZZ(\TT ,\cbo) = \left(\sum_{c^1} \la \ci\ra ^2\right)^{-|\TT_0|} \sum_{c^i} \prod_{\sigma^1} \la \ci\ra  \prod_{\sigma^2} \frac{1}{\la\thet \ra } \prod_{\sigma^1} \la \tet\ra  
\end{equation}
\end{defi}

If the evaluation $\la\cdot\ra$ furthermore satisfies the four moves of Diagram \ref{fig-dualmoves} this defines a TQFT on $\nCob^{PL}$.

\subsection{The Partition Function as a Dual Network}

In fact for a TQFT of the type defined in \ref{defi-graphicalstatesum} we can use the graphical calculus to evaluate the partition function of any triangulation of the ball $B^3$ in terms of the dual network of its boundary triangulation. As per our discussion above we can construct a triangulation of $S^2 \times I$ interpolating between the triangulation of the sphere induced by the triangulation of the 3-ball and the triangulation of the sphere given by the boundary of a single tetrahedron. We do this by gluing tetrahedra onto the 3-ball iteratively on one, two or three faces, generating the 2-dimensional Pachner moves. Now observe that the network obtained by eliminating the internal simplices created by the gluing is again the network of links dual to a planar projection of the surface of the new triangulation. Iterating this procedure we obtain that in general the basis element of $\ZZ(\TT^3, B^3)_{c}$ in a particular colouring $c: \partial\TT^3 \rightarrow C$ of the boundary is given by the number associated to the coloured network of links dual to boundary of $\TT$. That is, we have the following theorem:

\begin{theo}[Dual Network Evaluation of the Partition Function]\label{theo-dualevalofPF}
The partition function $\ZZ(\TT^3,B^3)_{c}$ of a TQFT given by Definition \ref{defi-graphicalstatesum} can be expressed as 
\be
\ZZ(\TT^3,B^3)_{c} = \la(\partial\TT^3)^*_{c}\ra\;,
\ee
where $(\partial\TT^3)^*_{c}$ is the coloured network obtained as the Poincar\'e dual of a mapping of $(\partial\TT^3)$ to the plane.
\end{theo}

This algorithm was first developed by Moussouris in \cite{Moussouris} and an extension to handle bodies for the case of the Ponzano-Regge model was introduced in \cite{Dowdall2010}.

\section{$\SU(2)$ and Spin Networks}

We will now show how to construct a diagrammatic calculus satisfying all the above conditions from the representation theory of $\SU(2)$. This is only a concrete example. The conditions discussed above have been translated completely into algebraic language and many categories satisfying them are known. The theory based on $\SU(2)$ is particularly interesting for physics in that we can find strong relations with geometry, as we will make explicit using coherent states in the next chapter.

The construction of a finite TQFT in terms of the $\SU(2)$ theory is difficult as the colouring set is not finite and the theory requires regularisation. We will not consider these problems here and refer the reader to the literature \cite{Barrett2006,Barrett2009,Freidel2004,Freidel2004a}.

\subsection{A Brief Review of $\SU(2)$ Representation Theory}\label{sec-su2reps}

The group is $\SU(2)$, the group of complex 2-dimensional unitary matrices $g$ of unit determinant. It is a 3-dimensional group generated by the Lie algebra $\su(2)$ with three generators $L^i$ and commutation relations $[L^i,L^j] = - \epsilon^{ijk} L^k$. The generators are a basis of the traceless anti-Hermitian 2-dimensional matrices. The fundamental representation on $\C^2$ is called the spin $\frac12$ representation and we denote states in $\C^2$ by $\alpha^A$ with $A$ running over $\{0,1\}$. We will at times find it convenient to use the Hermitian Pauli matrices $\sigma_P^i$ which are related to the $L^i$ by $L^i = \frac{i}{2} \sigma_P^i$.

The unitary irreducible representation are labelled by half integers $j$ and are given by the symmetric subspaces of the tensor product of $2j$ fundamentals. They are then $(2j+1)$-dimensional and we write their states as $$\alpha^{(A_1 A_2  \dots A_{2j})} = \alpha_j^A$$ where the $A_i = 0,1$ and $A$ now runs from $0$ to $2j$. The Lie algebra elements act as $$L^i_j = L^i\tensor 1 \tensor \cdots \tensor 1 + 1\tensor L^i \tensor \cdots \tensor 1 + 1\tensor 1 \tensor \cdots \tensor L^i.$$

We will drop the index $j$ when it is clear from context which representation we are talking about. By definition the Hermitian inner product on these representations is invariant under the group action: $\braket{g \alpha}{g \alpha'} = \braket{\alpha}{\alpha'}$. Explicitly we write $\braket{\alpha}{\alpha'} = \delta_{AB} \overline{\alpha^A} \alpha'^B$ where the overline denotes complex conjugation and $\delta_{AB}$ is the diagonal Kronecker tensor which is $1$ if $A=B$ and $0$ otherwise. Its inverse is the same and they contract to the Kronecker delta: $\delta^{AB} \delta_{BC} = \delta^A_C$. 

Furthermore, there is a bilinear inner product on the representations. In the fundamental representation it is given by the antisymmetric tensor $\epsilon_{AB}$. We also define its antisymmetric inverse through $\epsilon^{AB} \epsilon_{BC} = \epsilon_{CB} \epsilon^{BA} = \delta^A_B$. This is invariant as $g^A_{\phantom{A}A'} g^B_{\phantom{B}B'} \epsilon_{AB}$ is antisymmetric and thus $g^A_{\phantom{A}A'} g^B_{\phantom{B}B'} \epsilon_{AB} \propto \epsilon_{A'B'}$. This is in fact an equality as contracting both sides with $\epsilon^{A'B'}$ gives $-2 \det(g) = -2 \det(\delta) = -2$. Thus we have $g^A_{\phantom{A}A'} g^B_{\phantom{B}B'} \epsilon_{AB} = \epsilon_{A'B'}$. We write $$(\alpha, \alpha') = \alpha^A \alpha'^B \epsilon_{AB}.$$

This can be defined on any representation by symmetrizing indices, that is $$ (,)_j = \epsilon_{(A_1 A_2  \dots A_{2j}) (B_1 B_2  \dots B_{2j})} = \epsilon_{(A_1(B_1} \epsilon_{A_2 B_2} \dots \epsilon_{A_{2j})B_{2j})}.$$ Note that this is now graded anti-symmetric, that is, $(\alpha, \alpha')_j = (-1)^{2j} (\alpha',\alpha)_j$. We therefore have the graded trace $(O_j)^A_B {\epsilon_j}_{AC} {\epsilon_j}^{BC} = (-1)^{(2j)} \tr{(O_j)}.$

The bilinear inner product can be related to the Hermitian inner product through an anti-linear operator that commutes with the group action $J$. It is defined as $(J \alpha)^A = \overline{\delta^{AA'} \epsilon_{A'B} \alpha^{B}}$. We then have $$\braket{J \alpha}{\alpha'} = (\alpha,\alpha')$$ as $\braket{J \alpha}{\alpha'} = \delta^{AB} \epsilon_{BB'} \alpha^{B'} \delta_{AA'} {\alpha'}^{A'} = \epsilon_{A'B'} \alpha^{B'} {\alpha'}^{A'}$. It follows that $J$ has to commute with the group action. This can be checked explicitly as $$\overline{\delta^{AB} \epsilon_{BB'} g^{B'}_{\phantom{B'}C} \alpha^{C}} = \overline{\delta^{AB} (g^\dagger)^{C'''}_{\phantom{C'''}B} \delta_{C'''C''}}\; \overline{\delta^{C''C'} \epsilon_{C'C} \alpha^{C}} = g^{A}_{\phantom{A}C''} \overline{\delta^{C''C'} \epsilon_{C'C} \alpha^{C}}.$$

The action of $J$ immediately extends to all representations. In particular note that $J^2 = (-1)^{2j}$ and that using the graded antisymmetry of $(,)$ we have that $\braket{J \alpha }{J \alpha' } = \overline{\braket{\alpha}{\alpha'}}$.

Furthermore we will need the invariant subspaces of tensor products of representations: $\Inv(j_1 \tensor j_2 \tensor \dots \tensor j_n)$. Its elements $\iota \in \Inv(j_1 \tensor j_2 \tensor \dots \tensor j_n)$ satisfy $g \iota = \iota$ and are called intertwiners. $\Inv(j_1 \tensor j_2)$ is empty unless $j_1 = j_2$ in which case it is 1-dimensional with a basis given by $\epsilon^{AB}$. $\Inv(j_1 \tensor j_2 \tensor j_3)$ is empty unless $j_1$, $j_2$ and $j_3$ satisfy the triangle inequalities and $j_1 + j_2 + j_3$ is integer. In this case we say $j_1$, $j_2$ and $j_3$ are admissible. We define the function $\ad (j_1,j_2,j_3)$ to be $1$ if its arguments are admissible and $0$ otherwise.

Using the epsilon tensor we can construct the invariant maps between tensor products of Hilbert spaces $\Hom(j_1\tensor j_2 \dots \tensor j_n, j_{n+1}\tensor j_{n+2} \dots \tensor j_m)$ by contracting the first $n$ indices of an element in $\Inv(j_1 \tensor j_2 \dots \tensor j_m)$. As epsilon has an inverse this is a natural isomorphism of vector spaces. In this way the statement about $\Inv (j_1 \tensor j_2)$ is equivalent to Schur's Lemma for $\Hom(j_1,j_2)$.

Using the maps $\Hom(j_1\tensor j_2, j_3)$ for admissible spins we can decompose $j_1 \tensor j_2$ into representations $j_3$. Semi-simplicity of the group tells us that this gives the whole representation space and in the case of $\SU(2)$ each $j_3$ occurs only once in the decomposition. We can use the invariant tensors to project onto the subspace of $j_3\tensor j_2$ isomorphic to $j_1$: $$P^{j_3\tensor j_2}_{j_1} = N^{j_3\tensor j_2}_{j_1} \iota'^{B_3 B_2 B_1} \epsilon_{B_1 A_1} \iota^{A_1 A_2' A_3'} \epsilon_{A_3' A_3} \epsilon_{A_2' A_2} $$ where the normalisation can be calculated using Schur's Lemma and is given by $$N^{j_3\tensor j_2}_{j_1} = \frac{{\epsilon_{j_1}}_{AB} \epsilon_{j_1}^{AB}}{\iota'^{B_3 B_2 B_1} \epsilon_{B_1 A_1} \epsilon_{B_2 A_2} \epsilon_{B_3 A_3} \iota^{A_1 A_2 A_3}}.$$ Semi-simplicity is then the statement that the tensor product of representations can be decomposed into a direct sum of representations:

\begin{equation}\label{eq-semisimple}
\sum_{j_3: ad(j_1, j_2, j_3)=1} P^{j_1 \tensor j_2}_{j_3} = \id_{j_1\tensor j_2}.
\end{equation}

\subsection{Spin Networks and Graphical Calculus}\label{sec-SpinNets}

We will now turn to the question of how to use the representation theory of $\SU(2)$ to construct a graphical evaluation $\la\cdot\ra $ of the type described in Section \ref{sec-DualNets}. The graphical calculus will correspond to the one described in \cite{KauffmanLins} for parameter $A=-1$ and with generalised intertwiner normalisations.

\begin{figure}[htbp]
\begin{center}
\includegraphics[scale=1]{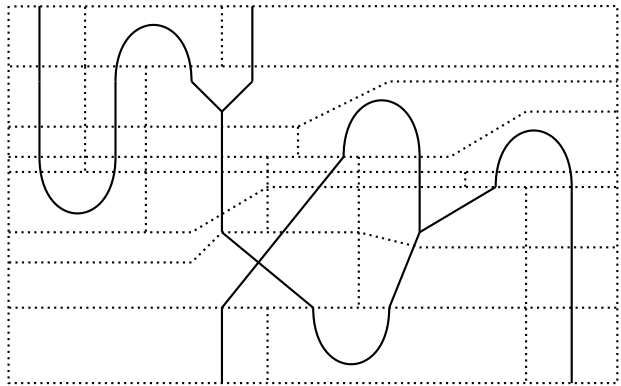}
\caption{The diagrammatic composition.}
\label{fig-diagramsinboxes}
\end{center}
\end{figure}

The construction we will use is in fact much more general than needed here. However, it is illustrative as it covers several other state sum TQFTs like the Turaev-Viro model \cite{Turaev1992} that are based on the graphical calculus for generic $A$. As we saw above it is actually sufficient to consider only the evaluation of links. We can then chop the link diagrams up into elementary blocks which can be combined in natural ways. That is, we draw a box around a particular part of the diagram in such a way that only the top and bottom of the box intersect with links. This defines a diagrammatic category where objects are points on the boundary of the square and morphisms are given by ways of connecting them up. These boxes can then be composed in a natural way. These kind of diagrammatic categories have been well studied in the literature \cite{MR1020583,MR1113284,Barrett1999}. We can arrange the diagram in the box in such a way that all vertices are either at the top or the bottom of the box. The composition of morphisms is then given by stacking boxes vertically and there is a natural tensor product given by lining boxes up horizontally and erasing the boundary between them, for an examples see Figure \ref{fig-diagramsinboxes}. Composition in this diagram can be thought of as erasing dotted lines.

If we want this calculus to be invariant under different ways of cutting the diagram into boxes, it follows that a box with no incoming or outgoing lines has to be a map from the identity of the tensor product of spaces to itself. If the set of such maps is abelian, such a category immediately defines an evaluation $\la\cdot\ra $.

In our case we want to associate boxes with linear operators between representation spaces of $\SU(2)$. The colouring set for links and therefore on the boundary of the boxes will be the set of irreducible representations $j$ of $\SU(2)$. Furthermore we will colour vertices with a basis of $\Inv(j_1 \tensor j_2 \tensor j_3)$ with an arbitrary but fixed ordering of spins. We require this basis to be orthogonal in the epsilon inner product. As $\Inv(j_1 \tensor j_2 \tensor j_3)$ is 1 or 0-dimensional this is simply given by a normalisation and phase in the former case. We will interpret the former case as setting the contribution of any non-admissible colouring to be identically zero. Thus the intertwiner colouring restricts the colouring of the edges of the network. The admissibility conditions and the ordering we have to choose will be crucial. The ordering is required as the different possible orderings will be related by nontrivial isomorphisms.

As argued above we will have to show that the different way of cutting up the diagrams are equivalent and the assignment is invariant under ambient isotopy of the sphere. We do this by chopping up every diagram into six elementary diagrams using a simple form of Morse theory. Then we associate particular $\SU(2)$ invariant operators to the elementary diagrams. We then give a set of relations that relate any two ways of chopping up any two regular isotopic diagrams and show that the elementary operators satisfy corresponding equations.

\begin{figure}[htbp]
\begin{center}
\includegraphics[scale=1]{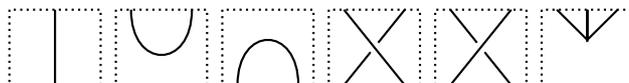}
\caption{The elementary diagrams.}
\label{fig-elementarydiagrams}
\end{center}
\end{figure}

To decompose an arbitrary diagram into boxes we use the standard height function on the plane of the diagram. We will require the diagram to have finitely many minima, maxima, intersections and vertices with respect to this height function. We choose a set of equipotential surfaces such that the interval between two consecutive surfaces contains at most one minimum, maximum, vertex or intersection. This can always be done by perturbing the height function slightly. Within each of these strips we will now have several vertical lines and exactly one of the other 5 types of diagrams in Figure \ref{fig-elementarydiagrams}. As per above we can thus consider this the tensor product between the operators associated to the vertical line and the elementary operators.

Now reading the diagrams from bottom to top, and associating lower indices with incoming representations, and upper indices with outgoing representations we make the following identifications:

\begin{itemize}
\item vertical line: ${\delta_j}^A_B$
\item cup: ${\epsilon_j}^{AB}$
\item cap: ${\epsilon_j}_{AB}$
\item overcrossing: $j_1 \tensor j_2 \rightarrow j_2 \tensor j_1: \alpha \tensor \beta \rightarrow (-1)^{4j_1j_2}\; \beta \tensor \alpha$
\item undercrossing: $j_1 \tensor j_2 \rightarrow j_2 \tensor j_1: \alpha \tensor \beta \rightarrow (-1)^{4j_1j_2}\; \beta \tensor \alpha$
\item vertex with all legs up and in order of the colouring: The basis element of $\Inv(j_1 \tensor j_2 \tensor j_3)$ it is coloured with.
\end{itemize}

We will return to the definition of the other forms of the vertex in a moment. Note that the over and under crossing need to carry these minus signs for consistency as the epsilon inner product is graded anti-symmetric. For example see the first Reidemeister move which would only be true up to sign with the trivial crossing.

\begin{figure}[htbp]
\begin{center}
\includegraphics[scale=.9]{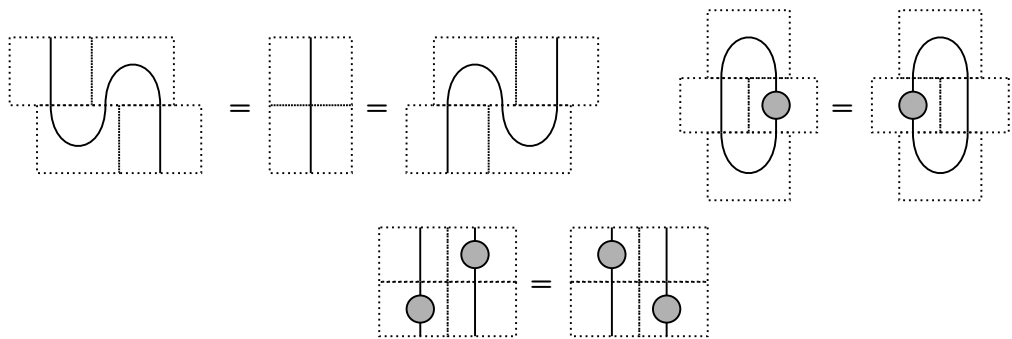}
\caption{Regular isotopy diffeomorphisms 1.}
\label{fig-regisodiff1}
\end{center}
\end{figure}

The conditions of ambient isotopy on the sphere now translate into various equations satisfied by these operators. Diffeomorphisms that don't introduce new minima or maxima can only translate operators in the diagram. They can only translate horizontally if there is no outgoing line on the boxes. In this case the fact that the Hilbert space associated to the box side without lines is $\C$, the identity of $\tensor$, ensures that these shifts don't change the overall operator. If they change the vertical ordering they change the order in which operators are multiplied, but since the operators will by necessity be acting on different spaces and be tensored with the identity on the others space this does not change the overall operator. It is clear then that diffeomorphisms of the graph away from the vertices and intersections do not change the associated operator unless they induce new maxima and minima or move a line past the point at infinity. This corresponds to the first two deformations in Figure \ref{fig-regisodiff1}.

The corresponding equations for the first equation in Figure \ref{fig-regisodiff1} are simply the inverse relations for $\epsilon^{AB}$. Moving a line past the point at infinity follows from the graded anti-symmetry of the cup and the cap. Note that this diagram evaluates to the graded trace of the unspecified section of the diagram. Next we consider how operators interact. The equations of the Reidemeister moves, including the one using the vertex, that were given in Figure \ref{fig-Reidemeistered} deal with the situation where we move intersections and lines past each other. It is a straightforward calculation that they are satisfied. As we are not distinguishing between under and overcrossings here the calculus is actually more constrained as no braiding can occur. Moving the vertex past a line only adds a set of minus signs which cancel due to the integrality condition of spins at an intertwiner.

\begin{figure}[htbp]
\begin{center}
\includegraphics[scale=1]{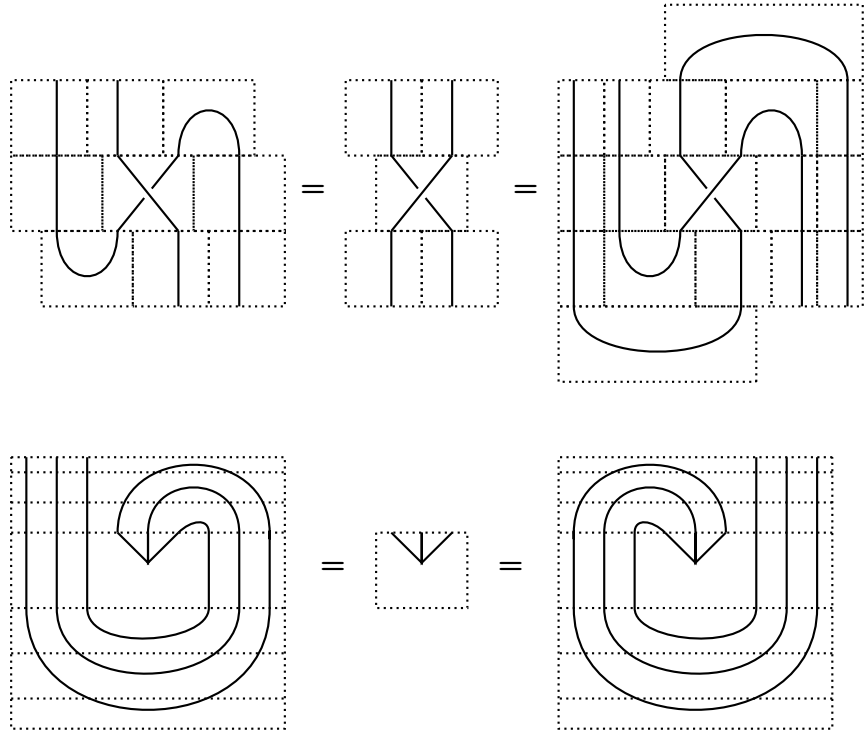}
\caption{Regular isotopy diffeomorphisms 2.}
\label{fig-regisodiff2}
\end{center}
\end{figure}

At the vertices and intersections diffeomorphisms can also act by a twist. This will introduce new minima and maxima. The twists that relate the elementary operators defined above are given in Figure \ref{fig-regisodiff2}. The ones involving the crossings are easily verified.

\begin{figure}[htbp]
\begin{center}
\includegraphics[scale=1]{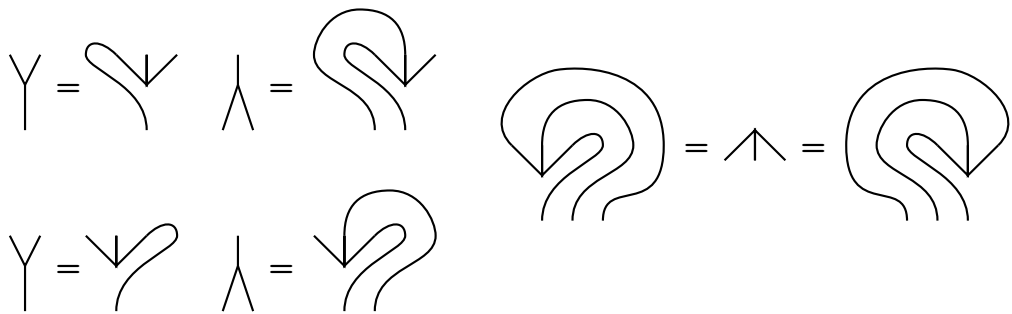}
\caption{The twisted vertices.}
\label{fig-generalvertices}
\end{center}
\end{figure}

To fully understand the twists acting at the vertices we need to define the operators associated to vertices which are not of the form for which we defined the elementary operator above but have links going down. Notice that as in the examples in Figure \ref{fig-generalvertices} we can use twists in the vertex and caps to generate all the possible vertices with all orderings. This corresponds to the natural isomorphism between $\Inv(j_1\tensor j_2\tensor j_3)$ and $\Hom(j_1\tensor j_2,j_3)$ defined above and the isomorphism between $\Inv(j_1\tensor j_2\tensor j_3)$ and $\Inv(j_2\tensor j_1\tensor j_3)$ defined by the crossing. We will then define a vertex coloured by an intertwiner to be the operator obtained by twisting the diagram until the vertex has all upward facing legs in order $j_1$, $j_2$, $j_3$ or $j_2$, $j_1$, $j_3$, and using the crossing map in the latter case.

It remains to show that all ways of twisting the vertex lead to the same result. We can twist a vertex either left or right or more than once. However, it follows from the second set of equalities of diagrams in Figure \ref{fig-regisodiff2} that any two ways to do so actually are the same. To see this note that a consecutive partial left and partial right twist cancel each other out. If we obtain a particular ordering of outgoing legs by twisting to the right we can thus replace the central vertex with its fully left twisted version and cancel twists to obtain the required ordering through left twists. The equalities of Figure \ref{fig-regisodiff2} are straightforward to show. Every full twist is $(-1)^{2j}$ times the identity, thus we simply have a sign factor of $(-1)^{2j_1 + 2j_2 + 2j_3}$ which due to the integrality of spins at an intertwiner is equal to $1$. As we have used the twists to define the general vertices it follows in particular that the operators associated to a diagram are invariant under twisting the vertices.

\subsubsection{The Ponzano-Regge Model}

This concludes the definition of the graphical calculus. We have established its invariance under regular isotopy. We can now associate a complex number to every closed diagram that is invariant under the way the diagram is presented. This in turn defines a state sum model $\ZZ_{PR}$ by Definition \ref{defi-graphicalstatesum}. Evaluating the diagrams for a particular colouring in which the intertwiners are ordered from left to right as they appear in the diagram we obtain the following numerical values in terms of $\SU(2)$ operators:

\begin{itemize}
\item $\la\tet\ra = \iota^{A_1 A_2 A_3} \iota^{A_3' A_4 A_5} \iota^{A_4' A_2' A_6} \iota^{A_6' A_5' A_1'}\times$

$\phantom{\la\tet\ra = \iota^{A_1 A_2 A_3}} \times \epsilon_{A_1 A_1'}\epsilon_{A_2 A_2'}\epsilon_{A_2 A_2'}\epsilon_{A_3 A_3'}\epsilon_{A_4 A_4'}\epsilon_{A_5 A_5'}\epsilon_{A_6 A_6'} $
\item $\la\thet\ra = (-1)^{4(j_1j_2+j_1j_3+j_2j_3}(\iota,\iota')_{j_1\tensor j_2\tensor j_3}$
\item $\la\ci\ra = (-1)^{2j} (2j+1)$
\end{itemize}

To arrive at the formula for the $\tet$ net we represent it in a particular way and use the associations defined above. It follows from our discussion that while this formula depends on the way we have proceeded, the result doesn't and in particular has all the symmetries of an actual tetrahedron. Note however that the signs here do depend on the ordering chosen at each vertex. The overall state sum does not depend on the orderings chosen at each vertex as the changes to signs induced by a change of choice of ordering in $\la\theta\ra^{-1}$ would compensate the change of signs in $\la\tet\ra$.

In the case of $\SU(2)$ there are actually canonical intertwiners for each ordering which can be used to eliminate the need for choosing an ordering and a particular normalisation. As our analysis in the following chapters will not use this canonical normalisation we will not discussed it here.

To show that these weights actually define a topological quantum field theory we need to establish the four additional moves, semi-simplicity, the two invariances, and orthogonality. These are simply graphical restatements of various group theoretic relations stated in Section \ref{sec-su2reps}. Semi-simplicity is just the sum over projector decomposition of the identity on the tensor product of representations given in equation \eqref{eq-semisimple}. Three-valent invariance is the observation that the pair of identical intertwiners projects onto the invariant subspace of the three representations but the state coming in is already invariant, being a contraction of invariant operators, thus it can be replaced by the identitiy. Two-valent invariance is equivalent to Schur's lemma, and orthogonality is a consequence of the orthogonality of basis elements chosen to colour vertices and is trivial in our case as there is only one such basis element. Thus we have the following definition:

\begin{defi}[The Ponzano-Regge State Sum]\label{defi-PRStatesum} The Ponzano-Regge state sum $\ZZ_{PR}$ is the state sum defined by formally applying the graphical calculus defined in Section \ref{sec-SpinNets} to the Definition \ref{defi-graphicalstatesum}.
\end{defi}

This state sum satisfies the requirements to define a TQFT up to finiteness and was first considered in \cite{ponzanoRegge}. This definition is only formal as in contrast to the requirements in the preceding sections the colouring set is not finite. In particular note that the normalisation factor is divergent and it is necessary to regularize the state sum \cite{Freidel2004, Barrett2009}.

\subsubsection{4 Dimensional Amplitudes from Spin Networks}

\begin{figure}[htbp]
\begin{center}
\psfrag{r}{$j$}
\psfrag{i}{$j$}
\includegraphics[scale=1]{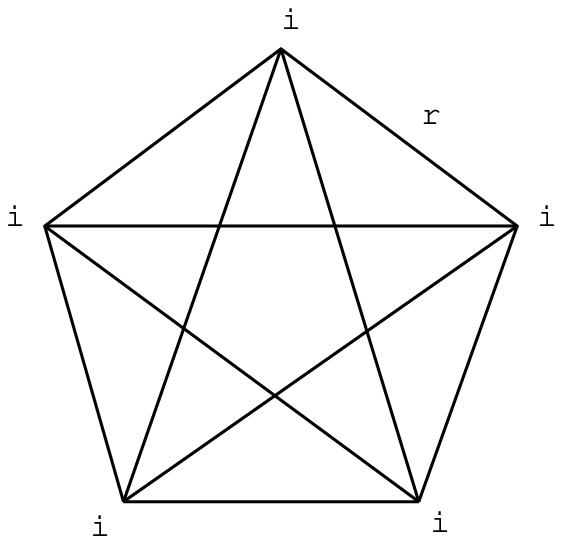}
\caption{4-simplex.}
\label{fig-4simplexnet}
\end{center}
\end{figure}

We will not discuss the construction of 4-dimensional state sums and TQFTs in detail here but refer the reader to the literature \cite{Ooguri1992a, Crane:1993if}. We will merely point out that the same construction of a tetrahedral amplitude from the dual spin network of the surface of the tetrahedron can be carried out in 4d. In this case we obtain a four-valent spin network based on four-valent intertwiners dual to tetrahedra and links dual to triangles. The evaluation of this spin network is well known in recoupling theory of $\SU(2)$ as the $\ftj$-symbol. Its name stems from the fact that ten spins label the edges and five spins label the intertwiners. The graphical representation of such a spin network is given in Figure \ref{fig-4simplexnet}.

\subsection{Spin Foam Models}

As in the Ponzano-Regge model the data can be entirely given in terms of the 2-skeleton of the dual of the triangulation and due to the use of spin networks as a basis of cylindrical functions in Loop Quantum Gravity that, evaluated on the 0-connection evaluate up to sign to the graphical calculus defined above, these models are also known as spin foam models \cite{Baez2000a,Baez1998b}, and several alternative approaches to their definition have been proposed. In particular for the case of the graphical calculus defined by the representations of semi-simple Lie groups they can be understood heuristically as quantisations of so-called $BF$-theories. We refer the interested reader to the review articles \cite{Baez2000a,Baez1998b,Oeckl2003}.

\chapter{The Geometry of Representations}

In Chapter \ref{chap-statesum} we showed how to construct a state sum invariant using the representation theory of $\SU(2)$. A similar construction can be given for $\Spin(4)$, the covering group of $\SO(4)$, and $\SL(2,\C)$, the covering group of $\SO(3,1)^+$ the identity connected component of $\SO(3,1)$. In four dimensions the analogue construction is called the Ooguri model \cite{Ooguri1992a}. In order to understand the geometry of these models we will begin by reviewing the geometric states arising in the representation theory of $\SU(2)$, $\Spin(4)$ and $\SL(2,\C)$.

\section{The Geometry of $\SU(2)$}

One way to the representation theory of $\SU(2)$ is by the quantization of the sphere, or alternatively of coadjoint orbits in the Lie algebra (see for example \cite{GuilleminSternberg}). In an inverse ``correspondence principle"-type logic the large quantum number behaviour should thus recover spherical geometry. This is in fact the case and can be seen easily by using Perelomov coherent states \cite{perelomov}.

\subsection{Coherent States}

Coherent states are defined as the eigenstates of Lie algebra elements. Given a 3-dimensional unit vector $\nb \in S^2$ the associated coherent states $\alpha_j(\nb)$ are defined by:
\begin{eqnarray}\label{eq-costa-def}
\nb\cdot L_j \, \alpha_j(\nb) &=& i\, j\, \alpha_j(\nb)\nn\\
\la \alpha_j(\nb)| L_j^m \, \alpha_j(\nb)\ra &=& i\, j\, \nb^m
\end{eqnarray}

This fixes the states $\alpha_j(\nb)$ up to a phase. As a matter of fact we can show that every state in the fundamental representation is proportional to a coherent state. Writing $\alpha(\nb) := \alpha_{\frac{1}{2}}(\nb)$ and using the Pauli matrices $\frac{i}{2}\sigma_P^i = L^i_{\frac12}$ the first line of \eqref{eq-costa-def} reads $\nb\cdot \sigma_P \, \alpha_{\frac12}(\nb) = \alpha_{\frac12}(\nb)$. Now note that the Pauli matrices together with the identity $\id$ form a basis of the two by two dimensional Hermitian matrices. The matrix $N^0 \id + N^i \sigma_P^i$ has eigenvalues equal to $N^0 \pm |N^i|$. Thus in particular the matrices $\frac{1}{2} \left(\id + \nb \cdot \sigma_P\right)$ are the Hermitian matrices with eigenvalues $1$ and $0$. Thus they are of the form $\alpha\tensor\alpha^\dagger$. Conversely every normalized state $\alpha'$ defines such a Hermitian matrix, which can therefore be written as $\frac{1}{2} \left(\id + \nb(\alpha') \cdot \sigma_P\right)$. It follows immediately that $\alpha'$ is an eigenstate of $\nb(\alpha') \cdot \sigma_P$. The fact that it is the $+1$ and not the $-1$ eigenstate implies the second line of the coherent state definition. Thus every normalized state $\alpha'$ is a coherent state $\alpha'(\nb(\alpha'))$. In particular we have that
\be\label{eq-costa-projectorform}
\alpha(\nb)\tensor\alpha(\nb)^\dagger = \frac{1}{2} \left(\id + \nb \cdot \sigma_P\right)
\ee

Note that the anti-linear map $J$ transforms a coherent state to one associated to the opposite direction:
\be\label{eq-costa-J}J \alpha_j(\nb) = \alpha_j(-\nb).\ee

From the definition of $L_j$ it is immediate to see that the coherent states satisfy an exponential property. We have that for every $\alpha_j(\nb)$ there is an $\alpha(\nb)$ in the fundamental representation such that
\begin{equation}\label{eq-costa-expo}
\alpha_j(\nb) = \alpha(\nb) \tensor \alpha(\nb) \tensor \dots \tensor \alpha(\nb) = \bigotimes^{2j} \alpha(\nb).
\end{equation}

Lie algebra elements transform under the vector representation of $\SU(2)$. We denote the vector representation of a group element $g$ by $\hat{g}$. This is the standard double cover of $\SO(3)$ by $\SU(2)$, $\pm g$ define the same $\hat{g}$. From the fact that $gL_j^lg^{-1} = \hat{g}^l_{\phantom{l}m} L_j^m$ we then have that $g \alpha_j(\nb) = \alpha_j(\hat{g}\nb)$ for some $\alpha_j(\hat{g}\nb)$. From the fact that the $\alpha_j(\nb)$ are eigenstates of the Lie algebra elements that are parallel to $\nb$ we obtain their transformation behaviour under rotations that stabilize $\nb$, $$g = \exp(\phi \nb\cdot L_j)\; :\; g \alpha_j(\nb) = e^{ij\phi} \alpha_j(\nb).$$

Using \ref{eq-costa-projectorform} we can easily calculate the modulus square of the inner product between two coherent states by noting that in the fundamental representation we have: \bea|\braket{\alpha(\nb)}{\alpha'(\nb')}|^2 &=& \tr\left(\alpha(\nb)\tensor\alpha(\nb)^\dagger \alpha'(\nb')\tensor\alpha'(\nb')^\dagger  \right) \nn\\&=& \frac{1}{4} \tr\left( \left(\id + \nb \cdot \sigma_P\right)\left(\id + \nb' \cdot \sigma_P\right) \right)\eea
The last can be calculated easily using that $\sigma_P^k \sigma_P^l = \delta^{kl} \id + i \sigma_P^m \epsilon^{klm}$ and the fact that the $\sigma_P$ are traceless. We obtain $$|\braket{\alpha(\nb)}{\alpha'(\nb')}|^2= \frac{1}{2} (1 + \nb \cdot \nb').$$  The modulus square of the inner product between two coherent states in arbitrary representations is then immediate thanks to the exponential property:
\be \label{eq-costa-innerprod}
  |\la \alpha_j(\nb)| \alpha'_j(\nb') \ra|^2 = \left( \frac{1}{2} (1 + \nb \cdot \nb') \right)^{2j}.
\ee

This shows in particular that in the large quantum number limit $j \rightarrow \infty$ coherent states become orthogonal. In this way they capture the ``correspondence principle"-type geometry of representations in which we are interested. Away from the asymptotic regime coherent states still form an over-complete basis of representation spaces. A resolution of the identity is given by integrating over the sphere:
\be\label{eq-costa-reso1}
(2j+1) \int_{S^2} \dd \nb \ket{\alpha_j(\nb)}\bra{\alpha_j(\nb)} = \id_j
\ee
where $\dd \nb$ is the normalized measure on the sphere.

It will furthermore be interesting to calculate the expectation value of $L^i$ between different coherent states. In order to avoid having to deal with the arbitrary phases we will calculate $\braket{\alpha(\nb_1)}{\alpha'(\nb_2} \bra{\alpha'(\nb_2)}L^i\ket{\alpha(\nb_1}$. To do so we again use equation \eqref{eq-costa-projectorform} and $\sigma_P^k \sigma_P^l = \delta^{kl} \id + i \sigma_P^m \epsilon^{klm}$ and we have:

\bea\label{eq-costa-diffstates}
\braket{\alpha(\nb_1)}{\alpha'(\nb_2} \bra{\alpha'(\nb_2)}L^i\ket{\alpha(\nb_1} &=& \frac{i}{2}\bra{\alpha(\nb_1)}\frac12 (\id + \nb_2 \cdot \sigma_P) \sigma_P^i\ket{\alpha(\nb_1}\nn\\
&=& \frac{i}{4} \bra{\alpha(\nb_1)} \sigma_P^i \ket{\alpha(\nb_1} \nn\\&&\;\;\;\;\;\;\;\;\;+ \nb_2^k \bra{\alpha(\nb_1)} \sigma_P^k \sigma_P^i\ket{\alpha(\nb_1}\nn\\
&=& \frac{i}{4} (\nb_1^i + \nb_2^i + i \epsilon^{kim}\nb_2^k \bra{\alpha(\nb_1)} \sigma_P^m\ket{\alpha(\nb_1}\nn\\
&=& \frac{i}{4} (\nb_1^i + \nb_2^i - i \epsilon^{ikm}\nb_2^k\, \nb_1^m).\nn
\eea

\subsection{Coherent Triangles}\label{sec-CohTria}

Coherent states give us the geometry of a sphere associated to particular representations $j$. Extending this we can give a geometric interpretation to the invariant subspace of the tensor product of representations $\Inv(j_1 \tensor j_2 \tensor \dots \tensor j_n)$. In this section we will discuss the case of the three-valent intertwiners which we will associate to triangles and in the next section we will discuss four-valent intertwiners and their relationship with tetrahedra.

To see this relationship first note that the shape space of triangles can be described as a constrained space of three spheres of radius $j_i$. This space has the product symplectic structure of that of the sphere used in the geometric quantisation \cite{GuilleminSternberg}. Given three vectors $j_a \nb_a$ of length $j_a$ that are on the spheres the constraint \be\label{eq-closuretrian}\sum_{a=1}^3 j_a \nb_a = 0\ee forces them to describe the edge vectors of a triangle with edge lengths $j_a$. Furthermore, with the given symplectic form this constraint generates rotations on these spheres. Thus the space obtained by symplectic reduction of the space of three spheres by the constraint \eqref{eq-closuretrian} is the shape space of triangles of edge length $j_a$. Of course this space is just a single point.

To quantize this state space we take the quantized unconstrained state space, that is the representation space $j_1\tensor j_2\tensor j_3$, and project on the gauge invariant subspace by averaging over the gauge orbit, that is, the rotation acting diagonally on all three state spaces:
\be\label{eq-cohtri}
\iota(\nb_1, \nb_2, \nb_3) = \int_{\SU(2)} dX \, X\alpha_{j_1}(\nb_1) \tensor X\alpha_{j_2}(\nb_2) \tensor X\alpha_{j_3}(\nb_3) \in \Inv(j_1\tensor j_2\tensor j_3).
\ee

The state $\iota$ depends on the arbitrary phases of the coherent states used to construct it. As we mentioned above the space $\Inv(j_1\tensor j_2\tensor j_3)$ is actually 1-dimensional which is in keeping with the observation that the classical reduced phase space is just a point. It is in fact guaranteed by an abstract theorem of Guillemin and Sternberg called ``quantisation commutes with reduction" \cite{1982InMat..67..515G} that by quantizing first and constraining after we end up in the same reduced Hilbert space. The over-parametrisation of this state space through the intertwiners associated to three vectors $\nb_a$ will nevertheless be extremely useful in what is to follow. Note that $\alpha_{j_1}(\nb_1) \tensor \alpha_{j_2}(\nb_2) \tensor \alpha_{j_3}(\nb_3)$ is not in $\Inv(j_1\tensor j_2\tensor j_3)$ even if the classical labels here close in the sense of \eqref{eq-closuretrian}. However the geometry of the classical phase space can again be seen explicitly asymptotically. If we introduce a scaling $\lambda$ and look at the elements $\iota(\nb_1, \nb_2, \nb_3)$ in $\lambda j_1 \tensor \lambda j_2 \tensor \lambda j_3$ we find they are exponentially suppressed for large $\lambda$ unless \eqref{eq-closuretrian} is satisfied \cite{livine-2007-76}.

\subsection{Coherent Tetrahedra}\label{sec-CohTet}

Just as three-valent intertwiners can be interpreted as quantized triangles we can interpret four-valent intertwiners as quantized tetrahedra \cite{Baez1999a,Barbieri:1997ks}. The use of coherent states for their study was introduced in the context of state sum models in \cite{livine-2007-76}. Consider a set of four vectors in directions $\nb_a$ of lengths $j_a$ satisfying closure \be\label{eq-closureTet}\sum_{a=1}^4 j_a \nb_a = 0.\ee The space of such closing vectors that are non-degenerate is the shape space of non-degenerate tetrahedra.

\begin{lem}[Tetrahedra from Closing Vectors]\label{lem-GeometryOfClosure}
Four vectors $j_a\nb_a$ of length $j_a$ that span 3-dimensional space and satisfy \eqref{eq-closureTet} are the outward normals of a non-degenerate tetrahedron with areas $j_a$ embedded in $\R^3$ which is unique up to translation. 
\end{lem}

\begin{proof} If they span 3-dimensional space and satisfy closure, any subset of them is linearly independent. It is immediate that every non-degenerate tetrahedron defines such a set of normals satisfying closure due to Gauss' theorem. Conversely, interpreting the $\nb_a$ as normal vectors is sufficient to determine the shape of the tetrahedron. Then the normal vectors satisfy closure with the areas determined up to scale. This closure condition cannot be independent of \eqref{eq-closureTet} as we otherwise could scale the geometric closure condition such that one area matches a $j_a$, subtract one equation from the other and obtain linear dependence between some set of $\nb_a$, which is in contradiction to our assumption. Thus the areas of the tetrahedron have to be proportional to $j_a$. By scaling the tetrahedron we obtain two tetrahedra with areas matching $j_a$, one with the $\nb_a$ as inward normals and the other with the $\nb_a$ as outward normals. \end{proof}

Again the constraint \eqref{eq-closureTet} generates rotations of the tetrahedron in question, and the non-degenerate sector of the reduced phase space is the shape space of tetrahedra. Implementing the constraint quantum mechanically again we obtain the over-parametrisation of the space of four-valent intertwiners by:

\bea\label{eq-cohtet}
\iota(\nb_1, \nb_2, \nb_3, \nb_4) = \int_{\SU(2)} dX \, X\alpha_{j_1}(\nb_1) \tensor X\alpha_{j_2}(\nb_2) \tensor X\alpha_{j_3}(\nb_3) \tensor X\alpha_{j_4}(\nb_4)&&\nn\\ \in \Inv(j_1\tensor j_2\tensor j_3 \tensor j_4)&&\nn\\
\eea

Again the phase of $\iota$ is not fixed by the labels $\nb_a$. Now this space is not 1-dimensional any more and it follows by the theorem of Guillemin and Sternberg mentioned above that this construction does indeed cover all of $\Inv(j_1\tensor j_2\tensor j_3 \tensor j_4)$. Again a state with classical data that closes does not immediately live in the invariant subspace but the length of the projection of states that don't close is exponentially suppressed asymptotically. Note that any two intertwiners $\iota$, $\iota'$ corresponding to sets of vectors $\nb_a$ related by a rotation are related by a phase. Furthermore the overlap between intertwiners corresponding to different tetrahedra is also suppressed exponentially asymptotically, showing that in the asymptotic regime we recover the shape space of tetrahedra in the same ``correspondence principle" way as we obtain the sphere from the individual representations \cite{livine-2007-76}.

\subsection{Coherent Boundary Manifolds}

To use the coherent intertwiners defined in the preceding sections for the analysis of state sums we need to extend the construction from triangles and tetrahedra to closed triangulated 2 and 3-dimensional manifolds. The states we will define will then be in the boundary state spaces of the state sums. In the 3-dimensional state sum $\ZZ_{PR}$ constructed above, the boundary state space was given by the span of colourings of edges of the boundary triangulation with spins and its triangles with intertwiners. In the 4-dimensional state sums we will consider, the difference will be only that the spins will colour triangles, and the intertwiners will colour tetrahedra. In this section we will give coherent states in these spaces. If the data in these states furthermore defines a consistent geometry on the boundary manifold we will give a canonical phase choice for these states.

\subsubsection{2-Dimensional}

Consider a closed oriented 2-dimensional manifold $\bo^2$ with a triangulation $\TT^2$. We define a flat geometry on the 2-simplices of the manifold by defining a linear and orientation preserving map from each simplex into $\R^2$: $\phi_a: \sigma^2_a \rightarrow \R^2$ where $a$ is a label on the set of 2-simplices $\sigma^2_a \in \TT_2$. Linear in this context will always mean affine linear, but we will mostly not be interested in the translational part of the maps, and will use the same symbol $\phi_a$ to mean the rotational parts of the affine linear maps.

The pullback of the standard metric on $\R^2$ then provides a metric on the simplex; the pullback of a basis defines a frame. The triangles $\sigma^2_a \in \TT_2$ get mapped to triangles $\phi_a(\sigma^2_a)$ in $\R^2$. 

The set of maps $\phi_a$ defines a simplex-wise linear metric on $\bo^2$ if the metric induced on edges of the triangulation $\sigma^1$ by the two triangles that contain them agrees. In the 2-dimensional case this simply means that the edge lengths have to agree. In this case we call $\bo^2$ with this triangulation and metric a Regge manifold \cite{Regge1961}. We will assume this is the case from here on.

It will actually be convenient for us to consider the trivial extension of the maps $\phi_a$ to $\R^3$, that is, consider $\phi_a$ as a map to the subspace $\no^\perp$ with $\no = (1,0,0)$ and orientation induced by considering $\no$ as outward. With this convention the $\phi_a$ define a set of three 3-dimensional edge vectors associated to each triangle. Every edge of the triangulation can be labelled by the two triangles that it borders and we will denote the oriented edges $\sigma^1_{ab} \in \TT_1$. We think of $\sigma^1_{ab}$ as belonging to the boundary of $\sigma^2_a$ with the boundary orientation. We then have that the maps $\phi$ determine two vectors per edge called $\nb_{ab}$ and $\nb_{ba}$. $\nb_{ab}$ is the unit vector in the direction of the edge $\sigma^1_{ab}$ in the triangle $\phi_a(\sigma^2_a)$, and $\nb_{ba}$ is its counterpart in the triangle $\phi_b(\sigma^2_b)$. There now exists a set of $\SO(3)$ maps $\hat{g}_{ba}$ that map the edge vectors that are identified to each other:
\bea\label{eq-2dReggeGluing}
\hat{g}_{ba} \no &=& \no\nn\\
\hat{g}_{ba} \nb_{ab} &=& - \nb_{ba}.
\eea

These are given by the rotational part of $\phi_b \circ (\phi_a)^{-1}$. It also follows that $\hat{g}_{ba}$ is the parallel transport when going from the frame on $\sigma^2_a$ to $\sigma^2_b$. The frame transforms by $\hat{g}_{ba}^{-1}$, thus the same vector expressed in this frame is transformed by $\hat{g}_{ba}$. Thus the $\hat{g}_{ba}$ are actually the discrete connection compatible with the simplex-wise flat geometry. A discrete connection is defined in Appendix \ref{sec-DiscConn} by Definition \ref{defi-DiscConn}. Given a spin structure on the boundary manifold we can then pick an $\SU(2)$ connection $g_{ab}$ that is a covering lift of $\hat{g}_{ba}$ according to Definition \ref{defi-CoveringLift}.

For a Regge boundary manifold with half-integer edge lengths that are admissible per triangle we can now define a state in $\ZZ(\bo^2, \TT^2)$ by choosing the representations $j_{ab}$ colouring the edges $\sigma^1_{ab}$ to be the edge lengths and the intertwiners colouring the triangles $\sigma^2_a$ to be given by $\iota(\nb_{ab_1}, \nb_{ab_2}, \nb_{ab_3})$ of equation \eqref{eq-cohtri}. As we pointed out in Section \ref{sec-CohTria} the intertwiner thus chosen actually depends only on the geometry of the triangle, not on the particular vectors. In this context this implies that the boundary state is independent of the particular maps $\phi_a$ and only depends on the induced geometry. We can furthermore use the boundary connection $g_{ba}$ to fix a particular phase for this state by choosing the coherent states entering the $\iota(\nb_{ab_1}, \nb_{ab_2}, \nb_{ab_3})$ to satisfy
\bea\label{eq-2dReggePhase}
g_{ba} J \alpha_{j_{ab}}(\nb_{ab}) = \alpha_{j_{ab}}(\nb_{ba})\; \forall a<b
\eea
where $a<b$ is a, so far, arbitrary ordering of the $\sigma^2_a \in \TT^2_2$, for $b<a$ the equation holds with opposite sign as $J^2 = -\id$. As $J$ is anti-linear this fixes the phase of $\alpha_{j_{ab}}(\nb_{ab}) \tensor \alpha_{j_{ab}}(\nb_{ba})$. We will denote states satisfying this phase choice $\ket{\nb_{ab}}_{j_{ab}}$, and their image under the complex structure $J$ by $\ket{-\nb_{ab}}_{j_{ab}}$. The overall boundary state, which we will call the Regge state can then be defined as follows:

\begin{defi}[2d Regge State]\label{defi-2dReggeState}

Given an oriented 2-dimensional manifold $\bo^2$ with a triangulation $\TT^2$ with triangles $\sigma^2_a \in \TT^2_2$ and oriented edges $\sigma^1_{ab} \in \TT^2_1$ and a collection of simplex-wise linear maps $\phi_a: \sigma^2_a \rightarrow \no^\perp \subset \R^3$ that are orientation preserving with respect to the orientation on $\no^\perp$ inherited from the standard orientation from $\R^3$ by considering $\no$ as outward normal, we obtain a set of edge lengths $j_{ab} = |\phi_a(\sigma^1_{ab})|$ and normalized edge vectors $\nb_{ab} \in \no^\perp$. If $j_{ab} = j_{ba}$ the $\phi_a$ define a simplex-wise flat metric as well as frames on $\bo^2$. Then we can chose a covering lift of the metric compatible discrete connection $\hat{g}_{ba}$ parallel transporting $\eqref{eq-2dReggeGluing}$ between these frames to an $\SU(2)$ connection $g_{ba}$ compatible with the spin structure. If $j_{ab}$ are half integer we define the associated Regge state to be

\bea\label{eq-2dReggeState}
\Psi = \left(\prod_{\sigma^2_a} \int_{\SU(2)} \dd X_a\right) \prod_{\sigma^1_{bc}} X_b \ket{\nb_{bc}}_{j_{bc}} \tensor X_c \ket{\nb_{cb}}_{j_{bc}},
\eea
where $\ket{\nb_{cb}}_{j_{bc}}$ are coherent states satisfying the phase choice \eqref{eq-2dReggePhase} defined in terms of the covering lift $g_{ba}$.
\end{defi}

This state is in $\ZZ_{PR}(\bo^2, \TT^2)$. Furthermore we can see that the state we obtain depends only on the metric and the spin structure\footnote{Note that while the spin structure induces an orientation on the $\SO(3)$ bundle, this can induce different orientations on the base manifold. This can be seen, for example, from the different ways to map the tangent bundle into the principal $\SO(3)$ bundle. This induced orientation plays no role in our analysis and we have a seperately specified orientation on the manifold. Our edge vectors are based on this orientation, not that of the bundle.}:

\begin{lem}[2d Regge State Symmetries]\label{lem-2dReggeStateSymmetries}
The state $\Psi$ of \eqref{eq-2dReggeState} is independent of the orientation of the manifold, the frames and the choice of the $\SU(2)$ connection. It only depends on the spin structure on the $\SO(3)$ bundle and the metric on $\bo^2$.
\end{lem}

\begin{proof}
This follows from the symmetries of the intertwiners. In particular if we change the maps $\phi_a$ to $\hat{g}_a \phi_a$, this would induce the following transformation on the data we have defined:

$$\nb_{ab} \rightarrow \hat{g}_a \nb_{ab}$$

and

$$\hat{g}_{ba} \rightarrow \hat{g}_b \hat{g}_{ba} \hat{g}_b^{-1}.$$

This is a gauge transformation on the discrete connection $\hat{g}_{ba}$ which can be lifted to a gauge transformation $g_a$ on the spin connection $g_{ba}$ by Lemma \ref{lem-CommDiagLift}. If the set $\ket{\nb_{cb}}_{j_{ab}}$ satisfies the Regge phase choice with respect to $g_{ba}$ then the set $g_c \ket{\nb_{cb}}_{j_{bc}}$ satisfies the Regge phase choice with respect to the gauge transformed connection $g_b g_{ba} g_a^{\dagger}$. However the state $\Psi$ defined with $g_c \ket{\nb_{cb}}_{j_{bc}}$ is the same as the state defined with $\ket{\nb_{cb}}_{j_{bc}}$ as the group elements $g_c$ can be absorbed into the integration over $X_c$. The gauge transformation includes the map $diag(-1, -1, 1)$ which is orientation reversing on $\no^\perp$. Finally the choices of lift of the $\SO(3)$ connection differ by $g_{ba} \rightarrow \epsilon_b \epsilon_a g_{ba}$. The group elements $g_{ab}$ acting on the representations $j_{ab}$ change by $\epsilon_a^{2j_{ab}} \epsilon_b^{2j_{ab}}$. Thus $\Psi$ changes by $\epsilon_a^{2(j_{ab_1} + j_{ab_2} + j_{ab_3})}$. By the admissibility condition that the sum of spins at an intertwiner has to be an integer this is equal to $1$ and again the state is left unchanged.\end{proof}

In fact $\ZZ_{PR}(\bo^2, \TT^2)$ is spanned by states of this form. This follows from the fact that states in $\ZZ_{PR}(\bo^2, \TT^2)$ have to satisfy the admissibility conditions. These implement the triangle inequalities on the colours $j_{ab}$. Thus we immediately have a piecewise flat metric on $\TT^2$. Furthermore every intertwiner in $\Inv(j_{ab_1}, j_{ab_2}, j_{ab_3})$ is proportional to the coherent triangle made with fitting closing data. The choice of spin structure only changes the overall state by a sign, thus the Regge states are indeed an over-complete basis of $\ZZ_{PR}(\bo^2, \TT^2)$.

Finally note that the action of $J$ on $\Psi$ is to change all $\nb_{ab}$ to $-\nb_{ab}$. Note that as $\ket{\nb_{ab}}$ satisfy the phase condition in equation \eqref{eq-2dReggePhase}, so do the states $J\ket{\nb_{ab}}$. Thus $J\Psi$ is a Regge state again. This is again merely a gauge transformation that does not change the metric as the $\SO(3)$ element $diag(1,-1,-1)$ also changes all $\nb_{ab}$ to $-\nb_{ab}$. It follows that $J \Psi = \Psi$.

\subsubsection{3-Dimensional}

The construction for 3-dimensional Regge states operates very similarly. The key difference is that we interpret the $\nb$ as the outward normals of tetrahedra as by Lemma \ref{lem-GeometryOfClosure}. Thus given a 3-dimensional manifold $\bo^3$ with triangulation $\TT^3$ we again obtain a metric and frame on the simplices by a set of maps $\phi_a: \sigma^3_a \rightarrow \R^3$.

If the geometry of the triangles $\phi_a (\sigma^2_{ab})$ and $\phi_b (\sigma^2_{ba})$ does not agree, the rotational parts of the connecting maps $\phi_b \circ (\phi_a)^{-1}$ are not in $\SO(3)$ and do not have a straightforward geometric interpretation. We still obtain four normals $\nb_{ab}$ and four areas $j_{ab}$ per tetrahedron $a$ that, taken together, satisfy closure. If the $j_{ab}$ are half integer we can again take coherent states in the appropriate representations. We then have coherent boundary states without a specified phase given by:

\begin{defi}[3d Coherent Manifold States]\label{defi-3dCoherentBoundaryStates}

Let $\bo^3$ be an oriented, 3-dimensional manifold with a triangulation $\TT^3$ with tetrahedra $\sigma^3_a \in \TT^3_3$ and oriented triangles $\sigma^2_{ab} \in \TT^3_2$ and a collection of simplex-wise linear maps $\phi_a: \sigma^3_a \rightarrow \R^3$ that are orientation preserving we obtain a set of areas $j_{ab} = |\phi_a(\sigma^2_{ab})|$ and face normals $\nb_{ab} \in \R^3$. If $j_{ab} = j_{ba}$ is half integer we have coherent manifold states of the form

\bea\label{eq-3dCoBoundaryState}
\Psi_C = \left(\prod_{\sigma^3_a} \int_{\SU(2)} \dd X_a\right) \prod_{\sigma^2_{bc}} X_b \alpha({\nb_{bc}})_{j_{bc}} \tensor X_c \alpha({\nb_{cb}}_{j_{bc}}).
\eea

\end{defi}

This state is again invariant under the gauge transformations $g_a$ acting by $\alpha({\nb_{cb}})_{j_{bc}} \rightarrow g_b \alpha({\nb_{bc}})_{j_{bc}}$. Thus all states arising from $\phi_a$ defining the same simplex-wise geometry differ only by a phase. However, this symmetry now no longer undoes the orientation choice.

If our maps $\phi_a$ satisfy the Regge condition that $\phi_a (\sigma^2_{ab})$ and $\phi_b (\sigma^2_{ba})$ have the same geometry, then our connecting maps $\hat{g}_{ba}$ are the discrete metric compatible $\SO(3)$ connection again. They are again given by the rotational parts of $\phi_b \circ \phi_a^{-1}$ and satisfy

$$\hat{g}_{ba} \nb_{ab} = - \nb_{ba}.$$

This follows as the $\nb$ are outward facing and the $\phi_a$ are orientation preserving. We again have a lift of the metric compatible connection to $\SU(2)$ by $g_{ba}$ and we can define a geometric phase choice by \eqref{eq-2dReggePhase} and thus a Regge state:

\begin{defi}[3d Regge States]\label{defi-3dReggeStates}

Given an oriented 3-dimensional manifold $\bo^3$ with a spin structure and a triangulation $\TT^3$ with tetrahedra $\sigma^3_a \in \TT^3_3$ and oriented triangles $\sigma^2_{ab} \in \TT^3_2$ and a collection of simplex-wise linear maps $\phi_a: \sigma^3_a \rightarrow \R^3$ that are orientation preserving we obtain a set of areas $j_{ab} = |\phi_a(\sigma^2_{ab})|$ and face normals $\nb_{ab} \in \R^3$. If $j_{ab} = j_{ba}$ and the geometry of $\phi_a (\sigma^2_{ab})$ and $\phi_b (\sigma^2_{ba})$ agree we have a metric compatible discrete connection. We can choose a lift of this discrete connection compatible with the spin structure. Then we can define the 3-dimensional Regge state as

\bea\label{eq-3dReggeState}
\Psi_R = \left(\prod_{\sigma^3_a} \int_{\SU(2)} \dd X_a\right) \prod_{\sigma^2_{bc}} X_b \ket{\nb_{bc}}_{j_{ab}} \tensor X_c \ket{\nb_{cb}}_{j_{ab}},
\eea

where $\ket{\nb_{cb}}_{j_{bc}}$ are coherent states satisfying the phase choice \eqref{eq-2dReggePhase} with respect to the covering lift $g_{ab}$.
\end{defi}

This state depends only on the metric, the spin-structure and the orientation:

\begin{lem}[3d Regge State Symmetries]\label{lem-3dReggeStateSymmetries}
The state $\Psi_R$ of \eqref{eq-3dReggeState} is independent of the frames and the choice of the $\SU(2)$ connection. It only depends on the orientation, spin structure and metric on $\bo^3$, $\TT^3$.
\end{lem}
\begin{proof}
This follows from similar arguments as in the 2-dimensional case above.\end{proof}

Note that in the 3-dimensional case the action of $J$ on these states is again to reverse all $\nb \rightarrow -\nb$. However, in the 3-dimensional case the $\nb$ are interpreted as outward normals, thus this action has the same effect as inverting the orientation of the 3 manifold. Thus $J\Psi_R$ is the Regge state for the same spin structure and metric but opposite orientation as $\Psi_R$.

The formulation of constraints on $j$ and $\nb$, or equivalent data, which force the geometries of $\phi_a (\sigma^2_{ab})$ and $\phi_b (\sigma^2_{ba})$ to agree, has received considerable attention recently (e.g. \cite{Dittrich2010a,Dittrich2008a}). Alternatively it has been proposed to directly interpret the simplex-wise geometries of the states defined in Definition $\ref{defi-3dCoherentBoundaryStates}$ (e.g. \cite{Freidel2010}).

\section{The Geometry of $\Spin(4)$} \label{sec-Spin4Geometry}

For the 4-dimensional models we will consider, we will need to understand the geometry of the representation theory of $\Spin(4)$. To do so we will consider the Lie algebra $\spin(4)$ which is isomorphic to $\so(4)$, the Lie algebra of 4-dimensional rotations. We can understand this Lie algebra as arising from bivectors. We will then show how it decomposes into a left and right sector under the action of the Hodge star. This will allow us to give the representation theory of $\Spin(4)$ in terms of $\SU(2)$ representations and define coherent bivectors. Finally we will give a necessary and sufficient set of conditions for a set of bivectors to define a geometric 4-simplex $\sigma^4$.

\subsection{Bivectors}

Bivectors in $\R^4$ are elements of $\Lambda^2(\R^4)$, that is, 2-dimensional antisymmetric tensors $B^{IJ} = - B^{JI}$, $I,J = 0,\dots, 3$. $\Lambda^2(\R^4)$ is 6-dimensional. We define the norm of a bivector by $|B|^2 = \frac{1}{2} B^{IJ} B_{IJ}$. Bivectors can be written as linear combinations of so-called simple bivectors, that is, bivectors of the form $B^{IJ} = (N\wedge M)^{IJ} = N^I M^J - M^I N^J$. These have norm $|N\wedge M|^2 = |N|^2|M|^2 - (N\cdot M)^2$. The Lie algebra $\so(4)$ is the algebra of antisymmetric matrices with Lie product given by the commutator. The space of bivectors is isomorphic to $\so(4)$ by lowering one of the indices: $L^I_{\phantom{I}J} = B^{IK} \delta_{KJ}$. We can give a basis of the Lie algebra through a basis of bivectors $B_{\alpha \beta}^{IJ} = 2 \delta_{[\alpha}^I \delta_{\beta]}^J = (\delta_{\alpha}^I \delta_{\beta}^J - \delta_{\beta}^I \delta_{\alpha}^J)$ satisfying $|B_{\alpha\beta}| = 1$. We then have the basis ${L_{\alpha\beta}}^I_{\phantom{I}J} = 2 \delta_{[\alpha}^I \delta_{\beta]}^K \delta^\ph{I}_{KJ} = 2 \delta_{[\alpha}^I \delta^\ph{I}_{\beta ]J}$ for $\alpha < \beta$, $\alpha, \beta = 0,\dots, 3$. It is then easy to calculate the Lie algebra structure constants:

\bea\label{eq-so4structure} [L_{\alpha\beta},L_{\gamma\delta}] &=&4\left( \delta_{[\alpha}^I \delta^\ph{I}_{\beta]I'} \delta_{[\gamma}^{I'} \delta_{\delta]K} - \delta_{[\gamma}^I \delta^\ph{I}_{\delta]I'} \delta_{[\alpha}^{I'} \delta^\ph{I}_{\beta]K}\right)\nn\\
&=& 4\left(\delta_{[\alpha}^I \delta^\ph{I}_{\beta][\gamma} \delta^\ph{I}_{\delta]K} - \delta_{[\gamma}^I \delta^\ph{I}_{\delta][\alpha} \delta^\ph{I}_{\beta]K}\right)\nn\\ 
&=& +\delta^\ph{I}_{\beta\gamma}\left(\delta_{\alpha}^I \delta^\ph{I}_{\delta K} - \delta_{\delta}^I \delta^\ph{I}_{\alpha K}\right) - \delta^\ph{I}_{\alpha\gamma}\left(\delta_{\beta}^I \delta^\ph{I}_{\delta K} - \delta_{\delta}^I \delta^\ph{I}_{\beta K}\right)\nn\\
&& +\delta^\ph{I}_{\alpha\delta}\left(\delta_{\beta}^I \delta^\ph{I}_{\gamma K} - \delta_{\gamma}^I \delta^\ph{I}_{\beta K}\right)-\delta^\ph{I}_{\beta\delta}\left(\delta_{\alpha}^I \delta^\ph{I}_{\gamma K} - \delta_{\gamma}^I \delta^\ph{I}_{\alpha K}\right)\nn\\
&=& \delta^\ph{I}_{\beta\gamma}L^\ph{I}_{\alpha\delta} - \delta^\ph{I}_{\alpha\gamma}L^\ph{I}_{\beta\delta}
+\delta^\ph{I}_{\alpha\delta}L^\ph{I}_{\beta\gamma}-\delta^\ph{I}_{\beta\delta}L^\ph{I}_{\alpha\gamma}.
\eea

\subsubsection{(Anti-)Self Dual Decomposition and Representations}

The Hodge star acts naturally on $\Lambda^2(\R^4)$ by $(\star B)^{IJ} = \frac12 \epsilon^{IJ}_{\phantom{IJ}I'J'} B^{I'J'}$. Its square is $\star^2 = \id$. It decomposes the space of bivectors into two 3-dimensional subspaces corresponding to its eigenvalues $\pm 1$. In terms of the basis elements $B_{\alpha\beta}$ it is now immediate that $\star B_{\alpha \beta} = \epsilon^{IJ}_{\ph{IJ}\alpha\beta} = \frac12 \epsilon^{\alpha \beta \delta \gamma} B_{\delta \gamma}$.

To understand the commutation relations in this space we will first calculate the commutation relations for $L_{0i}$ and $\hodge L_{0i}$. The key identity is that $\frac12 \epsilon^{ijk}\epsilon^{ki'j'} =  \delta^{i[i'} \delta^{j,j']}$ where $\epsilon^{ijk} = \epsilon^{0ijk}$ is the 3-dimensional $\epsilon$ and $i,j,k = 1,\ldots, 3$. It follows that $$B_{ij} = \delta^{i[i'} \delta^{j,j']} B_{i'j'} = \frac12 \epsilon^{ijk}\epsilon^{ki'j'} B_{i'j'} = \epsilon^{ijk} \hodge B_{0k}.$$

We can then immediately read of that
$$[L_{0i},L_{0j}] = - L_{ij} = - \epsilon^{ijk} \hodge L_{0k},$$
and
$$[L_{0i},\hodge L_{0j}] = - L_{ij} = - \epsilon^{ijk} L_{0k}.$$
A small calculation shows that
$$[\hodge L_{0i}, \hodge L_{0j}] = - \epsilon^{ijk} \hodge L_{0k}.$$

This means the $\hodge L_{0i}$ generate an $\SU(2)$ subgroup. Denoting the north pole as $\pole = (1,0,0,0)$ we have $\hodge L_{0i} \cdot \pole = 0$, and thus this is the subgroup that stabilises the pole $\pole$.  It also shows that with this norm the length of the bivector is indeed the angle of rotation generated by it. We introduce the shorthand notation $L_i = \hodge L_{0i}$ and $K_i = L_{0i}$ in terms of which we have:
$$[K_i,K_j] = - \epsilon^{ijk} L_k,$$
$$[K_i, L_j] = - \epsilon^{ijk} K_k,$$
and
$$[L_i, L_j] = - \epsilon^{ijk} L_k.$$

The Hodge star is invariant under the action of $\SO(4)$, thus it allows us to decompose the lie algebra into two copies. We write $L^\pm_{i} = \frac12(L_i \pm K_i) = \frac12(\hodge \pm \id) L_{0i}$. The operator $\frac12(\hodge \pm \id)$ is of course $\pm$ the projector on the $\mp 1$ eigenbivectors of $\hodge$ and thus should give a seperation of the $\so(4)$ Lie algebra. The commutation relations are indeed simply
$$[L^\pm_i, L^\pm_j] = - \epsilon^{ijk} L^\pm_j,$$
and
$$[L^+_i, L^-_j] = 0.$$

This is the decomposition of $\so(4)$ into $\su(2)\oplus\su(2)$. We can then in general write a bivector or Lie algebra element in this basis as $\bb^+_i L^+_i + \bb^-_i L^-_i = (\bb^+, \bb^-)$. We then have $\hodge(\bb^+,\bb^-) = (\bb^+,- \bb^-)$. Note that a simple bivector of the form $\pole\wedge\nb$ we have $$\pole\wedge\nb = \nb_i L_{0i} = \nb_i K_i =  \nb_i (L^+_i - L^-_i) = (\nb_i,\nb_i).$$ A simple bivector orthogonal to $\pole$ is of the form $(\nb,\nb)$. Thus we also have that for simple bivectors $|\bb^+|^2 = |\bb^-|^2$, and as $\pole\wedge\nb$ is normalized, we have in general that $2|B|^2 = |\bb^+|^2 + |\bb^-|^2$. In particular this again shows that $|\bb^\pm|$ is the angle of rotation generated by a simple bivector. The group element generated by a general bivector $(\bb^+, \bb^-)$ in $\SU(2)\times\SU(2)$ is written as $(g^+, g^-)$ and it covers a group element in $\SO(4)$. As $L_i = L^+_i + L^-_i$ the $\Spin(4)$ elements stabilising the North pole are of the form $(g,g)$.

This decomposition also shows that the universal covering group of $\SO(4)$ is $\SU(2)\otimes\SU(2) = \Spin(4)$. We immediately have the representation theory of $\Spin(4)$ as a direct product of $\SU(2)$ representations. A unitary irreducible representation of $\Spin(4)$ can be labelled by two spins $j^+$,$j^-$ and is given as the tensor product of the two $\SU(2)$ irreps. Given two coherent states we can then define a coherent bivector in the $(j_+, j_-)$ representation by $\alpha_{j_+}(\nb_+)\tensor \alpha_{j_-}(\nb_-)$. These are then still eigenstates of $\so(4)$ Lie algebra generators and inherit all the properties of $\SU(2)$ coherent states.

The covering of $\SO(4)$ can be seen explicitly by use of a convenient diffeomorphism from the sphere $S^3$ to $\SU(2)$: \be\label{eq-S3toSU2}\zeta: N^I\rightarrow N^0 \id + 2 N^i L^i.\ee A group element $G = (g_+, g_-)$ acts as $\zeta(GN) = g_+\zeta(N)g_-^{-1}$. Thus it is for example immediate that $G=(g,g)$ leaves the north pole invariant.

Note that the parity operation $P = diag(1,-1,-1,-1)$, which is in $\OO(4)$, anti-commutes with the Hodge star, that is, $$\star P = - P \star.$$ It can then be checked that $P(\bb_+, \bb_-) = (\bb_-, \bb_+)$.

\subsection{Geometric Bivectors}

This subsection will pertain to metrics of signature $(++++)$ and $(-+++)$. In the latter case we will require the bivectors to be spacelike in the sense that they are orthogonal to timelike normal vectors $N$, this will cover all cases occuring in this thesis. We call a set of bivectors geometric if they are the bivectors of the faces of a geometric 4-simplex $\sigma^4_\imm$ in $\R^4$. Using $a,b = 1,\dots ,5$ to denote the tetrahedral faces $\sigma^3_\imm$ of the 4-simplex we denote its outward facing normal vectors by $N_a$. The bivectors can then be written as 

\be\label{eq-GeomBivs}
B_{ab} (\sigma^4_\imm) = |\sigma^2_{(ab)}|\hodge\frac{N_a \wedge N_b}{|N_a \wedge N_b|}
\ee

With $B_{ab} (\sigma^4_\imm)= - B_{ba} (\sigma^4_\imm)$. The bivector $B_{ab}(\sigma^4_\imm)$ is simple and orthogonal to $N_a$ and $N_b$ and therefore describes the plane of $\sigma^2_{ab}$. By its explicit normalisation it has the area of the triangular face $\sigma^2_{(ab)}$. Therefore it is indeed the geometric bivector associated to this face. This definition encodes the standard orientation of $\R^4$ in the Hodge star.

Now consider a set of bivectors $B_{ab}$ satisfying the following list of geometricity conditions:

\begin{itemize}
\item Simplicity: $\exists N_a$ such that $N_a \cdot B_{ab} = 0 \; \forall\, b$.
\item Orientation: $B_{ab} = -B_{ba}$.
\item Closure: $\sum_{b:b\neq a} B_{ab} = 0$. 
\item 4d Non-degeneracy: The $B_{ab}$ with $a,b\neq c$ span $\Lambda^2(\R^4)$ $\forall\, c.$
\end{itemize}

Note that simplicity implies that $B_{ab}$ is indeed simple according to our definition above and closure together with the simplicity condition imply that the $N_a\cdot\star B_{ab}$, if they span the orthogonal subspace, are the normals of a geometric tetrahedron orthogonal to $N_a$. We then have the following theorem.

\begin{theo}[Bivector Reconstruction]\label{Theo-BivReconstruction}
The bivectors $B_{ab}(\sigma^4_\imm)$ associated to the triangles of a non-degenerate geometric 4-simplex $\sigma^4_\imm$ satisfy the geometricity conditions. Conversely a set of bivectors $B_{ab}$ satisfying the geometricity conditions define a geometric 4-simplex $\sigma^4_\imm$, unique up to inversion and translation, and with $\mu_\imm B_{ab} = B_{ab}(\sigma^4_\imm)$, $\mu_\imm = \pm 1.$
\end{theo}
\begin{proof}
Due to 4d non-degeneracy, $N_a$ and $N_b$ cannot be proportional. Otherwise $B_{ab}$,  $B_{ac}$ and $B_{bd}$ with $a,b,c,d \neq e$ would all lie in the same 3-dimensional hyperplane and could not span $\Lambda^2(\R^4)$ together with $B_{cd}$.

By the simplicity and orientation conditions the bivectors $B_{ab}$ are orthogonal to the normals $N_a$, $N_b$. Therefore they are of the form $B_{ab} \approx \star N_{a}\wedge N_{b}$. Thus they are proportional to the bivectors of a non-degenerate 4-simplex with outward normals $N_a$. As they satisfy a set of linear relations in closure the proportionality in fact has to be the same for all bivectors. Thus we can scale the geometric 4-simplex to match the areas. As the areas are quadratic in the scaling a freedom of sign remains which is accounted for in $\mu_{\imm}$.\end{proof}

A stronger version of this theorem, in the sense that the characterisation of geometricity was done entirely in terms of the bivectors without referring to the normals, was given in \cite{Barrett1998,Barrett2000}. The formulation here is similar to \cite{Freidel:2007py}. We will furthermore use the following lemma which characterises the possible departure from geometricity if we replace the 4d non-degeneracy condition with a weaker 3-dimensional non-degeneracy condition that $N_a\cdot\star B_{ab}$ span the space orthogonal to $N_a$. The proof given here is identical to that of \cite{Barrett2009a}.

\begin{lem}[Non-degeneracy of the Bivectors]\label{lem-BivNonDeg} 
Given a set of bivectors satisfying simplicity orientation and closure as well as 3-dimensional non-degeneracy in any non-degenerate metric, the normals $N_a$ are either all parallel or three of them are linearly independent. In the latter case the bivectors satisfy the 4-dimensional non-degeneracy condition.
\end{lem}
\begin{proof}
First note that if three of the normals $\{N_a, N_b, N_c\}$ are proportional to $N$ then the bivectors $B_{ad}$, $B_{bd}$ and $B_{cd}$ are orthogonal to $N$ and by closure so is $B_{ed}$. Due to 3-dimensional non-degeneracy they define a tetrahedron in the plane orthogonal to $N$. The vector $N_d$ is orthogonal to this tetrahedron and thus parallel to $N$.

In all other cases three of the normals $\{N_a, N_b , N_c\}$ must be pairwise linearly independent. The intersection of the hyperplanes $N_a^\perp$ and $N_b^\perp$ and the intersection of $N_a^\perp$ and $N_c^\perp$ can then not be parallel due to the 3-dimensional non-degeneracy condition, and thus the set $\{N_a, N_b , N_c\}$ must be linearly independent.

Now given these we will construct six linearly independent bivectors from the $B_{ab}$. Take two vectors $x^1_{ab}$ and $x^2_{ab}$ spanning the plane of $B_{ab}$ as well as a vector $x_A$, $x_B$ independent of them in $N_a^\perp$ and $N_b^\perp$ respectively. Furthermore choose $x_B$ such that it lies in $N_b^\perp\cap N_c^\perp$. We then have the five linearly independent bivectors $$\{x^1_{ab}\wedge x^2_{ab}, ~x^1_{ab}\wedge x_A, ~x^1_{ab}\wedge x_B,~ x^2_{ab}\wedge x_A,~ x^2_{ab}\wedge x_B\}.$$

As each of these is in a hyperplanes orthogonal to some $N$, 3-dimensional non-degeneracy implies that they are all linear combinations of some $B_{ab}$. Together with $x_A \wedge x_B$ they span the entire space of bivectors.

To show that $x_A \wedge x_B$ can be generated by the $B_{ab}$ take vectors $x^1_{bc},x^2_{bc}$ generating the intersection $N_b^\perp\cap N_c^\perp$ and an independent vector $x_C$ in $N_c^\perp$. Writing $x_C=\alpha_1 x^1_{ab}+ \alpha_2 x^2_{ab} + \alpha_3 x_B + \alpha_4 x_A$ we have $\alpha_4 \neq 0$ as otherwise $N_c^\perp$ and $N_b^\perp$ would coincide. Finally $x^1_{bc}=\beta_1 x^1_{ab}+ \beta_2 x^2_{ab} +
\beta_3 x_B$ can always be chosen in such a way that $\beta_3 \neq 0$.

Then bivector $x^1_{bc}\wedge x_C$ contains the nonzero term $\alpha_4 \beta_3 x_A \wedge x_B$ and the other 5 basis bivectors given above and can again be written in terms of the $B_{ab}$. Thus the $B_{ab}$ do indeed satisfy the non-degeneracy condition. \end{proof}

\section{The Geometry of $\SL(2,\C)$}

In order to work with theories with Lorentz symmetry we need to consider representations of $\SO(3,1)$. More specifically we will consider the identity connected component $\SO(3,1)^+$ and its double cover $\SL(2,\C)$. That is, the part of $\SO(3,1)$ that takes future pointing normals to future pointing normals. We will begin again by showing how to obtain the Lie algebra $\so(3,1)$ from bivectors and then consider a few representations and their properties. Our presentation here is less complete than in the other chapters as we don't want to go into the subtleties related to the non-compactness of the group. For more background we refer to \cite{Barrett2010}.

\subsection{Bivectors}

Whereas the Euclidean metric $\delta_{IJ}$ provided an isomorphism from the bivectors to the generators of the Euclidean rotation group $\so(4)$, the Minkowski metric $\eta_{IJ}$ (signature $(-,+,+,+)$) provides us with an isomorphism to the generators of the rotations in Minkowski space $\R^{1,3}$. We can again give a specific basis for this space through the bivectors $B_{\alpha\beta}$ defined above. The corresponding Lorentzian generators are then $(L_{\alpha\beta})^I_{\ph{I}J} = \delta^I_{[\alpha}\eta^\ph{I}_{\beta]J}$ and the structure constants are given by the same calculation as in the $\spin(4)$ case:

\be\label{eq-sl2cstructure} [L_{\alpha\beta},L_{\gamma\delta}]= \eta^\ph{I}_{\beta\gamma}L^\ph{I}_{\alpha\delta} - \eta^\ph{I}_{\alpha\gamma}L^\ph{I}_{\beta\delta}
+\eta^\ph{I}_{\alpha\delta}L^\ph{I}_{\beta\gamma}-\eta^\ph{I}_{\beta\delta}L^\ph{I}_{\alpha\gamma}.
\ee

While we have a $\star$ operator given by $\epsilon^{IJ}_{\ph{IJ}KL}$ again, it now satisfies $\star^2 = -\id$, and has eigenvalues $\pm i$. Thus we no longer have the convenient decomposition of the Lie algebra into left and right parts unless we complexify the algebra. However, we still have the decomposition into boosts and rotations. As the spatial indices are unchanged from the Euclidean case this will again give us the rotation subgroup:
$$K^i = L_{0i}$$
and
$$L^i = \frac12 \epsilon^{ijk}L_{jk}.$$

By the same computation as above we have the following Lie brackets from \eqref{eq-sl2cstructure}: $$[L^i, L^j] = - \epsilon^{ijk} L^k,$$ $$[K^i, K^j] = \epsilon^{ijk} L^k,$$ and $$[L^i, K^j] = -\epsilon^{ijk} K^k.$$ Note that the only change to the Euclidean case is the sign for the boosts, this is the only place where the $\eta_{00}$ component of the metric enters. Thus the $L^i$ again generate an $\SU(2)$ subgroup. It follows that any unitary representation of the Lorentz group will split into a tower of irreducibles of $\SU(2)$. However, as the Lorentz group is not compact its unitary representations are infinite dimensional and we will obtain infinitely many $\SU(2)$ irreps. We will not work with the unitary representations but instead refer the reader to \cite{Barrett2009a} and references therein for details.

\subsection{Spinors}\label{sec-Spinor}

From the commutation relations above it also follows that we can obtain a non-unitary representation by setting $K^i = i L^i$. In particular for the fundamental representation of $\SU(2)$ this leads to the complex group generated by all traceless matrices. This is just the group of linear matrices of determinant 1, $\SL(2,\C)$. Thus we have shown that the Lie algebra associated to bivectors in Minkowski space is indeed that of $\SL(2,\C)$. We can define an action of $\SL(2,\C)$ on $\R^{1,3}$ by using an isomorphism between $\R^{1,3}$ and the space of Hermitian matrices given by $$\Gamma^{-1}(N) = N^0 \id + N^i \sigma_P^i.$$ This has the property that for $G \in \SL(2,\C)$, and $\hat{G}\in \SO(3,1)$ the corresponding element of the vector representation, we have $$G\Gamma^{-1}(N)G^\dagger = \Gamma^{-1}(\hat{G}N).$$

This clearly takes Hermitian matrices to Hermitian matrices. We can now give a new interpretation to our coherent state vectors $\alpha(\nb)$. Consider the Hermitian matrix $\alpha(\nb)\tensor \alpha(\nb)^\dagger$. As we used in the proof of equation \eqref{eq-costa-diffstates} this can be written in the form $\alpha(\nb)\tensor \alpha(\nb)^\dagger = \frac12 (\id + \nb^i \sigma_P^i)$. Thus we have immediately that $$\Gamma(\alpha(\nb)\tensor \alpha(\nb)^\dagger) = \frac12 (1,\nb)$$ maps spinors to null vectors. In a slight abuse of notation we will write  $$\Gamma(\alpha(\nb)) = \frac12 (1,\nb).$$ It follows immediately that we also have $\Gamma(J \alpha(\nb)) = \frac12 (1,-\nb)$. From these two vectors we can now build a set of bivectors again $\frac12 (1,\nb)\wedge(1,-\nb) = (1,0,0,0)\wedge (0,\nb).$

This is in fact consistent as the Hermitian matrix $$\frac{1}{2}\left(\alpha(\nb)\tensor\alpha(\nb)^\dagger - \alpha(-\nb)\tensor\alpha(-\nb)^\dagger\right)$$ actually generates the $\SL(2,\C)$ element that covers the $\SO(3,1)$ element generated by $(1,0,0,0)\wedge (0,\nb)$. In fact all boosts are of this form.

Note that the invariance of the epsilon inner product $(,)$ only relied on the unit determinant of the matrices, therefore this is also $\SL(2,\C)$ invariant. On the other hand $\braket{}{}$ is clearly not. As they are related by the anti-linear structure $J$ it follows that for $G\in\SL(2,\C)$ we have $JGJ^{-1} = (G^\dagger)^{-1}$: 

$$\la G\cdot , \cdot\ra =  \la \cdot , G^\dagger \cdot\ra = (J \cdot, G^\dagger \cdot) = ((G^\dagger)^{-1} J \cdot, \cdot) = \la J^{-1} (G^\dagger)^{-1} J\cdot,\cdot\ra.$$

\chapter{The Geometry of 3-Dimensional State Sums}

In this chapter we will analyse the geometry of the 3-dimensional state sum $\ZZ_{PR}$ defined in Definition \ref{defi-PRStatesum}. In particular we will evaluate $\ZZ_{PR}(B^3,\TT^3)$ with the geometric boundary states $\Psi$ defined in \ref{defi-2dReggeState}. To avoid the problems associated with the regularisation of the formal Definition \ref{defi-PRStatesum}, we will use the evaluation of the state sum as the evaluation of its boundary spin network given by Theorem \ref{theo-dualevalofPF} which is finite. We will then analyse the large spin behaviour of this evaluation using stationary phase and show that it has a geometric interpretation in terms of immersions of the boundary manifold into $\R^3$.

\section{Definition of the State Sum}

Recall that Theorem \ref{theo-dualevalofPF} gave the evaluation of $\ZZ(B^3,\TT^3)$ with a certain colouring $c$ of the boundary $\partial\TT^3$ as $$\ZZ(\TT^3,B^3)_{c} = \la(\partial\TT^3)^*_{c}\ra.$$ Here $(\partial\TT^3)^*_{c}$ is the dual of the boundary triangulation mapped onto the plane and $\la\cdot\ra$ is the network evaluation that defines $\ZZ$. We take the right hand side to define the regularised state sum $\ZZ_{PR}$. This implies that we need to take the spin network evaluation defined in Section \ref{sec-SpinNets} of the dual of the boundary coloured by the state $\Psi$. To give this we will need to give a certain convention for the ordering at the vertices. This can be obtained by considering the dual network of a particular mapping of the dual of $\partial\TT^3$ onto the plane. We then can straighten out the network such that every edge is a straight line. If necessary we can move the vertices slightly to ensure that no vertex is directly above or below another one. Then we can pull down the vertices slightly so they all have three legs facing upwards and use the ordering from left to right at the vertices in the definition of the state $\Psi$. The network evaluation is then simply given by contracting the intertwiners in $\Psi$ according to the network using the epsilon inner product $(\cdot,\cdot)=\braket{J\cdot}{\cdot}$. To keep track of the contractions we will label the vertices with numbers $a$, such that $a<b$ if and only if the vertex dual to $\sigma^2_a \in (\partial \TT^3)_2$ is left of the vertex dual to $\sigma^2_b$. We write $\ZZ_{PR}(\Psi)$ for the value of $\ZZ_{PR}(B^3,\TT^3)$ evaluated at the colouring $\Psi$ and have:

\bea\label{eq-PRAmplitude}
\ZZ_{PR}(\Psi) &=& \left(\prod_{\sigma^2_a} \int_{\SU(2)} \dd X_a\right) \prod_{\sigma^1_{bc}:\; b<c} ( X_b \ket{\nb_{bc}}_{j_{bc}} , X_c \ket{\nb_{cb}}_{j_{bc}})\nn\\ &=& \int\dd X \prod_{\sigma^1_{bc}:\; b<c} \bra{-\nb_{bc}} X_b^\dagger X_c \ket{\nb_{cb}}_{j_{bc}}.
\eea

Where we have introduced the convention $$\int\dd X = \left(\prod_{\sigma^2_a} \int_{\SU(2)} \dd X_a\right),$$ and $\ket{\nb_{ab}}_{j_{ab}}$ satisfy the Regge phase condition \eqref{eq-2dReggePhase} with respect to the ordering chosen here. We have used the convention $J\ket{\nb} = \ket{-\nb}$. Note that $S^2$ has only one spin structure, so taking into account the symmetries of the state $\Psi$ given in Lemma \ref{lem-2dReggeStateSymmetries}, the amplitude $\ZZ_{PR}(\Psi)$ is really defined up to sign in terms of the spins $j_{ab}$. The sign is fixed by the ordering convention.

An immediate consequence of these symmetries is that as $J \Psi = \Psi$ and $(J\cdot ,J\cdot) = \overline{(\cdot,\cdot)}$ the amplitude $\ZZ_{PR}(\Psi)$ is real.

\subsection{Exponential Form of the State Sum}

The exponential property of coherent states \eqref{eq-costa-expo} allows us to write the amplitude \eqref{eq-PRAmplitude} in exponential form. Note that the phase condition \eqref{eq-2dReggePhase} holds for every factor of the exponential decomposition. Writing $\ket{\nb} = \ket{\nb}_{\frac12}$ we thus have that $$\ket{\nb_{ab}}_{j_{ab}} = \bigotimes^{2j_{ab}} \ket{\nb_{ab}}.$$
Using this in the amplitude $\ZZ_{PR}(\Psi)$ we obtain its exponential form:

\be\label{eq-PRAmpExpo}
\ZZ_{PR}(\Psi)=\int\dd X \prod_{\sigma^1_{bc}\; b<c} \bra{-\nb_{bc}} X_b^\dagger X_c \ket{\nb_{cb}}^{2j_{bc}} = \int\dd X e^{S_{PR}}
\ee

with \be\label{eq-PRAction}S_{PR}= \sum_{\sigma^1_{bc}\; b<c} 2j_{bc}\ln(\bra{-\nb_{bc}} X_b^\dagger X_c \ket{\nb_{cb}}).\ee

$S_{PR}$ and the logarithms in its definition are only defined up to factors of $2\pi$. This is now linear in the spins $j$, that is $S_{PR}(X,\nb,\lambda j) = \lambda S_{PR}(X,\nb,\lambda j)$. Thus the amplitude is now in a form where we can apply stationary phase analysis to analyse its asymptotic behaviour for large $j$. As we are analysing the behaviour of the integral over the $X$ variables we consider $\nb_{ab}$ and $j_{ab}$ to be fixed boundary data.

\subsubsection{Symmetries of the Action}\label{sec-SymmPRAction}

Keeping $\nb$ and $j$ fixed, the $X$ dependence of the action \eqref{eq-PRAction} allows for two kinds of symmetries:

\begin{itemize}
\item Continuous: The transformation $X_a \rightarrow Y X_a$ for all $Y\in \SU(2)$ leaves the action invariant.
\item Discrete: The transformation $X_a \rightarrow \epsilon_a X_a$ for $\epsilon = \pm 1$ leaves the action invariant due to the integrality condition on the spins $j_{ab}$.
\end{itemize}

\section{Asymptotic Geometry of the State Sum}

Given a coherent boundary state $\Psi$ we can write $\Psi_\lambda$ for the same state with rescaled spins $\lambda j_{ab}$. We then want to use stationary phase analysis on the integral \eqref{eq-PRAmpExpo} to analyse the leading order behaviour of $\ZZ_{PR}(\Psi_\lambda)$ in $\frac{1}{\lambda}$. A brief review of stationary phase methods is given in Appendix \ref{sec-StatPhase}. To find the leading order contribution we need to find the critical and stationary points of the action $S_{PR}$, that is, the group elements $X_a$ such that $\Reel(S_{PR})$ is maximal and $\delta_X S_{PR} = 0$. We will give a geometric characterisation of these points and evaluate the action $S_{PR}$ at them.

\subsection{Asymptotic Equations of Motion}\label{sec-AsymptoticEOM3d}

\subsubsection{Critical Points}

We begin with analysing the condition that $\Reel(S_{PR})$ is maximal. Note first that as the modulus square of the inner product between two coherent states,

\be\tag{\ref{eq-costa-innerprod}}
  |\la \alpha_j(\nb)| \alpha'_j(\nb') \ra|^2 = \left( \frac{1}{2} (1 + \nb \cdot \nb') \right)^{2j},
\ee
is smaller than or equal to $1$ we have that $\Reel(S_{PR}) \leq 0$. We are thus interested in the cases where it attains this maximum $\Reel(S_{PR}) = 0$. If the boundary data is such that the maximum is smaller than $0$ the amplitude is exponentially suppressed. It is more convenient to work with the exponentiated condition. We then obtain that $$\prod_{\sigma^1_{bc}\; b<c} |\bra{-\nb_{bc}} X_b^\dagger X_c \ket{\nb_{cb}}|^{2\lambda j_{bc}} = 1.$$

This implies that $|\bra{-\nb_{bc}} X_b^\dagger X_c \ket{\nb_{cb}}| = 1$ for all $bc$. Noting that $X_c \ket{\nb_{bc}} = \alpha(\hat{X}_c \nb_{bc})$ is a coherent state and using equation \eqref{eq-costa-innerprod} we obtain the following critical point equations:

\be\label{eq-PRCritPoint}
\hat{X}_b \nb_{bc} = - \hat{X}_c \nb_{cb}.
\ee

In the following we will find it convenient to introduce the shorthand notation $\nb'_{bc} = \hat{X}_b \nb_{bc}$.

\subsubsection{Stationary points}

We next give the equations for the variation of the action $\delta_X S_{PR}$. To do so we will use the fact that the variation of the matrix elements of the group are given by the Lie algebra. That is $$\delta^i_{X} \bra{\alpha} X \ket{\alpha'} = \frac{d}{ds} \bra{\alpha} \exp(s L^i) X \ket{\alpha'}|_{s=0} = \bra{\alpha} L^i X \ket{\alpha'}.$$

Similarly we have $$\delta^i_{X} \bra{\alpha} X^\dagger \ket{\alpha'} = \frac{d}{ds} \bra{\alpha} X^\dagger \exp(- s L^i)  \ket{\alpha'}|_{s=0} = - \bra{\alpha} X L^i \ket{\alpha'}.$$

The variation of the action is then 
\bea \delta_{X_a}^i S_{PR} &=& \sum_{b:\, \exists \sigma^1_{ba},\; b<a} 2j_{ab} \delta_{X_a}^i \ln(\bra{-\nb_{ba}} X_b^\dagger X_a \ket{\nb_{ab}}) \nn\\
&&+ \sum_{b:\, \exists \sigma^1_{ab},\; a<b} 2j_{ab} \delta_{X_a}^i \ln(\bra{-\nb_{ab}} X_a^\dagger X_b \ket{\nb_{ba}}) \nn\\
&=& \sum_{b:\, \exists \sigma^1_{ba},\; b<a} 2j_{ab} \frac{\delta_{X_a}^i \bra{-\nb_{ba}} X_b^\dagger X_a \ket{\nb_{ab}}}{\bra{-\nb_{ba}} X_b^\dagger X_a \ket{\nb_{ab}}} \nn\\
&&+ \sum_{b:\, \exists \sigma^1_{ab},\; a<b} 2j_{ab} \frac{\delta_{X_a}^i \bra{-\nb_{ab}} X_a^\dagger X_b \ket{\nb_{ba}}}{\bra{-\nb_{ab}} X_a^\dagger X_b \ket{\nb_{ba}}} \nn\\
&=& \sum_{b:\, \exists \sigma^1_{ba},\; b<a} 2j_{ab} \frac{\bra{-\nb_{ba}} X_b^\dagger L^i X_a \ket{\nb_{ab}}}{\bra{-\nb_{ba}} X_b^\dagger X_a \ket{\nb_{ab}}} \nn\\
&&- \sum_{b:\, \exists \sigma^1_{ab},\; a<b} 2j_{ab} \frac{\bra{-\nb_{ab}} X_a^\dagger L^i X_b \ket{\nb_{ba}}}{\bra{-\nb_{ab}} X_a^\dagger X_b \ket{\nb_{ba}}}.\nn
\eea

Using \eqref{eq-PRCritPoint} and the definition of coherent states, together with the observation that the denominator cancels the phase of the coherent states, we can easily evaluate these matrix elements:

$$\frac{\bra{-\nb_{ba}} X_b^\dagger L^i X_a \ket{\nb_{ab}}}{\bra{-\nb_{ba}} X_b^\dagger X_a \ket{\nb_{ab}}} = \frac{i}{2} \nb_{ab}.$$

Alternatively we could also use equation \eqref{eq-costa-diffstates} to give the full answer but we will not need this here.

Now combining these equations we obtain the stationary phase conditions:

\be
\delta_{X_a}^i S_{PR} = i \sum_{b:\, \exists \sigma^1_{ba},\; b<a} j_{ab} X_a \nb_{ab} - i \sum_{b:\, \exists \sigma^1_{ab},\; a<b} j_{ab} X_b \nb_{ba} = i \sum_{b:\, \exists \sigma^1_{ba}} j_{ab} X_a \nb_{ab}.\nn
\ee

Therefore the condition $\delta_{X_a}^i S_{PR} = 0$ is simply closure:

\be\label{eq-PRStatPoint}
\sum_{b:\, \exists \sigma^1_{ba}} j_{ab} X_a \nb_{ab} = \sum_{b:\, \exists \sigma^1_{ba}} j_{ab} \nb'_{ab} = 0.
\ee

\subsection{Geometricity}

We now can give a geometric interpretation for the solutions $X_a$ to equations \eqref{eq-PRCritPoint} and \eqref{eq-PRStatPoint}. Note that \eqref{eq-PRStatPoint} is actually a condition on the $j_{ab}$ and $\nb_{ab}$ only. It is automatically satisfied as the $j_{ab}$ and $\nb_{ab}$ are defined in terms of the geometry underlying $\Psi$. Thus they automatically close. However, this equation is also sufficient to force the $j_{ab}$ and $\nb_{ab}$ to define a boundary geometry. This shows that if we had chosen to allow non-geometric boundary data the boundary geometricity would be forced by the equations of motion.

The interpretation of \eqref{eq-PRCritPoint} is given by the following lemma.

\begin{lem}[3d Geometricity]\label{lem-3dGeometricity}

A set of vectors $\vb_{ab} \in \R^3$ associated to the edges $\sigma^1_{ab}$ of a triangulation $\TT^2$ of the oriented sphere $S^2$ satisfying \[\sum_{a:\, \exists \sigma^1_{ab}} \vb_{ab} = 0 \] and \[\vb_{ab} = - \vb_{ab}\] defines a simplex-wise affine linear, continuous and orientation preserving immersion $\imm: \, \TT^2 \rightarrow \R^3$ of the sphere $S^2$ up to translations.
\end{lem}
\begin{proof}
The idea is to define $\vb_{ab}$ as the vector associated to the boundary edge of the immersed triangle $\imm(\sigma_a)$. As $\imm_a$, the restriction $\imm$ to $\sigma^2_a$, is required to be affine linear, this fixes $\imm_a$ up to translation. It remains to be shown that the translation parts of $\imm_a$ can be chosen to make $\imm$ continuous.

To see this first note that as the $\sigma^2_a$ are oriented the triangles $\imm_a(\sigma_a)$ inherit an orientation and $\vb_{ab}$ can be seen as the vector associated to the edge $\imm_a (\sigma^1_{ab})$ with orientation inherited from $\sigma_a$, while $\vb_{ba}$ is the same edge vector with orientation inherited from $\sigma_b$, $\imm_b (\sigma^1_{ba})$.

We now chose an arbitrary reference vertex $\sigma_r^0$ of the triangulation and place it at the origin of $\R^3$: $\imm(\sigma_r^0) = (0,0,0)$. Then for any other vertex $\sigma^0$ pick an arbitrary collection of edges $\sigma^1$ that form a continuous oriented path $P_{\sigma^0}$ between the vertex mapped to the origin and $\sigma^0$. Then define the position of the other vertices by $\imm(\sigma^0) = \sum_{\sigma^1_{ab}\in P_{\sigma^0}} \vb_{ab}$ with $\vb_{ab}$ the vector associated with the appropriately oriented edges. Any two paths $P_{\sigma^0}$ and $P'_{\sigma^0}$ differ by a closed path, and on the sphere all closed paths are the sum of boundaries of disks. Closure is just the condition that the sum of $\vb_{ab}$ along the boundary of a disk is $0$, and therefore $\imm(\sigma^0)$ is independent of the paths chosen. 

The map $\imm$ then immediately extends to the edges and triangles of $\TT^2$. It is unique up to translation of the reference vertex $\imm{\sigma^0_r}$. A change of reference vertex merely induces a further translation.\end{proof}

This lemma applied to $\vb_{ab} = j_{ab} \nb'_{ab}$ immediately gives us an immersion $\imm$ of the boundary geometry encoded in our state $\Psi$. On the other hand every immersion $\imm$ defining the same geometry on $S^2$ gives us a solution to the equations \eqref{eq-PRCritPoint}. To see this note that $\phi_a$ of the Definition \ref{defi-2dReggeState} of $\Psi$ is defined to be simplex-wise linear. The rotation part of the map $\imm \circ \phi_a^{-1}$ then maps the vectors $j_{ab} \nb_{ab}$ to the $\vb_{ab}$ of the immersion. Thus if the geometries of $\phi_a(\sigma^2_a)$ and $\imm(\sigma^2_a)$ agree, the rotation part of $\imm \circ \phi_a^{-1}$ defines an $\SO(3)$ element $\hat{X}_a$. This now solves \eqref{eq-PRCritPoint} as $$\hat{X}_a \nb_{ab} = \imm \circ \phi_a^{-1} \nb_{ab} = \frac{1}{j_{ab}} \imm(\sigma^1_{ab}) = \frac{1}{j_{ab}} \vb_{ab} = - \frac{1}{j_{ab}} \vb_{ba} = - \hat{X}_{b} \nb_{ba}.$$

Thus we have the following corollary:

\begin{Cor}[Classification of the Solutions]\label{cor-PRClassification}
The simplex-wise linear, orientation preserving immersions up to translation of the triangulated sphere $S^2$ with geometry that defined by $j_{ab}$, $\nb_{ab}$ satisfying closure are in one to one correspondence with the solutions to equation \eqref{eq-PRCritPoint}.
\end{Cor}

Note that the requirement that $\hat{X}_a$ be in $\SO(3)$ fixes a normal ``outward" direction on the immersed surface $\imm(S^2)$ in terms of the fiducial normal direction to $\no = (1,0,0)$ used in Definition \ref{defi-2dReggeState}. This is given explicitly by $\nb_a = \hat{X_a}\no$.

To fully understand the geometry of our solutions we need to connect it to the geometry of our state. In particular we can see that $\hat{X}_a$ relates the boundary connection $\hat{g}_{ab}$ to the dihedral connection $\hat{d}_{ab}$ of the immersion $\imm$. We define the dihedral connection as $\hat{d}_{ab} \in \SO(3)$:

\bea\label{eq-3dDihedralConnect}
\hat{d}_{ab} \nb_b &=& \nb_a\nn\\
\hat{d}_{ab} \vb_{ab} &=& \vb_{ab}.
\eea

That is, it is the connection that maps the tangent spaces of the immersed surface $\imm(S^2)$ into each other. We then have the following commuting diagram:

\begin{equation}
\label{eq-3dCommDiagSO3}
\xymatrix{\ar @{} [dr]  \phi_a(\sigma^2_a) \ar[d]_{\hat{g}_{ba}} \ar[rr]^{\hat{X}_a} && ~\imm(\sigma^2_a) \ar[d]^{ \hat{d}_{ba}}   \\
\phi_b(\sigma^2_b)\ar[rr]_{\hat{X}_b} && ~\imm(\sigma^2_b)  }
\end{equation}

This follows from the various definitions and can be checked explicitly, $$\hat{g}_{ba}\hat{X}_a^{-1} \hat{d}_{ab} \hat{X}_b \nb_{ba} = \hat{g}_{ba}\hat{X}_a^{-1} \hat{d}_{ab} \frac{\vb_{ba}}{j_{ab}} = \hat{g}_{ba}\hat{X}_a^{-1} \frac{- \vb_{ab}}{j_{ab}} = -\hat{g}_{ba} \nb_{ab} =  \nb_{ba},$$
and
$$\hat{g}_{ba}\hat{X}_a^{-1} \hat{d}_{ab} \hat{X}_b \nb_0 = \hat{g}_{ba}\hat{X}_a^{-1} \hat{d}_{ab} \nb_{b} = \hat{g}_{ba}\hat{X}_a^{-1} \nb_{a} = \hat{g}_{ba} \nb_{0} = \nb_{0}.$$

This tells us that we can consider $\hat{X}_a$ as a gauge transformation that takes the boundary geometry connection associated to $\Psi$ to the dihedral connection associated to the immersion $\imm$. To complete the geometric characterisation of the solutions $X_a$ we have to lift this analysis to $\SU(2)$. Note first that using the spin structure on $S^2$ we can lift $\hat{d}_{ab}$ to an $\SU(2)$ connection $d_{ab}$ in the same way as we lifted $\hat{g}_{ab}$ up to the sign ambiguity $d_{ab} \rightarrow \epsilon_a \epsilon_b d_{ab}$.

This means using Lemma \ref{lem-CommDiagLift}  we immediately have the commuting diagram:

\begin{equation}\label{eq-3dCommDiagSU2}
\xymatrix{\ar @{} [dr]  \phi_a(\sigma^2_a) \ar[d]_{g_{ba}} \ar[rr]^{\epsilon'_a X_a} && ~\imm(\sigma^2_a) \ar[d]^{d_{ba}}   \\
\phi_b(\sigma^2_b)\ar[rr]_{\epsilon'_b {X}_b} && ~\imm(\sigma^2_b)  }
\end{equation}

\subsubsection{Symmetries of the Solutions}

As the action \eqref{eq-PRAction}, giving rise to the equations \eqref{eq-PRCritPoint} and \eqref{eq-PRStatPoint}, has the symmetries discussed in Section \ref{sec-SymmPRAction}, we will describe how these symmetries act on the solutions. Note first that by corollary \ref{cor-PRClassification} the solutions are in one-to-one correspondence with immersions $\imm$ with the same boundary geometry as that induced by the $\phi_a$ and are then given by $\hat{X}_{a} = \imm \circ \phi_a$. The continuous symmetries act by $\hat{X}_a \rightarrow \hat{Y}\hat{X_a}$ for $\hat{Y} \in \SO(3)$ and it follows immediately that the symmetry acts on the associated immersion by $\imm \rightarrow \hat{Y} \imm$.

Note that given $\imm$ we can obtain a new solution $\imm'$ with the same boundary geometry not just through $\SO(3)$ rotations but through $\OO(3)$ as well. In particular if $n'_{ab}$ defines a solution then so does $P n'_{ab} = -n'_{ab}$. This is not a symmetry of the action, thus we obtain a genuinely new solution this way.

Finally the discrete symmetries simply change the signs $\epsilon'_a$ in \eqref{eq-3dCommDiagSU2}.

\subsubsection{Immersible Geometries and the Generalised Regge Action}

Without giving proofs we will briefly consider the type of geometries and actions consistent with the geometricity condition derived above. Consider an arbitrary interior triangulation $\TT^3$ of $B^3$ inducing the boundary triangulation $\TT^2$ of $S^2$. We call a simplex-wise flat metric on $B^3$ immersible in $\R^3$ if there is a simplex-wise isometric, continuous immersion of $B^3$ into $\R^3$. Clearly all flat metrics are immersible, however there is a much wider class of immersible surfaces. Given an immersion of the surface triangulation we can extend it to an immersion of the interior triangulation simply by placing the vertices of the interior triangulation anywhere in $\R^3$. The pullback of the metric to the interior under this immersion will then obviously be immersible. Whereas flat metrics can be characterised by the fact that the sum of dihedral angles around an edge equals $2\pi$, $$\sum_{\sigma_a^3 \ni \sigma^1} \Theta^{\sigma^1}_{\sigma_a^3} = 2\pi,$$ the immersible surfaces can be characterised by the fact that there is a set of signs $\mu_a$ associated to the 3-simplices such that $$\sum_{\sigma_a^3 \ni \sigma^1} \mu_a \Theta^{\sigma^1}_{\sigma_a^3} = \mathbb{Z}2\pi.$$ In the case of the immersible metric defined by the pullback of the standard metric along an immersion, the signs are given by the pullback of the orientation relative to a fiducial standard orientation on $B^3$.

As an immersion of the manifold induces an immersion of the boundary geometry our geometricity theorem tells us that the boundary geometry encoded in $\Psi$ has to be the boundary of an immersible geometry for critical points to exist.

The Regge Action is the action of discretised GR on a triangulated manifold. Its solutions are the flat metrics. It is given by $$S_R = \sum_{\sigma^1} |\sigma^1| \left(\sum_{\sigma_a^3 \ni \sigma^1} \Theta^{\sigma^1}_{\sigma_a^3} - 2\pi\right).$$ On solutions it is zero except at the boundary. We then define the generalised Regge actions as $$S'_R (\mu_a, N_{\sigma^1}) = \sum_{\sigma^1} |\sigma^1| \left(\sum_{\sigma_a^3 \ni \sigma^1} \mu_a \Theta^{\sigma^1}_{\sigma_a^3} - N_{\sigma^1} 2\pi\right),$$ with $\mu_a = \pm$ and $N_{\sigma^1}\in \mathbb{Z}$. Every immersible metric is a solution to one of the generalised Regge actions. On solutions its only non-zero contribution is again given by the boundary terms and thus depends only on the immersed boundary geometry.

\subsection{The Regge Action of an Immersed Surface}

Taking the discusion of the preceding subsection as motivation we will define the Regge action for an oriented, immersed surface $\imm(S^2)$ described by a set of vectors $\vb_{ab} = -\vb_{ba}$ directly. We will call the normalized vector $\hat{\vb}_{ab} = \frac{\vb_{ab}}{|\vb_{ab}|}$. For a convex polyhedron the boundary Regge action is defined as $\sum_{\sigma^1_{ab}} |\vb_{ab}| \Theta_{ab}$ where $\Theta_{ab} = \Theta_{ba}$ is the dihedral angle defined by $\nb_a\cdot\nb_{b} = \cos(\Theta_{ab})$ and $0\leq \Theta_{ab}< \pi$. As we don't have convexity we cannot confine the range of dihedral angles. To fix the sign of the dihedral angle we will instead use the orientation on the surface and the dihedral connection $\hat{d}_{ab}$. This has to be of the form $\hat{d}_{ab} = \exp{\Theta_{ab} \hat{\vb}_{ab} \cdot L_1}$ where $L_1$ are the generators of $\SO(3)$, and the spin $1$ generators of $\SU(2)$. We take this as the definition of the dihedral angles. As $\hat{\vb}_{ab} = -\hat{\vb}_{ba}$ this definition satisfies $\Theta_{ab} = \Theta_{ba}$. The Regge action is then again simply 

$$S_R(\imm) = \sum_{\sigma^1_{ab}} |\vb_{ab}| \Theta_{ab}.$$

If the polyhedron is convex and the orientation on the surface is that induced through the outward facing normals $\nb_{a}$, it furthermore follows that $0\leq \Theta_{ab}< \pi$. Thus this definition agrees with the standard one in this case.

Note that the oriented surface obtained through parity $P$ has the opposite dihedral angles as the definition of $\hat{d}_{ab}$ in equation \eqref{eq-3dDihedralConnect} is invariant under parity $\vb_{ab} \rightarrow \vb'_{ab} = P \vb_{ab} = -\vb_{ab}$. Thus we have that 

$$\hat{d}_{ab} = \exp{\left(\Theta_{ab} \hat{\vb}_{ab} \cdot L_1\right)} = \exp{\left(\Theta'_{ab} \hat{\vb}'_{ab} \cdot L_1\right)} = \exp{\left(- \Theta'_{ab} \hat{\vb}_{ab} \cdot L_1\right)}$$
and thus that $\Theta_{ab} = - \Theta'_{ab}$, and the boundary Regge action changes sign under parity.

\subsection{The Action}

We can now evaluate the action \be\tag{\ref{eq-PRAction}}S_{PR}(j_{ab}, n_{ab}, X_a) = \sum_{\sigma^1_{ab}\; a<b} 2j_{ab}\ln(\bra{-\nb_{ab}} X_a^\dagger X_b \ket{\nb_{ba}})\ee on the solutions in terms of their geometry. We will show in particular that it is simply equal to the Regge action of the immersed surface.

To do so first note that for any covering lift $d_{ab}$ of the dihedral connection have $$X_a J \ket{\nb_{ab}} = \epsilon'_a \epsilon'_b d_{ab} X_b g_{ba}^{(-1)} J \ket{\nb_{ab}} = \epsilon'_a \epsilon'_b d_{ab} X_b \ket{\nb_{ba}},$$ with the signs those of the commuting diagram \eqref{eq-3dCommDiagSU2}.

Now note that $X_b\ket{\nb_{ba}}$ is a coherent state $\alpha(\hat{X}_b \nb_{ba})$, and therefore it is a $+\frac{i}{2}$ eigenstate of $(\hat{X}_b \nb_{ba}) \cdot L = \nb'_{ab} \cdot L$. Now $d_{ba} = \nu_{ab} \exp\left( \Theta_{ab} \nb'_{ba} \cdot L \right) = \nu_{ab} \exp\left( - \Theta_{ab} \nb'_{ab} \cdot L \right)$ is a covering lift of $\hat{d}_{ab}$ compatible with the spin structure. As the discrete symmetries of the action range over the full set of signs $\epsilon'_a$ we can then evaluate the matrix elements appearing in the action at the point where $\epsilon'_a = 1$:

\bea
\bra{- \nb_{ab}} X^\dagger_a X_b \ket{\nb_{ba}} &=& \bra{-\nb'_{ab}} X_b g_{ba} J \ket{\nb_{ab}} \nn\\&=& \bra{-\nb'_{ab}} d_{ba} X_a J \ket{\nb_{ab}} \nn\\&=& \bra{-\nb'_{ab}} d_{ba} \ket{-\nb'_{ab}} \nn\\&=& \nu_{ab} e^{\frac{i}{2} \Theta_{ab}}\nn
\eea
where we have used the Regge phase choice in the first line, the commuting diagram in the second line and the explicit form of the dihedral connection in the last. Thus we finally obtain that

\be\label{eq-PRActionEvalToRegge}
S_{PR} = i \sum_{\sigma^1_{ab}} j_{ab} \Theta_{ab} = i S_R.
\ee

\subsection{The Asymptotic Formula}

We can now combine the analysis of the preceding sections to give the leading order behaviour of $\ZZ_{PR}(\Psi_{\lambda})$. We write $\nu_{crit} = \prod_{a<b} \nu_{ab}^{2 j_{ab}}$ at each critical point. To apply the asymptotic analysis of Section \ref{sec-StatPhase} it is necessary to factor out the symmetries of the action. We can easily do so by fixing one of the group elements $X_a$ to be $\pm \id$. We are then left with $2^{|\TT^2_2|}$ points related by the discrete symmetries. As these are isolated from each other we do not need to fix them. However, it is possible for the immersion of a surface to allow flexing, that is, continuous deformations that do not change the boundary geometry. If we have such a case and the space of flexing solutions is a submanifold of the configuration space we call it $\mathfrak{I}$ and call it a flexifold. Note that the Regge action is constant on these flexifolds. In this case we have to apply the generalized theorem of Appendix \ref{sec-StatPhase}. We call $(sc) = \frac{3|\TT^2_2| - 1}{2}$ the scaling induced by the $3|\TT^2_2| - 1$ integrations we have after fixing the continuous symmetry, and $d_{max}$ the dimension of the maximal dimension flexifold. We thus have:

\begin{theo}[Asymptotic Behaviour of the Ponzano Regge Amplitude]\label{theo-PRAsymptotics}
If the boundary geometry induced by $\Psi$ on the $\TT^2$ triangulated sphere $S^2$ allows no flexible immersions we have that
\be\label{eq-PRAsymptoticsNoFlex}
\ZZ_{PR}(\Psi_{\lambda}) = \lambda^{- (sc)} N \sum_\imm \nu_\imm^\lambda \frac{e^{i S_R (\imm)}}{\sqrt{\det(-H_{|\imm})}} + o(\lambda^{- (sc) -1}),
\ee
where $H_{|\imm}$ is the Hessian of $S_{PR}$ evaluated at the critical point associated to $\imm$, and 
$$N = (2 \pi)^{-(sc)} \left(\frac{2}{(4 \pi)^2}\right)^{- (sc)}.$$

If there are no immersions of the boundary geometry and the sum is empty the amplitude is exponentially suppressed and we have that
$$\ZZ_{PR}(\Psi_{\lambda}) = o(\lambda^N) \,\, \forall N.$$

If the boundary geometry induced by $\Psi$ on the $\TT^2$ triangulated sphere $S^2$ allows flexible immersions, and the immersions form a manifold $\mathfrak{I}$ we have that
\be\label{eq-PRAsymptoticsFlex}
\ZZ_{PR}(\Psi_{\lambda}) = \lambda^{-(sc) + d_{max}} \sum_{\mathfrak{I}} \nu_{\mathfrak{I}}^\lambda N_{\mathfrak{I}} e^{i S_R (\mathfrak{I})} + o(\lambda^{-(sc) + d_{max} - 1}),
\ee
where the sum runs only over manifolds $\mathfrak{I}$ with dimension $d_{max}$, and $$N_{\mathfrak{I}} = (2 \pi)^{-(sc) + d_{max}} \left(\frac{2}{(4 \pi)^2}\right)^{- (sc) + d_{max}} \int_{\mathfrak{I}} \frac {1}{\sqrt{ \det (-H^\perp)}}.$$

Here the integration measure and $H^\perp$ are defined as in Appendix \ref{sec-StatPhase}.

\end{theo}

\chapter{The Geometry of Euclidean 4-Dimensional State Sum Amplitudes}

In this chapter we will analyse the amplitudes for the 4-dimensional Ooguri model. As a corollary we can then easily give the asymptotics of the models of \cite{Engle:2007wy,Engle:2007qf,Freidel:2007py,Livine:2007ya,livine-2007-76} mentioned at the end of Section \ref{sec-SpinNets}. The Ooguri model is the state sum corresponding to $\SU(2)\, BF$-theory. As for $\ZZ_{PR}$, its amplitude is given by the graphical evaluation of a spin network dual to the surface triangulation, in this case that of a 4-simplex. The evaluation of this spin network is known in recoupling theory as the $\ftj$ symbol as we have ten edges that can be coloured by spins $j$ and five intertwiners, a basis of which can also be coloured by spins. We will colour the $\ftj$ symbol with a coherent boundary state $\Psi$ and analyse its large spin behaviour. The key complication compared to the last chapter will be the significantly more tricky analysis of geometricty.

This will allow us to give a straightforward evaluation of the geometry arising in various new spin foam models proposed recently.

\section{Definition of the $\SU(2)\, BF$ Amplitude}

The 4-simplex amplitude of the Ooguri model is, up to sign, the $\ftj$ symbol which is, again up to sign, the contraction of intertwiners according to the combinatorics of a 4-simplex, see e.g. Figure \ref{fig-15j}.

\begin{figure}[htbp]
\begin{center}
\psfrag{r}{$j$}
\psfrag{i}{$\iota$}
\includegraphics[scale=1]{b-4simplex}
\caption{$\ftj$ combinatorics.}
\label{fig-15j}
\end{center}
\end{figure}

If the colouring of the spin network is given by a coherent boundary state $\Psi_C$ of Definition \ref{defi-3dCoherentBoundaryStates} we call its evaluation $\ftj(\Psi_C)$. This can then be given explicitly again by contracting $\Psi_C$ with the epsilon inner product $(,)$. We use an arbitrary labelling of the five vertices of the network $a,b,c = 1,\dots, 5$ and we obtain:

\bea\label{eq-ftjCoherentBoundaryState}
\ftj(\Psi_C) = (-1)^s \left(\prod_{\sigma^3_a} \int_{\SU(2)} \dd X_a\right) \prod_{b<c} (X_b \alpha({\nb_{bc}})_{j_{bc}}, X_c \alpha({\nb_{cb}})_{j_{bc}}).
\eea

Here $s$ is a function of $j$, and its form depends on the particular representation of the diagram chosen, as well as on the arbitrary ordering of the vertices. We will again abbreviate $$\int \dd X = \left(\prod_{\sigma^3_a} \int_{\SU(2)} \dd X_a\right).$$ Using the exponential properties of coherent states this immediately takes on exponential form:

\bea\label{eq-ftjCoBoStaexponential}
\ftj(\Psi_C) &=& (-1)^s \int \dd X \prod_{a<b} (X_a \alpha({\nb_{ab}}), X_b \alpha({\nb_{ba}}))^{2j_{bc}}\nn\\
&=& (-1)^s \int \dd X \prod_{a<b} \la J \alpha({\nb_{ab}})| X_a^{\dagger} X_b | \alpha({\nb_{ba}})\ra^{2j_{bc}}.
\eea

In the case of a Regge state $\Psi_R$ of Definition \ref{defi-3dReggeStates} using the same ordering $a<b$,  we again can use the coherent states $\ket{\nb_{ab}}$, $J\ket{\nb_{ab}} = \ket{-\nb_{ab}}$ with phase fixed by the boundary connection $g_{ba}$ and we have:

\bea\label{eq-ftjRegeStaexponential}
\ftj(\Psi_R) = (-1)^s \int \dd X \prod_{a<b} \la -\nb_{ab}| X_a^{\dagger} X_b |\nb_{ba}\ra^{2j_{bc}} = (-1)^s \int \dd X \exp({S_{\ftj}}
)\eea

with \be\label{eq-ftjActionRegState} S_{\ftj}(X_a,\nb_{ab}, j_{ab}) = i 2 \sum_{a<b} j_{ab} \ln\la -\nb_{ab}| X_a^{\dagger} X_b |\nb_{ba}\ra.\ee

The action is again only defined up to $2\pi$. We can also define an action for the more general case of a coherent boundary state:

\be\label{eq-ftjActionCoBoState} S_{\ftj}(X_a,\alpha(\nb_{ab}), j_{ab}) = i 2 \sum_{a<b} j_{ab} \ln\la J\alpha(\nb_{ab})| X_a^{\dagger} X_b |\alpha(\nb_{ba})\ra.\ee

This has the disadvantage of not only depending on the direction $\nb_{ab}$ associated to the coherent state but also on its phase. Nevertheless we will call $\nb_{ab}$ and $j_{ab}$ the boundary data for both actions. Both of these actions are again linear under rescaling of the spins $S_{\ftj}(\lambda j) = \lambda S_{\ftj}(j)$ and we can use stationary phase techniques in the same way as in the last chapter.

\subsubsection{Symmetries of the Action}
The symmetries of the actions are again the same discrete and continuous symmetries as in the 3-dimensional case:

\begin{itemize}
\item Continuous: The transformation $X_a \rightarrow Y X_a$ for all $Y\in \SU(2)$ leaves the actions invariant.
\item Discrete: The transformation $X_a \rightarrow \epsilon_a X_a$ for $\epsilon = \pm 1$ leaves the actions invariant due to the integrality condition on the spins $j_{ab}$.
\end{itemize}

\section{Asymptotic Geometry of the $\SU(2)\, BF$ Amplitude}

Going forward we will assume throughout that the tetrahedra in our boundary states $\Psi$ are non-degenerate. The calculation of the asymptotic equations proceeds completely analogously to the last chapter.

\subsection{Asymptotic Equations of Motion}

By the same calculation as in Section \ref{sec-AsymptoticEOM3d} we obtain the critical point equations

\be\label{eq-CritPointSU24d}
\hat{X}_a \nb_{ab} = - \hat{X}_b \nb_{ba}
\ee
and the stationary point equations evaluated when the critical point equations hold
\be\label{eq-StatPointSU24d}
\sum_{b: b\neq a}j_{ab} \nb_{ab} = 0.
\ee

Note that this calculation does not depend on the arbitrary phase of the coherent states $\alpha(\nb_{ab})$.

\subsection{Geometric Interpretation, General State}

Even though the critical and stationary point equations have remained the same the combinatorics of the data is now given by a genuinely 4-dimensional object, the 4-simplex, instead of a triangulated 2-surface. Interpreting the equations as a 3-dimensional geometry can be done but is unenlightening. Rather we will look for interpretations that are specific to the 4-simplex. In the most general case of the coherent boundary state $\Psi_C$ this is done by the following lemma:

\begin{lem}[Moduli Space of Constant b-fields on a 4-simplex]\label{lem-ConstbField}
The space of constant $\SU(2)$ Lie algebra-valued 2-forms $\bb$ on an oriented 4-simplex is parametrized by their surface integrals on the oriented faces $\sigma^2_{ab}$ of the 4-simplex, $$\bb_{ab} = \int_{\sigma^2_{ab}} \bb.$$

These surface integrals satisfy $$\bb_{ab} = -\bb_{ba}$$ and $$\sum_{b: b\neq a} \bb_{ab} = 0.$$ Conversely every set of vectors satisfying these conditions occurs as the surface integrals of a Lie algebra-valued 2-form.
\end{lem}
\begin{proof}
For the proof simple counting shows that both, the space of Lie algebra-valued 2-forms and the space of vectors satisfying the equations of the lemma are 18-dimensional linear spaces. The integrals $\int_{\sigma^2_{ab}}$ form a linear map from the space of Lie algebra-valued two forms to the space of such vectors. Its kernel is equal to the 2-form that is identically $0$. Thus the map is a vector space isomorphism and the lemma follows.\end{proof}

By setting $\bb_{ab} = j_{ab} \hat{X}_a \nb_{ab}$ this immediately gives us an interpretation consistent with the idea that the amplitude under consideration is a quantisation of $\SU(2)\, BF$ theory. The continuous symmetry acts as a rotation of the ``$\su(2)$-frame'' on the 4-simplex and reduces the the configuration spaces to 15 dimensions. However we cannot easily classify how many such $\bb$-fields are compatible with a given set of boundary data $\nb_{ab}$, $j_{ab}$. To do so we will now consider the special case when the boundary state is Regge.

\subsection{Geometric Interpretation, Regge state}\label{sec-SU24d GeomInt, Regge State}

\subsubsection{Solutions from Geometry}
As the geometry of a Regge state is that of a closed metric 3-manifold triangulated as the surface of a 4-simplex we can classify the states according to these geometries:

\begin{lem}[4-Simplex Boundary Geometries]\label{lem-4SimplexBoundary}
Every 3-sphere triangulated with the boundary triangulation of the 4-simplex and equipped with a simplex-wise flat non-degenerate Euclidean 3-geometry occurs as the boundary of a 4-simplex with flat 4d Euclidean, 4d Minkowski, or degenerate 3d Euclidean metric.
\end{lem}
\begin{proof}
To see this take some fiducial 4-simplex with edge vectors $V_{1a}^\mu = \delta^{\mu a}, \; a = 2,\ldots,4$. Note that given a 4-simplex with constant metric $g(\cdot,\cdot)$ on it, $g(V_{1a},V_{1b})$ can be evaluated using only the pullback of the metric to the boundary. Thus the boundary geometry determines the 10 numbers $g_{ab} = g(V_{1a},V_{1b})$ and the map $g(\cdot,\cdot) \rightarrow g(V_{1a},V_{1b})$ is linear and has zero kernel and is thus a vector space isomorphism. Finally note that the $g(V_{1a},V_{1b})$ uniquely determine the boundary geometry as they determine all the edge lengths.

As the metric on the boundary simplices is obtained by projecting onto the boundary tetrahedra and these have an Euclidean metric we can have at most one timelike or null direction in the metric. Thus we have the lemma. \end{proof}

We can then obtain solutions for the critical point equations from the linear immersions of these metric geometries into flat space. We again call such an immersion $\imm$. It is an affine linear map from the abstract 4-simplex to $\R^4$, and it induces a metric and orientation by pulling back either the Minkowski or the Euclidean metric and the standard orientation. By Lemma \ref{lem-4SimplexBoundary} every Regge state corresponds to a flat geometric 4-simplex. Now note that  every geometry on $\sigma^4$ with at most one timelike direction can be obtained from the pullback of the Euclidean or Minkowski metric. Thus there always exists an immersion $\imm$ that induces the same geometry on the boundary as the $\phi_a$ defining the Regge state. From now on we will consider the case where the boundary data is that of a Euclidean 4-simplex and the immersion is into Euclidean space. It will turn out that this is sufficient to fully classify the solutions.

We will find it convenient to again extend the $\phi_a$ to maps $\Phi_a$ into $\R^4$. We do so by identifying the plane orthogonal to $\pole$, $\pole^\perp$, and equipped with the orientation inherited by considering $\pole$ outward, with $\R^3$ and its standard orientation, and using $\phi_a$ to map into this subspace, that is, $\Phi_a = (0, \phi_a)$. Thus $\Phi_a(\sigma^3_a)$ is a tetrahedron orthogonal to $\pole$ with outward normals in $\pole^\perp$ given by $(0,\nb_{ab})$ and bivectors

\be\label{eq-TetBivectors} B_{ab}(\Phi_a(\sigma^3_a)) = \star j_{ab} \pole \wedge (0,\nb_{ab}) = j_{ab} (\nb_{ab}, - \nb_{ab}),\ee
where we have used that $\pole\wedge \nb = (\nb,\nb)$ as shown in Section \ref{sec-Spin4Geometry}. Note that the stationary and critical point equations are again only equations for the $\SO(3)$ elements covered by the $\SU(2)$ variables.

Note that as described in Section \ref{sec-Spin4Geometry}, two $\SU(2)$ elements $g^\pm$ combine into a $\Spin(4)$ element $(g^+, g^-)$. We write $(g^+, g^-)_{\SO(4)} = (-g^+, -g^-)_{\SO(4)}$ for the $\SO(4)$ element covered by them and $(\hat{g}^+, \hat{g}^-)$ for the corresponding element in $\SO(3)\times \SO(3)$. As the $\SO(4)$ action happens by adjoint we have that $-\id_{\SO(4)} = (-\id,\id)_{\SO(4)}$. Note that bivectors are actually only sensitive to the $\SO(3)\times\SO(3)$ action. We then have the following lemma.

\begin{lem}[Solutions from Immersed 4-Simplices]\label{lem-Immersed4simpSols}
A 4-simplex $\sigma^4_\imm$ immersed into Euclidean $\R^4$ with boundary geometry that of the Regge state gives two solutions to the critical and stationary point equations \eqref{eq-CritPointSU24d} and \eqref{eq-StatPointSU24d} $\hat{X}^\pm_a$. If the orientation induced on the boundary of $\sigma^4$ by $\imm$ is the same as the one used in the Regge state, the rotational part of $\imm\circ\Phi_a^{-1}$ is orientation preserving and we have

\be(X^+_a, X^-_a)_{\SO(4)} = \imm\circ\Phi_a^{-1}.\label{eq-SolutionsFromImmersions1}\ee

Otherwise $\imm\circ\Phi_a^{-1}\circ P$, where $P$ is parity, is orientation preserving and we have

\be(X^+_a, X^-_a)_{\SO(4)} = \imm\circ\Phi_a^{-1}\circ P.\label{eq-SolutionsFromImmersions2}\ee

In both cases the right-hand side means the rotational part of the affine map. The bivectors of $\sigma^3_a (\imm)$ are given by

$$B_{ab} (\sigma^4_{\imm}) = \mu_\imm j_{ab} (\hat{X}^+_a, \hat{X}^-_a) (\nb_{ab}, - \nb_{ab}),$$
with $\mu_\imm = +1$ in the first case and $\mu_\imm = -1$ in the second. The outward normals of the immersed 4-simplex are given by $N_a = (X^+_a, X^-_a)_{\SO(4)}\pole$
\end{lem}
\begin{proof}
By Theorem \ref{Theo-BivReconstruction} the bivectors $B_{ab}(\sigma^4_i)$ satisfy closure and orientation. If the orientations agree, the map \eqref{eq-SolutionsFromImmersions1} maps the geometric tetrahedron of the boundary state onto the geometric tetrahedron of the 4-simplex and thus maps its bivectors \eqref{eq-TetBivectors} to the geometric ones. If they disagree they map the tetrahedron with $\nb_{ab}$ inward normals to the boundary tetrahedra of the 4-simplex, accounting for the factor $\mu$. Putting the form of the $B_{ab}(\sigma^4_\imm)$ into the orientation equation of Theorem \ref{Theo-BivReconstruction} immediately shows that the $\hat{X_a^\pm}$ solve \eqref{eq-CritPointSU24d} and \eqref{eq-StatPointSU24d}. As $P \pole = \pole$ and the orientations were chosen with $\pole$ as outwards to $\pole^\perp$ the form of the outward normals $N_a$ also immediately follows.\end{proof}

It is possible that these solutions can be related by the symmetries. Say $Y \epsilon_a X^-_a = X^+_a$. In this case we have $$N_a = \epsilon_a (Y X^-_a, X^-_a)_{\SO(4)}\pole = \epsilon_a (Y,\id)_{\SO(4)}( X^-_a, X^-_a)_{\SO(4)}\pole = \epsilon_a (Y,\id)_{\SO(4)}\pole.$$ Thus if the two solutions are related by symmetries all normal vectors are parallel to each other. Conversely if the $N_a$ are all parallel then there is a $(Y,\id)$ that has $N_a = \epsilon_a (Y,\id)_{\SO(4)} \pole$ and $\epsilon_a (Y,\id)_{\SO(4)}^{-1} (X^+_{a}, X^-_a)_{\SO(4)}\pole = \pole$. This implies that this group element is in the diagonal subgroup and thus that $\pm \epsilon_a Y^{-1} X^+_{a} = X^-_a$, that is they are related by symmetry.

We will now further discuss the 3-dimensional geometry one obtains in this case. Without loss of generality we will assume that the solution has $N_a = \epsilon_a \pole$. Then the solution Lemma \ref{lem-Immersed4simpSols} gives us is of the form $(\epsilon_a X_a, X_a)_{\SO(4)}$. Consider the rotated tetrahedra $(\epsilon_a X_a, X_a)_{\SO(4)} \Phi_a (\sigma^3_a)$. Note that the outward normals of these tetrahedra do not satisfy orientation but instead $\epsilon_a \nb'_{ab} = -  \epsilon_b \nb'_{ba}$. This follows as the bivectors satisfy orientation and these are built with the ``outward" normals $N_a = \epsilon_a \pole$. That is, in this immersion of the 4-simplex into a 3-dimensional hyperplane we have some of the tetrahedra facing ``downward" and the rest ``upward". Note that the downward facing one are not 3-dimensional rotations of the tetrahedra $\Phi_a(\sigma^3_a)$ as the 4-dimensional rotation in the solution turns them ``upside down''. Instead they are related by rotation to the tetrahedra $-\Phi_a(\sigma^3_a)$ which have $\nb_{ab}$ as inward normals. This can be seen geometrically by looking at the 1-4 Pachner move in diagram \ref{fig-PAprovediagrams1-4}. There is one large tetrahedron and four small tetrahedra inside it. Where the small and the large tetrahedra are glued their outward normals are parallel rather than antiparallel. Taking the small tetrahedra as downward, rotating them up in four dimensions around the face on which they are glued to the large terahedron creates a four pointed star. This is equivalent to flipping them along the triangle on which they are glued. 

\subsubsection{Geometry from Solutions}

Crucially we can now go the other way and reconstruct a geometry given solutions. In particular Theorem \ref{Theo-BivReconstruction} tells us that given two solutions $\hat{X^+}$, $\hat{X^-}$ not related by any of the continuous symmetries, we obtain an embedded 4-simplex $\sigma^4_E$ as the bivectors defined by $j_{ab} (\hat{X}^+_a, \hat{X}^-_a) (\nb_{ab}, -\nb_{ab})$ satisfy all the conditions on geometric bivectors. This is immediate for simplicity, as they are orthogonal to $N_a = \pm (X^+_a, X^-_a)_{\SO(4)}\pole$, orientation, which is our critical point equations and closure. In Lemma \ref{lem-BivNonDeg} it was furthermore shown that either all normals are parallel, in which case by the discussion in the preceding chapter the solutions $\hat{X^+}$, $\hat{X^-}$ have to be related by symmetry, or we have to have full non-degeneracy. Thus two solutions reconstruct an embedded 4-simplex up to inversion.

We will see later that three solutions cannot occur. This leaves the case where we only have one solution. We already saw in the last section that one solution occurs when the boundary data is that of a 3d Euclidean 4-simplex. We thus need to show that it doesn't occur otherwise, that is, that no solutions exist for Lorentzian boundary data. We do this by giving a duality transformation that takes any solution into a second one and show that if the second solution is related to the first by symmetry we are in the case of the 3-dimensional Euclidean 4-simplex.

\begin{lem}[Second Solution from Involution]\label{lem-2ndSolFromInvol}
Given a solution $\hat{X}_a$ of \eqref{eq-CritPointSU24d} and the boundary connection $\hat{g}_{ab}$ there always exists a solution $\hat{X}'_a$ with \be\label{eq-Involution}{\hat{X'}_a}^{-1} \hat{X}'_b = \hat{g}_{ab} \hat{X}_b^{-1} \hat{X}_a \hat{g}_{ab}.\ee If $\hat{X'}_a$ and $\hat{X}_a$ are related by symmetries we have the case of a 3-dimensional Euclidean 4-simplex. 
\end{lem}
\begin{proof}
Without loss of generality consider the case $g_{1b} = \id$. This can always be done using the symmetries of the 3d Regge state given in Lemma \ref{lem-3dReggeStateSymmetries}. The tetrahedra $\phi_b(\sigma^3_b)$ are now glued around $\phi_1(\sigma^3_1)$ in a star shape. Now given a solution $\hat{X}_a$ use the continuous symmetries to set $\hat{X}_1 = \id$. Now equation \eqref{eq-Involution} reads: $$\hat{X'}_b^{-1} = \hat{X}_b,\;\; b = 2,\dots, 5.$$ Note that from $g_{1b} = \id$ we have $\nb_{1b} = -\nb_{b1}$. It follows that $\hat{X}_b \nb_{1b} = \nb_{1b}$, that is, $\hat{X}_b$ are rotations around $\nb_{1b}$ such that $\hat{X}_b \nb_{bc} = - \hat{X}_c \nb_{bc}$. Now note further $\nb_{1b}$, $\nb_{b1}$, $\nb_{bc}$ and $\nb_{cb}$ are all orthogonal to the edge $\phi_1(\sigma^1_{1bc})$ of tetrahedron $1$. This is because we have chosen a configuration for which the Regge gluing is already satisfied and thus $\phi_1(\sigma^1_{1bc})$ is parallel to $\phi_b(\sigma^1_{b1c})$ which is orthogonal to $\nb_{bc}$. Now using the shorthand $\nb_{bc}' = \hat{X}_b \nb_{bc}$ we write the critical equation $\nb_{bc}' = - \nb_{cb}'$. Clearly the vectors $\nb^r_{bc}$ obtained by reflecting $\nb'_{bc}$ along $\phi_1(\sigma^1_{1bc})^\perp$ still satisfy this equation. As $\nb'_{bc}$ is obtained from $\nb_{bc}$ by rotating it out of the plane $\phi_1(\sigma^1_{1bc})^\perp$ by a rotation around another vector in that plane, $\nb^r_{bc}$ can be obtained by rotating around the same vector with a negative angle, that is, by the inverse of the rotation. It follows that $\hat{X'}_b$ is indeed an alternative solution to the critical equations. The full form of the involution in the lemma reduces to this in a particular gauge and is gauge invariant and therefore has to hold as stated\footnote{For a fully gauge invariant and thoroughly incomprehensible proof see \cite{Barrett2010a}.}. 

We then need to show that if the resulting second solution is gauge related to the first we have the case of a 4-simplex immersed in $\R^3$.

Taking the gauge fixed form of the equations we can see that in this case we have to have $\hat{X}_b = \hat{X}_b^{-1}$ and thus $\hat{X}_b = 1$ or the rotation by $\pi$ around $\nb_{1b}$. In the latter case define $\epsilon'_b$ as $\hat{X}_b = \exp(\epsilon'_b \pi \nb_{1b} \cdot L_1)$. Then use the symmetries of the action with $Y = \epsilon'_b \id_{\SO(4)}$ and consider the solution $Y (X_b, X_b)_{\SO(4)} = (\epsilon'_b X_b, X_b)_{\SO(4)}$ for some lift $X_b$. This rotation now has the property that it leaves $\Phi_1 (\sigma^2_{1b})$ invariant and maps the outward normals of the tetrahedra to $\epsilon'_b \nb_{bc}$ and $N_b = \epsilon_b \pole$. This is exactly the structure of normals of a 4-simplex immersed in 3d Euclidean space.\end{proof}

\subsubsection{Action of the Symmetries on the Solutions}

Recall that for any solution $X_a$ the symmetries act as $X_a \rightarrow \epsilon_a Y X_a$. Given a pair of solutions $X^+$, $X^-$ not related by the symmetries we thus have $X^\pm_a \rightarrow \epsilon^\pm_a Y^\pm X^\pm_a$. Given the associated non-degenerate immersion $\imm$ it is immediate that this symmetry gives a solution associated to the immersion $$\imm' = (\epsilon^+_a Y^+,\epsilon^-_a Y^- )_{\SO(4)} \circ \imm.$$ Conversely we can act with any element of $\OO^4$ to obtain a new solution $\imm'$. This way we obtain all pairs of solutions related by the continuous symmetries.

Additionally, we will obtain solutions not related by the symmetries of the action if we act with an element of the non-identity connected component of $\OO(4)$, e.g. parity $P = diag(1,-1,-1,-1)$. As the pullback of the orientation under $P\circ \imm$ is the inverse of the orientation pulled back by $\imm$ this induces the transformation $(X'^+_a, X'^-_a)_{\SO(4)} = P (X^+_a, X^-_a)_{\SO(4)} P = (X^-_a, X^+_a)_{\SO(4)}$. Remember that the bivectors of $\imm$ were given by $\mu_\imm j_{ab} (\hat{X}^+_a, \hat{X}^-_a) (\nb_{ab}, - \nb_{ab})$, and $P$ switches the self-dual and anti-self-dual parts of the bivector. Thus the solution associated to $P\circ \imm$ has $\mu_{P \imm} = -\mu_\imm$ and $\hat{X}'^\pm_a = \hat{X}^\mp_a$.

\subsection{Classification of Solutions}

We can now classify the solutions to the equations \eqref{eq-CritPointSU24d} and \eqref{eq-StatPointSU24d}:

\begin{theo}[Classification of Solutions Euclidean 4d]\label{theo-ClassificationEuc4d}
The equations \eqref{eq-CritPointSU24d} and \eqref{eq-StatPointSU24d} with data $\nb_{ab}$, $j_{ab}$ arising from a coherent boundary state $\Psi$ with non-degenerate tetrahedra allow zero, one or two solutions up to symmetries. If they allow one solution only this solution corresponds to a constant $\su(2)$-valued 2-form on the 4-simplex. If they allow two solutions $\Psi$ is necessarily proportional to a Regge state  $\Psi_R$. The equations arising from a Regge state allow two solutions up to symmetries if its boundary geometry is that of a Euclidean 4-simplex, one solution up to symmetries if it is the boundary geometry of a 4-simplex immersed in 3d, and none if it is the boundary geometry of a Lorentzian 4-simplex.

In the first case the space of ordered pairs of $\SO(3)$ solutions is in 1-1 correspondence to the space of immersions $\imm$ of the 4-simplex $\sigma_4$ into $\R^4$ that induce the same boundary geometry on the 4-simplex as $\Psi_R$, up to translation and inversion. In the second case it is in correspondence to the immersions into $\R^3$.

The ordered pair of $\SU(2)$ solutions is, up to a spin lift sign, in one to one correspondence with 4-simplex immersions up to translation.

The correspondence is given by taking $(X^+_a, X^-_a)_{\SO(4)}$ as defined by equations \eqref{eq-SolutionsFromImmersions1} and \eqref{eq-SolutionsFromImmersions2}.
\end{theo}
\begin{proof}
The interpretation of the single solution is established by Lemma \ref{lem-ConstbField}. Lemma \ref{lem-BivNonDeg} shows that two solutions not related by the symmetry satisfy non-degeneracy and thus reconstruct a 4-simplex. Therefore two solutions exist if and only if the boundary data is that of a Regge manifold, then the coherent boundary state is proportional to a Regge state.

For a Regge state, given any solution $\hat{X}$, Lemma \ref{lem-2ndSolFromInvol} gives us a second solution $\hat{X}'$ not related by the symmetries to $\hat{X}$ or implies that $\hat{X}$ arises from an immersion $\imm$ of the 4-simplex into $\R^3 \subset \R^4$.

In the former case, Lemma \ref{lem-BivNonDeg} tells us that we have full non-degeneracy and Theorem \ref{Theo-BivReconstruction} give us an embedding $\imm$ of the 4-simplex.

Given a third solution $\hat{X}''$ we could obtain further immersions $\imm'$, but as it has the same geometry as $\imm$, it is related to $\imm$ by an $\OO(4)$ rotation and thus to $\imm$ or $P\imm$ by an $\SO(4)$ rotation. This induces an $\SO(3)$ symmetry relating $\hat{X}''$ to one of the existing solutions. Thus there can be no more than two solutions not related by the symmetries. There are two exactly if the boundary data is that of a non-degenerate 4-simplex, one if it is the boundary geometry of a 4-simplex immersed in 3 dimensions, and none otherwise.

Now given an immersion $\imm$ inducing the same boundary data, Lemma \ref{lem-Immersed4simpSols} gives us an ordered pair of solutions $\hat{X}^+$ and $\hat{X}^-$. It follows by the discussion of the symmetries that by acting with $\SO(4)$ we obtain $\imm$ providing all solutions $\hat{X'}^\pm$ related to $\hat{X}^\pm$ by symmetry, and by acting with the non-identity connected component of $\OO(4)$ we obtain all solutions $\hat{X'}^\pm$ related to the ordered pair $\hat{X}^\mp$ by the symmetries. On the other hand $\OO(4)$ also gives us all immersions $\imm$ defining the same boundary geometry. Thus the space of immersions is isomorphic to the space of solutions to \eqref{eq-CritPointSU24d} and \eqref{eq-StatPointSU24d} up to inversions and translations. Given $\SO(4)$ solutions the requirement that $N_a = (X_a^+, X_a^-)_{\SO(4)} \pole$ be outward fixes the inversion, too. The $\SO(4)$ solutions correspond to two $\SU(2)$ solutions up to a spin lifting sign. \end{proof}

\subsection{Regge State, Boundary Connection and Action}

To fully understand the geometric interpretation of the solutions $\hat{X}_a$ we will again show that they occur as the gauge transformations of a boundary geometry. For a 4-simplex $\sigma^4_\imm$ in $\R^4$ we can again define a dihedral connection $D_{ab} = (d^+_{ab}, d^-_{ab})_{\SO(4)}$ by

$$D_{ab} \imm(\sigma^2_{ab}) = \imm(\sigma^2_{ab}),$$
and
$$D_{ab} N_b = N_a.$$
Thus we have in particular
$$D_{ab} B_{ab}(\sigma^4_\imm) = B_{ab}(\sigma^4_\imm),$$
and writing $N_{ba}$ for the vector outward to $\imm(\sigma^2_{ba})$ in the 3-dimensional plane $\imm(\sigma^3_{b})$ we have
$$D_{ab} N_{ba} = - N_{ab}.$$

This implies that \be\label{eq-4dExpDihedralSO4}D_{ab} = \exp \left(\Theta_{ab} \star \frac{B_{ab}(\sigma^4_{\imm})}{|B_{ab}(\sigma^4_{\imm})|}\right)\ee where $0 < \Theta_{ab} < \pi$ is the dihedral angle. This follows as $\star B_{ab}(\sigma^4_\imm)$ generates the rotations leaving the plane of $B_{ab}(\sigma^4_\imm)$ invariant and $|\Theta_{ab}|$ is the angle of rotation between $N_a$ and $N_b$. We can then check by explicit calculation that the rotation angle with the choices made is between $0$ and $\frac{\pi}{2}$ rather than between $-\pi$ and $0$:

\bea
D_{ab} N_b &=& \exp \left(\Theta_{ab} \frac{N_a \wedge N_b}{|N_a \wedge N_b|}\right) N_b\nn\\
&\approx & \left(\id + \Theta_{ab} \frac{N_a \wedge N_b}{|N_a \wedge N_b|}\right) N_b\nn\\
&= & N_b + \Theta_{ab} \left(\frac{N_a}{|N_a \wedge N_b|} - N_b \frac{N_a\cdot N_b}{|N_a \wedge N_b|}\right)\nn\\
&\approx & N_b + \Theta_{ab} \left(\frac{N_a}{|\Theta_{ab}|} - \frac{N_b}{|\Theta_{ab}|}\right).\nn\\
\eea

Thus with the definitions as given $\Theta$ is indeed positive.

Note that Lemma \ref{lem-Immersed4simpSols} gives us not just an $\SO(3)\times\SO(3)$ element but actually a full $\SO(4)$ element $(X_a^+, X_a^-)_{\SO(4)}$, and that the covering lift $g_{ba}$ of $\hat{g}_{ba}$ gives us a full $\Spin(4)$ element $(g_{ba}, g_{ba})$. Being diagonal this satisfies $$(g_{ba}, g_{ba})_{\SO(4)} \pole = \pole,$$ and $$(g_{ba}, g_{ba})_{\SO(4)} (0,\nb_{ab}) = - (0,\nb_{ba}).$$ By definition it also satisfies $$(g_{ba}, g_{ba})_{\SO(4)} \Phi_a(\sigma^2_{ab}) = \Phi_b(\sigma^2_{ab})$$. We then obtain the following $\SO(4)$ commuting diagram:

\begin{equation}
\label{eq-ftjCommDiagSO4}
\xymatrixcolsep{4pc}\xymatrixrowsep{4pc}
\xymatrix{\ar @{} [dr]  \Phi_a(\sigma^3_a) \ar[d]_{(g_{ba}, g_{ba})_{\SO(4)}} \ar[rr]^{(X^+_a, X^-_a)_{\SO(4)}} && ~\imm(\sigma^3_a) \ar[d]^{ D_{ba}}   \\
\Phi_b(\sigma^3_b)\ar[rr]_{(X^+_b, X^-_b)_{\SO(4)}} && ~\imm(\sigma^3_b)  }
\end{equation}

Fixing the lift of either the self-dual or the anti-self-dual part of an $\SO(4)$ element fixes the lift to $\Spin(4)$. Next note that $$(d^+_{ab}, d^-_{ab}) = \left(\nu_{ab} \exp(\mu_{\imm} \Theta_{ab} (\hat{X}^+_{a} \nb_{ab}) \cdot L), \exp(- \mu_{\imm} \Theta_{ab} (\hat{X}^-_{a} \nb_{ab}) \cdot L)\right ) $$ is a covering lift of $D_{ab}$ compatible with the spin structure and we again have by Lemma \ref{lem-CommDiagLift} that any $(X^+_a, X^-_a)$ covering $(X^+_a, X^-_a)_{\SO(4)}$ is related by a sign $\epsilon'_{a}$ to the gauge transformation taking the discrete $\Spin(4)$ connection $(g_{ab}, g_{ab})$ to the discrete $\Spin(4)$ connection $(d^+_{ab}, d^-_{ab})$. Thus we have:

\begin{equation}
\label{eq-ftjCommDiagSpin4}
\xymatrixcolsep{4pc}\xymatrixrowsep{4pc}
\xymatrix{\ar @{} [dr]  \Phi_a(\sigma^3_a) \ar[d]_{(g_{ba}, g_{ba})} \ar[rr]^{\epsilon'_a(X^+_a, X^-_a)} && ~\imm(\sigma^3_a) \ar[d]^{ (d^+_{ba}, d^-_{ba})}   \\
\Phi_b(\sigma^3_b)\ar[rr]_{\epsilon'_b (X^+_b, X^-_b)} && ~\imm(\sigma^3_b)  }
\end{equation}

This again allows us to immediately evaluate the action $S_{\ftj}$ for a solution associated to a particular immersion. We consider the first element $X^+$ of the ordered pair that corresponds to an immersion to be associated to it. We then again look at the solution where $\epsilon_a = 1$ in diagram \ref{eq-ftjCommDiagSpin4}, all other solutions being related to this one by symmetry. Then we can evaluate the matrix elements entering the action:

\bea
\bra{J \nb_{ab}} {X^+_a}^\dagger X^+_b \ket{\nb_{ba}} &=& \bra{J \nb_{ab}} {X^+_a}^\dagger X^+_b g_{ba} \ket{J \nb_{ab}}\nn\\
&=& \bra{J \nb_{ab}} {X^+_a}^\dagger d^+_{ba} X^+_a \ket{J \nb_{ab}}\nn\\
&=& \nu_{ab} e^{\frac{i}{2} \mu_{\imm} \Theta_{ab}},
\eea
where we have used the commuting diagram in the first step, equation \eqref{eq-2dReggeGluing} in the second, and the fact that the coherent states $X^+_a \ket{J \nb_{ab}}$ are eigenvectors of the Lie algebra elements $(X^+_a \nb_{ab})\cdot L$ defining the lift of $d^+_{ab}$ with eigenvalues $- \frac{i}{2}$, and $d^+_{ab} = (d^+_{ba})^\dagger$ in the last.

Thus we obtain 

\be
S_{\ftj}(\imm) = i \sum_{a<b} \mu_{\imm} j_{ab} \Theta_{ab} = i \mu_{\imm} S_{Regge}(\sigma^4_\imm).
\ee

Note that as the action $S_{\ftj}$ depends on the arbitrary phases for the state $\Psi_C$, we do not give an asymptotic form for it here.

If the boundary geometry is that of a 3-dimensional 4-simplex we only have one solution related by the symmetries. In that case the $\Theta_{ab}$ are $\pi$ or $0$ as the case $\pi$ only occurs at the boundary of a set of 4-simplices the sum of $j_{ab}$ at this boundary has to sum to an integer and the action is equal to a multiple of $\pi$.

\subsection{The Asymptotic Formula}

Combining all the preceding discussion we can again give the full asymptotic formula now by fixing the continuous symmetry. We write $\nu_{crit} = \prod_{a<b} \nu_{ab}^{2j_{ab}}$ evaluated for a critical point. We then obtain

\begin{theo}[Asymptotic Behaviour of the 15j symbol for Regge states]\label{theo-ftjAsymptotics} Let $\Psi_R$ be a Regge state with non-degenerate tetrahedra and $\Psi_{\lambda}$ be the same state with spin $\lambda j_{ab}$. We write $H_{|\pm}$ for the Hessian of $S_{\ftj}$ evaluated at an immersion with $\mu_\imm = \pm$. Then if the boundary geometry induced by $\Psi_R$ on the $\sigma^4$ is that of a Euclidean 4-simplex in 4 dimensions we have

\bea\label{eq-ftj-4dEculidean}
\ftj(\Psi_{\lambda}) &=& (-1)^s \left( \frac{2 \pi}{ \lambda} \right)^6 \frac{2^4}{(4 \pi)^8}\times\nn\\&& \left( \nu_+^\lambda \frac{e^{i \lambda S_R (\sigma)}}{\sqrt{\det -H_
{|+}}} + \nu_-^\lambda \frac{e^{- i \lambda S_R (\sigma)}}{\sqrt{\det -H_
{|-}}}\right) + o(\lambda^{-7}).
\eea

If the geometry induced is 3d Euclidean we have

\be\label{eq-ftj-3dEculidean}
\ftj(\Psi_{\lambda}) = (-1)^s \nu^\lambda \left( \frac{2 \pi}{ \lambda} \right)^6 \frac{2^4}{(4 \pi)^8} \left( \frac{\pm}{\sqrt{\det -H}}\right) + o(\lambda^{-7}).
\ee

If the geometry is 4d Lorentzian we have

$$\ftj(\Psi_{\lambda}) = o(\lambda^N) \,\, \mbox{for all}\; N.$$

These are all possible cases.
\end{theo}

For general coherent boundary states we have that either the state is proportional to a Regge state or it allows only one solution. Thus the amplitude is either exponentially suppressed or proportional to the one for the 3d Euclidean case.

\section{The Asymptotics of EPRL-FK Type Amplitudes}

The amplitudes of the spin foam models proposed in \cite{Engle:2007wy,Engle:2007qf,Freidel:2007py} are based on the square of the $\ftj$ symbol. Therefore our analysis of the $\ftj$ also allows us to describe the geometry and asymptotics of these models.

There are three types of models. For the first two the 4-simplex amplitude is given simply by the square of the $\ftj$ with one of the factors rescaled by a factor $c_\gamma = \frac{|1-\gamma|}{1+\gamma}$.

The first type of amplitude is given simply by the square of the $\ftj$ symbol
\be
\ZZ(\Psi) = \ftj(\Psi) \times \ftj(\Psi_{c_\gamma}).
\ee

Both the EPRL and FK models for parameter $\gamma<1$, and the EPR model which has $\gamma=0$ (i.e., $c_\gamma=1$), are of this type.

The second type of model has 4-simplex amplitude given by
  \be
\ZZ^{cc}(\Psi) = \ftj(\Psi) \times \overline{\ftj(\Psi_{c_\gamma})}.
\ee
and includes the FK model with parameter $\gamma>1$, and for $\gamma=0$, the FK model without a parameter.

The third model, covering the case of the EPRL model for $\gamma > 1$ is not given directly in terms of the $\ftj$ but by inserting the state $\Psi$ into the lowest weight representation of $j \otimes c_{\gamma} j$. This can be accomplished by an integration over a new set of coherent states as described below.

Note that in the literature these models are usually written as functions of $k$ such that $j =\frac{1+\gamma}{2} k$ and $j' =\frac{|1-\gamma|}{2} k$. The possible values of $\gamma$ and $j$, $j'$ and thus $k$ are restricted as the state sum is identically zero unless $c_\gamma j$ takes on half integer values.

This formulation of the models was how the FK model \cite{Freidel:2007py} was initially defined, and for the EPRL model this formulation was developed in \cite{Barrett2009a}.

\subsection{The Exponential Form of EPRL-FK Type Amplitudes}

The exponential form of the 4-simplex amplitudes for the models without complex conjugation are then given by
\bea
&\ZZ(\Psi) &= (-1)^{s} \int_{\SU(2)^{10}} \left(\prod_{c=1...5} \dd X_c \dd X'_c\right) \prod_{a<b} \times\nn\\&&\times \bra{J \alpha(\nb_{ab})} X_a^\dagger X_b \ket{\alpha(\nb_{ba})}^{2 j_{ab}} \bra{J \alpha(\nb_{ab})} {X'}_a^\dagger X'_b \ket{\alpha(\nb_{ba})}^{2 c_\gamma j_{ab}}.\nn\\
\eea

For the models with complex conjugation, the $4$-simplex amplitudes read

\bea
&Z^{cc}(\Psi) &= (-1)^{s'} \int_{\SU(2)^{10}} \left(\prod_{c=1..5} \dd X_c \dd X'_c\right) \prod_{a<b}  \times\nn\\&&\times \bra{J \alpha(\nb_{ab})} X_a^\dagger X_b \ket{\alpha(\nb_{ba})}^{2 j_{ab}} \overline{\bra{J \alpha(\nb_{ab})} {X'}_a^\dagger X'_b \ket{\alpha(\nb_{ba})}}^{\, \,2 c_\gamma j_{ab}}.\nn\\
\eea

For the third type of model we have

\bea
&&Z^{3rd}(\Psi) = \nn\\&&(-1)^{s''} \int_{\SU(2)^{10}} \left(\prod_{c=1..5} dX_c dX'_c\right) \int_{{S^2}^20} \dd m_{ab} \prod_{a<b}  \times\nn\\&&\times \braket{\alpha(\mb_{ab})}{\alpha(\nb_{ab})}^{(1-c_{\gamma}) 2j_{ab}} \braket{\alpha(\mb_{ba})}{\alpha(\nb_{ba})}^{(1-c_{\gamma}) 2j_{ab}} \times\nn\\&&\times \bra{J \alpha(\mb_{ab})} X_a^\dagger X_b \ket{\alpha(\mb_{ba})}^{2 j_{ab}} \overline{\bra{J \alpha(\mb_{ab})} {X'}_a^\dagger X'_b \ket{\alpha(\mb_{ba})}}^{\, \,2 c_\gamma j_{ab}}.
\eea

The intermediate integrations have the effect of inserting the state $\alpha_{(1-c_{\gamma})j}(\nb)$ into the lowest weight subspace of the Clebsch-Gordon decomposition of the $\SU(2)$ representation $j \tensor c_{\gamma} j$.

The corresponding actions can be written easily in terms of the $\ftj$ action \eqref{eq-ftjActionCoBoState}:
\bea
S^\gamma (\alpha(\nb), j, X, X') &=& S_{\ftj}(\alpha(\nb), j, X) + c_\gamma S_{\ftj}(\alpha(\nb), j, X'),\nn\\
S^{\gamma,cc} (\alpha(\nb), j, X, X') &=& S_{\ftj}(\alpha(\nb), j, X) + c_\gamma \overline{S_{\ftj}(\alpha(\nb), j, X')},\nn\\
S^{\gamma,3rd} (\alpha(\nb), j, X, X', \mb) &=& S_{\ftj}(\alpha(\mb), j, X) + c_\gamma \overline{S_{\ftj}(\alpha(\mb), j, X')} \nn\\&&+ \sum_{a,b: a\neq b} 2(1-c_{\gamma})j_{ab} \ln \braket{\alpha(\mb_{ab})}{\alpha(\nb_{ab})}.\nn\\
\eea

The equations governing the stationary phase analysis of the first two models are identical and are simply closure \eqref{eq-StatPointSU24d} and two copies of \eqref{eq-CritPointSU24d}:
\bea \label{eq-EPRLFKTypeCritPoints}
&\displaystyle\sum_{b: b\ne a} j_{ab}\nb_{ab}=0,&\nn\\
&X_b \nb_{ba} = - X_a\nb_{ab},&\nn\\
&X'_b \nb_{ba} = - X'_a\nb_{ab}.&
\eea

The third type has the equivalent set of:
\bea \label{eq-3rdTypeCritPoints}
&\displaystyle\sum_{b: b\ne a} j_{ab}\nb_{ab}=0,&\nn\\
&X_b \mb_{ba} = - X_a\mb_{ab},&\nn\\
&X'_b \mb_{ba} = - X'_a\mb_{ab},&\nn\\
&\mb_{ab} = \nb_{ab}&
\eea

Therefore the geometric analysis of equations \eqref{eq-StatPointSU24d} and \eqref{eq-CritPointSU24d} applies immediately to these models. Furthermore, for the first two models the asymptotics are simply given by the product of the asymptotics of the $\ftj$ symbol. Due to the presence of the intermediate integrations the determinant of the Hessian entering the asymptotics of the third type of model will not be the product of the determinants of $S_{\ftj}$ Hessians. However the action evaluated at the critical points is just the sum of two $S_{\ftj}$. This is because at the critical points the $\alpha(\nb)$ differ from the $\alpha(\mb)$ by a phase only, however the amplitude does not depend on the phase of the $\alpha(\mb)$ and therefore we can choose $\alpha(\nb) = \alpha(\mb)$ at the critical points and the extra term in the $S^{\gamma,3rd}$ drops out, reducing it to $S^{\gamma,cc}$. Therefore going forward we will only consider the first two types of models. The Hessian of the third type is given explicitly in \cite{Barrett2009a}.

\subsubsection{Coherent Boundary States}

In general the critical point configurations can again be understood as constant $\su(2)$-valued 2-forms on the 4-simplex given in Lemma \ref{lem-ConstbField}. If two different solutions exist we are in the case of the Regge state therefore for coherent boundary states that do not admit a Regge state we have two copies of the same $\su(2)$-valued 2-form, possibly presented in different gauges in the two different factors.

As we are free to choose the phase in the coherent states the only interesting part of the asymptotics is the scaling. For the first two types of models this is obtained by multiplying two copies of the asymptotics of a single $\ftj$ scaled by $c_\gamma$:
\be \label{eq-ModelAsymCoBoSta}
|Z(\Psi_\lambda)| \sim \left( \frac{2 \pi}{ \lambda} \right)^{12} \frac{2^8}{(4 \pi)^{16}} \frac{1}{{c_\gamma}^6|\det -H|}.
\ee
Here $H$ is the Hessian of $S_{\ftj}$ evaluated on the critical points with unscaled spins given by $j$. As the Hessian is linear in the $j$ and 12-dimensional, the determinant evaluated at the critical point $j' = c_\gamma j$ differs by ${c_\gamma}^{12}$ from that evaluated at $j$. This accounts for the factor ${c_\gamma}^6$.

For the third type of model the above formula changes only by replacing the modulus of the Hessian by the square root of the Hessian of the action $S^{\gamma,3rd}$.

\subsubsection{Regge States}

For Regge states the solutions are again classified by the immersions of 4-simplices with the same boundary geometry. We will look at the cases of a 4-simplex immersed in 3d and in 4d separately.

\paragraph{4d Euclidean.}

Up to symmetries we have two solutions to the critical point equations. Taking two solutions not related by symmetries and calling them $X_a^\pm$ according to whether $\mu_\imm = \pm 1$ we have that the bivectors of the immersed 4-simplices corresponding to these solutions can be written in terms of $b^\pm_{ab} = j_{ab} X_a^\pm \nb_{ab}$ as $(b_{ab}^+, b_{ab}^-)$ and $- (b_{ab}^-, b_{ab}^+)$.

As before, given two inequivalent solutions $b^\pm_{ab} = j_{ab} X_a^\pm \nb_{ab}$ they necessarily correspond to the self-dual and anti-self-dual parts of geometric 4-simplex bivectors given by $B_{ab}(\sigma) = (b^+_{ab}, b^-_{ab})$ and $B_{ab}(P\sigma) = - (b^-_{ab}, b^+_{ab})$.

These combine to four solutions to the stationary and critical point equations \eqref{eq-EPRLFKTypeCritPoints}:
$$(b, {c_\gamma}^{-1} \, b') \in \{(b^+,b^+),(b^+,b^-),(b^-,b^+),(b^-,b^-)\}.$$
The $++$ and $--$ solutions are in some sense analogous to the solutions for non-geometric boundary data, as the solution is just the double of a pure $\SU(2)$ solution.

The full asymptotics is then given by the sum over these four critical points. The actions can be evaluated straightforwardly using the result that $S_{ftj}$ evaluates to $\mu_{\imm} S_R(\sigma^4_\imm)$. Then with $\epsilon, \epsilon' = \pm 1$ and $\nu$ defined as above we have
\be
Z^\gamma(\Psi_\lambda) \sim (-1)^s \left( \frac{2 \pi}{ \lambda} \right)^{12} \frac{2^8}{(4 \pi)^{16}} \sum_{\epsilon, \epsilon' = \pm 1} \nu_{\epsilon}^{\lambda} \nu_{\epsilon'}^{c_{\gamma} \lambda} \frac{e^{i \lambda (\epsilon + \epsilon' c_\gamma) S_R(\sigma^4)}}{{c_\gamma}^6\sqrt{\det -H_\epsilon \det -H_{\epsilon'}}},
\ee
for the models without complex conjugation and by
\be
Z^{cc}(\Psi_{\lambda}) \sim (-1)^{s'} \left( \frac{2 \pi}{ \lambda} \right)^{12} \frac{2^8}{(4 \pi)^{16}} \sum_{\epsilon, \epsilon'=\pm 1} \nu_{\epsilon}^{\lambda}\nu_{\epsilon'}^{c_{\gamma} \lambda} \frac{e^{i \lambda (\epsilon - \epsilon' c_\gamma)  S_R(\sigma^4)}}{{c_\gamma}^6\sqrt{\det -H_\epsilon \overline{\det -H_{\epsilon'}}}},
\ee
for those with complex conjugation. As the amplitudes leading to these are rescaled squares of the $\SU(2)$ $\ftj$ symbol this is of course just the rescaled square of the $\SU(2)$ asymptotics. $H_\pm$ is the Hessian of $S_{\ftj}$ evaluated on the critical points $X^\pm$. For the third type we would again obtain a different, more complicated, Hessian and different prefactors in the scaling.

We can rewrite the actions further by expressing them in terms of $k_{ab} = \frac{2}{(1+\gamma)} j_{ab}$. Writing $S_{\epsilon \epsilon'}$ for $S|_{(b^\epsilon, b^\epsilon)}$ we obtain 
\bea
S^\gamma_{\epsilon \epsilon'} &=& (\epsilon(1+\gamma) + \epsilon' |1-\gamma|) \frac{1}{2} \sum_{a<b} k_{ab} \Theta_{ab},\nn\\
S^{\gamma,cc}_{\epsilon \epsilon'} &=& (\epsilon (1+\gamma) - \epsilon' |1-\gamma|) \frac{1}{2} \sum_{a<b} k_{ab} \Theta_{ab},\nn
\eea
and thus 
\be
S^{\gamma > 1}_{\epsilon \epsilon'} = S^{\gamma < 1, cc}_{\epsilon \epsilon'} = ((\epsilon + \epsilon')\gamma + (\epsilon - \epsilon')) \frac{1}{2} \sum_{a<b} k_{ab} \Theta_{ab},\nn
\ee
\be
S^{\gamma > 1, cc}_{\epsilon \epsilon'} = S^{\gamma < 1}_{\epsilon \epsilon'} = ((\epsilon - \epsilon')\gamma + (\epsilon + \epsilon')) \frac{1}{2} \sum_{a<b} k_{ab} \Theta_{ab}.
\ee
Hence, we have in particular
$$S_{\mbox{{\tiny EPR}}}|_{\epsilon \epsilon'} = S^{\gamma<1}|_{\epsilon \epsilon' \, \gamma = 0} = \frac{\epsilon + \epsilon'}{2} \sum_{a<b} k_{ab} \Theta_{ab},$$
and 
$$S_{\mbox{{\tiny FK}}}|_{\epsilon \epsilon'} = S^{cc, \gamma<1}|_{\epsilon \epsilon'\, \gamma = 0} = \frac{\epsilon - \epsilon'}{2} \sum_{a<b} k_{ab} \Theta_{ab}.$$
This is the form in which the actions appeared in the literature so far, where the state sums were expressed in terms of $k$.

\paragraph{Single solution.}
As in the $\SU(2)\, BF$ case the action is zero or $\pi$ and the asymptotics contain a single term.

\paragraph{No solutions.}
In this case the boundary geometry is that of a 4-simplex with Lorentzian metric and we again have that the amplitude is suppressed exponentially for large spins.

\chapter{The Geometry of Lorentzian 4-Dimensional State Sum Amplitudes}

For the Lorentzian state sum proposed in \cite{Engle:2007wy,Pereira:2007nh} and as rigorously defined in \cite{Barrett2010} similar geometricity results hold. While the amplitudes before were all based on $\SU(2)$ recoupling theory the $\SL(2,\C)$ amplitudes require siginificantly more discussion for which we will refer the reader to the paper \cite{Barrett2010}. Here we will simply give the form of the exponential amplitude directly and show how the geometricity results of the last chapter can be generalised to the $\SL(2,\C)$ case. The boundary states for this amplitude are the same coherent boundary states as for the 4-dimensional $\SU(2)$ based model. The key difference to the last two chapters is that we will perform most calculations not in a unitary representation of $\SL(2,\C)$ but instead in the spinor representation discussed in Section \ref{sec-Spinor}.

\section{Definition of the Lorentzian Amplitude}

The irreducible unitary representations of $\SL(2,\C)$ are labelled by two numbers $p$ and $k$. We will be interested in the main sequence of irreps where $p\in\R$ and $k\in\N/2$. Every unitary irrep of $\SL(2,\C)$ is a unitary representation of its subgroup $\SU(2)$, and we therefore have the Clebsch-Gordon decomposition into infinitely many $\SU(2)$ irreps: $(p,k) = \bigoplus j$. It can be shown that every $\SU(2)$ irrep appears at most once in this decomposition. $k$ is the lowest weight $\SU(2)$ irrep that appears.

In \cite{Engle:2007wy} it was assumed that $p = \gamma k$ holds for all representations involved. We will instead add the real numbers $p$ and show that this relation must hold for stationary points to exist. Given a coherent boundary state $\Psi_c$ on the boundary triangulation of a 4-simplex and ten real numbers $p_{ab}$ associated to the triangles $\sigma^2_{ab}$ the Lorentzian amplitude is defined as an integration over $\CP^1$ and $\SL(2,\C)$:

\be\label{eq-LorAmplitude}
\ZZ(\Psi,p_{ab}) = (-1)^{\chi} \int_{(\SL(2,\C))^5} \delta(X_5) \,\prod_{a} dX_a \, \;\prod_{a<b} c_{ab} \; d_{k_{ab}} \; P_{ab},
\ee
with $c_{ab}=\frac{ \sqrt{k_{ab}^2+p_{ab}^2}}{\pi   ( k_{ab}-ip_{ab})}$, $d_k$ the $\SU(2)$ dimension of the irrep $k$ and

\bea\label{eq-LorPropagator}
P_{ab}=\int_{\CP^1}&& \la X_a^{\dagger}z|X_a^{\dagger}z\ra^{-1-ip_{ab}-k_{ab}}\la X_a^{\dagger}z|\alpha(\nb_{ab})\ra^{2k_{ab}}\nn\\
&\times&\la X_b^{\dagger}z|X_b^{\dagger} z\ra^{-1+ip_{ab}-k_{ab}}\la J \alpha(\nb_{ba})|X_b^{\dagger}z\ra^{2k_{ab}} \;\Omega_z.
\eea

Here $z \in \C^2$ are spinors and $X$ act naturally as $\SL(2,\C)$ matrices on the spinors $z$ and $\alpha(\nb)$.  $$\Omega_z=\frac i2(z_0\dd z_1-z_1\dd z_0)\wedge (\bar z_0\dd \bar z_1-\bar z_1\dd \bar z_0)$$ is the $\SL(2,\C)$  invariant measure on $\C^2$. Integrand and measure together are invariant under the transformation $z \rightarrow \kappa z$ with $\kappa\in\C$. Therefore we have to project the integration onto $\CP^1$.  Note that this reduces to the formula for the Lorentzian Barret-Crane model of \cite{Barrett2000} for $k_{ab}=0$.

For a derivation of the above formula we refer to the paper \cite{Barrett2010}. It can be understood as the insertion of the $SU(2)$ boundary state with $\SU(2)$ representation labels $k_{ab}$ into the lowest weight subspace of the representation $(p_{ab},k_{ab})$ contracted with the bilinear inner product on the representation $(p_{ab},k_{ab})$.

The amplitude can again be written as an action, allowing us to discuss its symmetries and asymptotic geometry. The presentation here closely follows section 4 of \cite{Barrett2010}.

First note that there is an internal variable, $z$, for each propagator. As it will be necessary to distinguish between these variables, we will denote the internal variable of the propagator $P_{ab}$ , $a<b$, by $z_{ab}$.

We will further introduce the shorthand notation
 $$\z_{ab} = X_a^{\dagger}z_{ab} \quad\text{ and }\quad \z_{ba} = X_b^{\dagger}z_{ab}$$ for the intermediate variable acted on by the group elements.

The propagator then is
\be
P_{ab} = \int_{\CP^1} \Omega_{ab} \left( \frac{\la \z_{ba}| \z_{ba} \ra}{\la \z_{ab}| \z_{ab} \ra} \right)^{i p_{ab}}
\left( \frac{\la \z_{ab}| \alpha(\nb_{ab}) \ra \la J \alpha(\nb_{ba})| \z_{ba} \ra}{\la \z_{ab}| \z_{ab} \ra^{1/2} \la \z_{ba}| \z_{ba} \ra^{1/2}} \right)^{2k_{ab}}, \nn
\ee
where
$$
\Omega_{ab} = \frac{\Omega_z}{\la \z_{ab}, \z_{ab} \ra \la \z_{ba}, \z_{ba} \ra},$$
which has the correct homogeneity to immediately project to a measure on $\CP^1$.

The amplitude is then again written in terms of an action as

\be
\ZZ(\Psi,p) = (-1)^{\chi} \int_{(\SL(2,\C))^5} \delta(X_5) \,\prod_{a} dX_a \,  \int_{(\CP^1)^{10}} \; e^{S_L}\;\prod_{a<b} c_{ab} d_{k_{ab}} \, \Omega_{ab}, \nn
\ee
with the action $S_{L}$ given by

\be
\label{eq-LorAction}
S[X,z] = \sum_{a<b} k_{ab} \log \frac{  \la \z_{ab}| \alpha(\nb_{ab}) \ra^2 \la J \alpha(\nb_{ba})| \z_{ba} \ra ^2}{\la \z_{ab}| \z_{ab} \ra \la \z_{ba}| \z_{ba} \ra} + i  p_{ab} \log \frac{\la \z_{ba}| \z_{ba} \ra}{\la \z_{ab}| \z_{ab} \ra} .
\ee

The first term is complex and defined mod $2\pi i$ whereas the second term is purely imaginery.

\subsection{Symmetries of the Action}\label{symmetriesaction}

We again have a number of symmetries of the integrand of the amplitude \eqref{eq-LorAmplitude} and the action \eqref{eq-LorAction} (modulo $2\pi i$):

\begin{itemize}
\item Continuous: A global $\SL(2,\C)$ transformation acting as $X_a \rightarrow Y X_a$, $z_{ab} \rightarrow (Y^{\dagger})^{-1} z_{ab}$, $Y \in \SL(2,\C)$.
\item Discrete: A local (at each tetrahedron) sign symmetry acting as $X_a \rightarrow \epsilon_a X_a$, $\epsilon_a = \pm$.
\item Rescaling: A local (at each triangle) rescaling acting as $z_{ab} \rightarrow \kappa_{ab} z_{ab}$ with $0\ne\kappa\in \mathbb{C}$.

\end{itemize}

We need to fix the continuous symmetry in the integral directly as the volume of its orbits is infinite and the integral would diverge otherwise. The same is true for the rescaling symmetry which is taken care of by projecting the integration to $\CP(1)$ instead of $\C^2$.

\section{Asymptotic Geometry of the Lorentzian Amplitude}

The amplitude \eqref{eq-LorAmplitude} is again exponential and we can analyse its asymptotic behaviour under the scaling $(\lambda p, \lambda k)$. To do so we will again look at the asymptotic equations.

\subsection{Asymptotic Equations of Motion}

The equations of motions are again given by the stationary points of $S_L$ for which $\mathrm{Re} \, S_L$ is maximal. It is assumed from now on that $(k,p)\ne0$.

\subsubsection{Critical points}

The real part of the action
\be
\mathrm{Re} \, S_L = \sum_{a<b} k_{ab} \log \frac{ |\la \z_{ab}| \alpha(\nb_{ab}) \ra |^2 |\la J \alpha(\nb_{ba})| \z_{ba} \ra |^2}{\la \z_{ab}| \z_{ab} \ra \la \z_{ba}| \z_{ba} \ra} \nn
\ee
satisfies $\mathrm{Re} \, S\le0$ and hence is at a maximum where it
vanishes. It vanishes if and only if, on each triangle $\sigma^2_{ab}$, the following condition holds
\be
\label{real}
\frac{ \la \alpha(\nb_{ab})| \z_{ab} \ra \la \z_{ab}| \alpha(\nb_{ab})\ra}{\la \z_{ab}| \z_{ab} \ra}\frac{\la J \alpha(\nb_{ba})| \z_{ba} \ra \la \z_{ba}| J \alpha(\nb_{ba}) \ra }{ \la \z_{ba}| \z_{ba} \ra} = 1.
\ee
As $| \z_{ba} \ra \la \z_{ba}|/\la \z_{ba}| \z_{ba} \ra$ is the projector in the direction of $\z_{ba}$, this equation implies that $\z_{ab}$ and $\z_{ba}$ are proportional to the coherent states $\alpha(\nb_{ab})$ and $J \alpha(\nb_{ba})$ respectively. Therefore we have:

\be
\label{eq-LorRealityEquation}
\alpha(\nb_{ab}) = \frac{e^{i \phi_{ab}}}{\parallel \z_{ab} \parallel} X_a^{\dagger} \, z_{ab}, \;\;\;\; \mbox{and} \;\;\;\; J \alpha(\nb_{ba}) = \frac{e^{i \phi_{ba}}}{\parallel \z_{ba} \parallel} X_b^{\dagger} \, z_{ab},
\ee
where $\parallel \z \parallel = |\braket{\z}{\z}|^{\frac12}$ is the Hermitian norm of $\z$, and $\phi_{ab}$ and $\phi_{ba}$ are phases defined by this equation. Eliminating $z_{ab}$, and introducing the notation $\theta_{ab} = \phi_{ab} - \phi_{ba}$, we obtain the equations for a critical point
\be
\label{eq-LorCritPoint1}
(X_a^{\dagger})^{-1} \, \alpha(\nb_{ab}) = \frac{\parallel \z_{ba} \parallel}{\parallel \z_{ab} \parallel} e^{i \theta_{ab}} (X_b^{\dagger})^{-1} J \, \alpha(\nb_{ba}),
\ee
for each $a<b$. The action can now be written in terms of $\frac{\parallel \z_{ba} \parallel}{\parallel \z_{ab} \parallel}$ and $\theta_{ab}$:

\be
S_L[\imm] = i \sum_{a<b} p_{ab} \, \log \, \frac{\parallel \z_{ba} \parallel^2}{\parallel \z_{ab} \parallel^2} + 2 k_{ab} \, \theta_{ab},\nn
\ee

\subsubsection{Stationary points}

We now look at the variation of the action with respect to the $z_{ab}$ and the $X_a$. We begin with the former.

\paragraph{Spinor Variation}

There is a spinor $z_{ab}$ for each triangle $ab$, $a < b$, and the variation of the action with respect to these complex variables gives two spinor equations for each triangle. For clarity will first treat the $\ket{z_{ab}}$ and the $\bra{z_{ab}}$ as independent variables and then add the variations

$$\delta_{z_{ab}} S_L = \delta_{\ket{z_{ab}}} S_L + (\delta_{\bra{z_{ab}}} S_L)^\dagger.$$

We have:

\bea
\delta_{\ket{z_{ab}}} S_L &=& i p_{ab} \left( \frac{\bra{\z_{ba}}X_b^{\dagger}}{\la \z_{ba}| \z_{ba} \ra} - \frac{ \bra{\z_{ab}} X_a^{\dagger}}{\la \z_{ab}| \z_{ab} \ra} \right) \nn \\
&& + k_{ab} \left( \frac{2 \bra{J \alpha(\nb_{ba})}X_b^{\dagger}}{\la J \alpha(\nb_{ba})| \z_{ba} \ra} - \frac{ \bra{\z_{ab}}X_a^{\dagger}}{\la \z_{ab}| \z_{ab} \ra}  - \frac{\bra{\z_{ba}} X_b^{\dagger}}{\la \z_{ba}| \z_{ba} \ra} \right), \nn
\eea
and
\bea
\delta_{\bra{z_{ab}}} S &=& i p_{ab} \left( \frac{X_b \ket{\z_{ba}}}{\la \z_{ba}| \z_{ba} \ra} - \frac{X_a \ket{\z_{ab}}}{\la \z_{ab}| \z_{ab} \ra} \right)
\nn \\ && + k_{ab} \left( \frac{2 X_a \ket{\alpha(\nb_{ab})}}{\la \z_{ab}| \alpha(\nb_{ab}) \ra} - \frac{X_a \ket{\z_{ab}}}{\la \z_{ab}| \z_{ab} \ra}  - \frac{X_b \ket{\z_{ba}}}{\la \z_{ba}| \z_{ba} \ra} \right). \nn
\eea

Adding these, and using equations \eqref{eq-LorRealityEquation}, we see that the terms proportional to $k$ cancel and we are left with

\be
\delta_{z_{ab}} S_L = i 2 p_{ab} \left( \frac{\bra{\z_{ba}}X_b^{\dagger}}{\la \z_{ba}| \z_{ba} \ra} - \frac{ \bra{\z_{ab}} X_a^{\dagger}}{\la \z_{ab}| \z_{ab} \ra} \right).\nn
\ee

Using the assumption that $p,k \neq 0$ and using \eqref{eq-LorRealityEquation} to express $\z_{ab}$ in terms of $\alpha(\nb)$ we finally obtain that $\delta_{z_{ab}} S_L = 0$ implies
\be
(X_a \, \alpha(\nb_{ab})^{\dagger} = \frac{\parallel \z_{ab} \parallel}{\parallel \z_{ba} \parallel} e^{- i \theta_{ab}} (X_b \, J \, \alpha(\nb_{ba})^{\dagger}. \nn \ee

Thus, taking the hermitian conjugate, our first stationarity equation is 
\be
\label{eq-LorCritPoint2}
X_a \, \alpha(\nb_{ab}) = \frac{\parallel \z_{ab} \parallel}{\parallel \z_{ba} \parallel} e^{i \theta_{ab}} X_b \, J \, \alpha(\nb_{ba}).
\ee

\paragraph{Group Variation}

Finally, we consider the variation with respect to the group variables. The right variation of an arbitrary $\SL(2,\C)$ element $X$ and its Hermitian conjugate are given by
\be
\delta X = X  L, \;\;\;\; \mbox{and} \;\;\;\; \delta X^{\dagger} = L^{\dagger}  X^{\dagger}
\ee
where $L$ is an arbitrary element of the real Lie algebra $\sl(2,\C)_{\R}$. As per the discussion of Section \ref{sec-Spinor} a basis for the Lie algebra

$$L = \alpha_i J^i + \beta_i K^i$$ with $\alpha_i, \beta_i$ in $\R$ and $i=1,\dots,3$ is given by 

$$J^i = L^i_{\SU(2)} =  \frac{i}{2} \sigma^i_P$$ and $$K^i = i L^i_{\SU(2)} = - \frac{1}{2} \sigma^i_P.$$ Here $\sigma^i_{P}$ are again the Pauli matrices and $L^i_{\SU(2)}$ are the generators in the fundamental representation of $\SU(2)$. Thus $J$ generate pure rotations and $K$ pure boosts.

The variation of the action with respect to the group variable $X_a$, $a=1,\dots,4$, yields
\bea
\delta_{X_a} S_L &=& -\sum_{b : b \neq a} \left[ i p_{ab} \left( \frac{\la \z_{ab}| L \, \z_{ab} \ra}{\la \z_{ab}| \z_{ab} \ra} + \frac{\la \z_{ab}| L^{\dagger} \, \z_{ab} \ra}{\la \z_{ab}| \z_{ab} \ra} \right) \right. \nn \\&& \left. + k_{ab} \left( \frac{\la \z_{ab}| L \, \z_{ab} \ra}{\la \z_{ab}| \z_{ab} \ra} + \frac{\la \z_{ab}| L^{\dagger} \, \z_{ab} \ra}{\la \z_{ab}| \z_{ab} \ra} - 2 \frac{\la \z_{ab}| L \, \alpha(\nb_{ab}) \ra}{\la \z_{ab}| \alpha(\nb_{ab}) \ra} \right) \right]. \nn
\eea
for the case where $a<b$. In this case $a$ always occurs to the left in the action. For the other cases similar equations hold.

Using equation \eqref{eq-LorRealityEquation}, $\delta_{X_a} S_L = 0$ becomes

\bea
0&=&\sum_{b : b \neq a} i p_{ab} \left( \la \alpha(\nb_{ab})| L \, \alpha(\nb_{ab}) \ra + \la \alpha(\nb_{ab})| L^{\dagger} \, \alpha(\nb_{ab}) \ra \right) \nn\\&&+ k_{ab} \left( - \la \alpha(\nb_{ab})| L \, \alpha(\nb_{ab}) \ra + \la \alpha(\nb_{ab}), L^{\dagger} \, \alpha(\nb_{ab}) \ra \right) = 0. \nn
\eea

Now using that the expectation values of $L$ in the state $\alpha(\nb_{ab})$ are given by
\be\tag{\ref{eq-costa-def}}
\frac{1}{i} \la \alpha(\nb_{ab})| \mathbf{K} \, \alpha(\nb_{ab}) \ra = \la \alpha(\nb_{ab})| \mathbf{J} \, \alpha(\nb_{ab}) \ra = \frac{i}{2} \, \mathbf{n}_{ab},
\ee
we immediately obtain
\be
\sum_{b : b \neq a} p_{ab} \mathbf{n}_{ab} = 0 \;\;\;\; \mbox{and} \;\;\;\;  \sum_{b : b \neq a} k_{ab} \mathbf{n}_{ab} = 0 \nn
\ee
as the six variational equations. If the tetrahedra defining defining $\Psi_c$, and hence the $\nb_{ab}$ are non-degenerate, these two equations can only hold simultaneously if we have that $p_{ab} = \gamma_a k_{ab}$ for some arbitrary constant $\gamma_a$ at the $a$-th tetrahedron. The equations hold for each tetrahedron, $\gamma_a = \gamma_b = \gamma$ and there is a global parameter $\gamma$ relating the representations:

\begin{equation}p_{ab} = \gamma k_{ab}.\label{eq-LorImmirzi}\end{equation}

This is the simplicity constraints given in \cite{Engle:2007wy}. Therefore we finally again have closure as a stationary point equation:

\be
\label{eq-LorStationary}
\sum_{b:b \neq a} k_{ab} {\bf n}_{ab} = 0.
\ee

\subsection{Geometric Interpretation, Bivector Equations}

We will begin by combining the equations \eqref{eq-LorCritPoint1} and \eqref{eq-LorCritPoint2} into a single bivector equation. We do so using the linear isomorphism $\Gamma$ of Section \ref{sec-Spinor} to take the Hermitian matrices associated to the spinors to vectors:  $\Gamma(X \alpha(\nb) \tensor (X \alpha(\nb))^\dagger) = \hat{X}(1,\nb)$, where $\hat{X}$ is the $\SO(3,1)^+$ element covered by $X$. Applying this to both sides we obtain the equations:

\be
\hat{X}_a \, (1,\nb_{ab}) = \frac{\parallel \z_{ab} \parallel^2}{\parallel \z_{ba} \parallel^2} \hat{X}_b \, (1, -\nb_{ba}) \nn
\ee
for \eqref{eq-LorCritPoint2}, and using that $J(X^{\dagger})^{-1} J = X$ we obtain

\be
\hat{X}_a \, (1,-\nb_{ab}) = \frac{\parallel \z_{ba} \parallel^2}{\parallel \z_{ab} \parallel^2} \hat{X}_b \, (1,\nb_{ba}). \nn
\ee

Wedging these equations together we obtain

\be
\hat{X}_a \triangleright (1,\nb_{ab})\wedge (1,- \nb_{ab}) = \hat{X}_b \triangleright (1,\nb_{ba})\wedge (1,- \nb_{ba}), \nn
\ee
where $\triangleright$ is the standard action on bivectors, or writing $\pole = (1,0,0,0)$ again

\be
\label{eq-LorBivOrientation}
\hat{X}_a \triangleright \pole \wedge (0,\nb_{ab}) = - \hat{X}_b \triangleright \pole \wedge (0, \nb_{ba}).
\ee

Note that as $\hodge \pole \wedge (0,\nb_{ab})$ is orthogonal in the Minkowski metric to $\pole$, we have that $\hodge \hat{X}_a \triangleright \pole \wedge (0,\nb_{ab})$ is orthogonal to the future pointing normal $F_a = \hat{X}_a \triangleright \pole$. Using the isomorphism $\Gamma$ we find that $F_a = \Gamma(X_a X^\dagger_a)$.

Therefore we also have

\be\label{eq-LorBivSimplicity}
F_a \cdot \left(\hodge \hat{X}_a \triangleright \pole \wedge (0,\nb_{ab})\right) = 0\,\,\,\,\,\; \mbox{for all}\; b\neq a.
\ee

\subsection{Geometric Interpretation, $\SU(2)$ Solutions.}

As the bivectors $k_{ab} \hat{X}_a \triangleright \pole \wedge (0,\nb_{ab})$ satisfy simplicity \eqref{eq-LorBivSimplicity}, orientation \eqref{eq-LorBivOrientation} and closure \eqref{eq-LorStationary}, we have by Lemma \ref{lem-BivNonDeg} that they either also satisfy non-degeneracy or they all lie in the same 3-dimensional plane and thus $F_a = F_b \;\; \mbox{for all}\; a,b$. In this case we can choose a gauge in which $F_a = \pole$, that is, $X_a^\dagger X_a = \id$, and the $X_a$ are in $\SU(2)$. Then the above equations reduce again to the $\SO(3)$ equations
\be\tag{\ref{eq-CritPointSU24d}}
\hat{X}_a \nb_{ab} = - \hat{X}_b \nb_{ba},
\ee
and
\be\tag{\ref{eq-StatPointSU24d}}
\sum_{b: b\neq a}k_{ab} \nb_{ab} = 0.
\ee

Therefore the classification Theorem \ref{theo-ClassificationEuc4d} immediately applies to these solutions.

\subsection{Geometric Interpretation, Lorentzian Solutions}

We now consider the case where the bivectors furthermore satisy 4-dimensional non-degeneracy. Then there is a Lorentzian 4-simplex with the tetrahedra of the boundary state as boundary tetrahedra and we can take the boundary state to be a Regge state. We will use the same extension of the boundary geometry maps $\phi_a$ of the Regge state to maps $\Phi_a$ into $\R^{(1,3)}$ as introduced in Section \ref{sec-SU24d GeomInt, Regge State}. We again have the bivectors of the tetrahedra $\Phi_a(\sigma^3_a)$ given by 

\be\label{eq-LorTetBivectors}
B_{ab}(\Phi_a(\sigma^3_a)) = \hodge k_{ab} \pole \wedge (0, \nb_{ab}),
\ee
with the only difference to equation \eqref{eq-TetBivectors} being that the Hodge operator here is the Minkowski version. We then immediately have the analogue of Lemma \ref{lem-Immersed4simpSols}, that is any embedding of the 4-simplex into Minkowski space that has as boundary geometry the geometry of the Regge state provides us with solutions.

\begin{lem}[Solutions from Immersed 4-Simplices]\label{lem-Lor Sols from 4-simplex}
A 4-simplex $\sigma^4_\imm$ immersed into Minkowski space $\R^{(1,3)}$ with boundary geometry that of the Regge state gives a solution to the critical and stationary point equations \eqref{eq-LorBivOrientation} and \eqref{eq-LorStationary} $\hat{X}_a$. If the orientation induced on the boundary of $\sigma^4$ by $\imm$ is the same as the one used in the Regge state, the rotational part of $\imm\circ\Phi_a^{-1}$ is orientation preserving and thus in $\SO(3,1)$. Thus either $\imm\circ\Phi_a^{-1}$ or $-\imm\circ\Phi_a^{-1}$ are in $\SO(3,1)^+$ and we define the sign $\varsigma_a = \pm$ such that $\varsigma_a \imm\circ\Phi_a^{-1} \in \SO(3,1)^+$. Then the solutions $\hat{X}_a$ are given by

\be\hat{X}_a = \varsigma_a \imm\circ\Phi_a^{-1}.\label{eq-LorSolutionsFromImmersions1}\ee

Otherwise $\imm\circ\Phi_a^{-1}\circ P$, where $P$ is parity, is orientation preserving and we have

\be\hat{X}_a = \varsigma_a  \imm\circ\Phi_a^{-1} P.\label{eq-LorSolutionsFromImmersions2}\ee

In both cases the right-hand side means the rotational part of the affine map. The bivectors are given by

$$B_{ab} (\sigma^4_{\imm}) = \mu_\imm k_{ab} (\varsigma_a \hat{X}_a) \triangleright \left(\pole \wedge (0, \nb_{ab})\right) = \mu_\imm k_{ab} \hat{X}_a \triangleright \left(\pole \wedge (0, \nb_{ab})\right).$$
with $\mu_\imm = +1$ in for the first case and $\mu_\imm = -1$ for the second one including parity. The future pointing normals of the immersed 4-simplex are given by $F_a = \hat{X}_a \pole$ and the outward pointing normals by $N_a = \varsigma_a F_a = \varsigma_a \hat{X}_a \pole$. Note that the immersion $\imm' = -\imm$ leads to the same solution $\hat{X}_a$ but signs $\varsigma'_a = - \varsigma_a$.
\end{lem}
\begin{proof}
By Theorem \ref{Theo-BivReconstruction} the bivectors $B_{ab}(\sigma^4_i)$ satisfy closure and orientation. As bivectors don't register inversions $\varsigma_a$ drops out. If the orientations agree the map \eqref{eq-LorSolutionsFromImmersions1} maps the geometric tetrahedron of the boundary state onto the geometric tetrahedron of the 4-simplex and thus maps its bivectors \eqref{eq-TetBivectors} to the geometric ones. If they disagree they map the tetrahedron with $\nb_{ab}$ inward normals to the bivectors of the 4-simplex, accounting for the factor $\mu$. Putting the form of the $B_{ab}(\sigma^4_\imm)$ into the orientation equation of Theorem \ref{Theo-BivReconstruction} immediately shows that the $\hat{X}_a^\pm$ solve \eqref{eq-LorBivOrientation} and \eqref{eq-LorBivSimplicity}. As $P \pole = \pole$ and the orientations were chosen with $\pole$ as outwards to $\pole^\perp$ the form of the future and outward normals $F_a$, $N_a$ also immediately follows.\end{proof}

Conversely, the bivectors $k_{ab} \hat{X}_a \triangleright \left(\pole \wedge (0, \nb_{ab})\right)$ satisfy simplicity, orientation, closure and non-degeneracy and therefore by theorem $\ref{Theo-BivReconstruction}$ define a pair of inversion related 4-simplices $\sigma^4_\imm$ immersed in $\R^(1,3)$ and a sign $\mu_{\imm}$. For a particular $\sigma^4_\imm$ we also have the signs $\varsigma_a$ from $N_a = \varsigma_a F_a$. The inversion related 4-simplex clearly has $\varsigma'_a = - \varsigma_a$. We can also see that the solution reconstructed from either of these 4-simplices is in fact the the solution $\hat{X}_a$ with which we began.

Together with the previous section this completely classifies the solutions. Furthermore it is also immediate that the discrete and scaling symmetries do not act on the solution and the $\SL(2,\C)$ symmetry $X_a \rightarrow Y X_a$ acts via $\SO(3,1)^+$ on the immersion $\imm \rightarrow \hat{Y} \imm$.

From the geometry we again can induce symmetries via all of $\OO(3,1)$. This means in particular that we have inversion which leaves the solution invariant but switches the signs $\varsigma_a$ and parity $P$. By the orientation definition $P\imm$ again has the opposite parameter $\mu$, that is, $\mu_\imm = - \mu_{P\imm}$.

\subsection{The Action}

We now again evaluate the action $\eqref{eq-LorAction}$ on the critical points. To do so we will begin by discussing the geometry and Regge action of a Lorentzian 4-simplex. The key difficulty here will be that whereas in the Euclidean there always is a rotation that takes any outward pointing normal of the 4-simplex to any other outward pointing normal, this is not the case in Minkowski space. This is due to the fact that we are working with $\SO(3,1)^+$ and thus future pointing normals can only be rotated to future pointing normals, but at least some of the normals must be past pointing.

As in the case for immersed surface in 3d, we will define the dihedral angle through the dihedral connection $\hat{D}_{ab} \in \SO(3,1)^+$ defined by

$$\hat{D}_{ab} F_b = F_a$$
and

$$\hat{D}_{ab} \; \imm(\sigma^2_{ab}) = \imm(\sigma^2_{ab}).$$

This is then of the form

\be\label{eq-LorDiConn}
\hat{D}_{ab} = \exp\left(\Theta_{ab} \hodge \frac{B_{ab}(\sigma^4_{\imm})}{|B_{ab}(\sigma^4_{\imm})|} \right).
\ee
where $\Theta_{ab}$ is defined as the dihedral angle. It is positive if $N_a$ and $N_b$ are both future or both past pointing, and negative if one is future and one is past pointing. The former case is called a thick wedge, the latter a thin wedge. We will again work with a covering lift of this connection in $\SL(2,\C)$.

We would now like to establish a commuting diagram of the type \eqref{eq-ftjCommDiagSpin4} in order to again evaluate the action, given a solution associated to the immersion $\imm$. However, the straightforward diagram based on $X_a$, $g_{ab}$ and $D_{ab}$ fails, as can be seen from the definition of $X_a$ in \eqref{eq-LorSolutionsFromImmersions1} and \eqref{eq-LorSolutionsFromImmersions2}. The problem is the presence of the signs $\varsigma_{a}$ that ensure our solution is in $\SO(3,1)^+$. Thus while we have that
$$ \hat{X}_b^{-1} \hat{D}_{ba} \hat{X}_a \hat{g}_{ab} \pole = \pole,$$

we have that
$$ \hat{X}_b^{-1} \hat{D}_{ba} \hat{X}_a \hat{g}_{ab}\; \Phi_b(\sigma^2_{ab}) = \varsigma_a \varsigma_b \Phi_b(\sigma^2_{ab}).$$

That is, the naive diagram would commute for thick wedges but not for thin wedges. To fix this we introduce a rotation by $|\varsigma_a + \varsigma_b| \frac{\pi}{2}$ in the plane of $\imm{\sigma^2_{ab}}$: $$\hat{R}_{ab} = \exp\left(|\varsigma_a + \varsigma_b| \frac{\pi}{2} B_{ab}(\sigma^4_{\imm}) \right).$$ That is, for thick wedges $R_{ab} = \id$ and for thin wedges $R_{ab}$ is a rotation by $\pi$ in the plane of the triangle. Thus  $$ R_{ab} \imm{\sigma^2_{ab}} = \varsigma_a \varsigma_b \imm{\sigma^2_{ab}}$$ and $$R_{ab} N_a = N_a.$$
Furthermore we have $\hat{D}_{ab} \hat{R}_{ab} = \hat{R}_{ab} \hat{D}_{ab}$. We then obtain
$$ \hat{X}_b^{-1} \hat{D}_{ba} \hat{R}_{ab} \hat{X}_a \hat{g}_{ab} \pole = \pole$$
and
$$ \hat{X}_b^{-1} \hat{D}_{ba} \hat{R}_{ab} \hat{X}_a \hat{g}_{ab}\; \Phi_b(\sigma^2_{ab}) = \Phi_b(\sigma^2_{ab}).$$

Thus we again have the commuting diagram

\begin{equation}
\label{eq-ftjCommDiagLor}
\xymatrixcolsep{4pc}\xymatrixrowsep{4pc}
\xymatrix{\ar @{} [dr]  \Phi_a(\sigma^3_a) \ar[d]_{\hat{g}_{ba}} \ar[rr]^{\hat{X}_a} && ~ \varsigma_a \imm(\sigma^3_a) \ar[d]^{ \hat{D}_{ba} \hat{R}_{ab}}   \\
\Phi_b(\sigma^3_b)\ar[rr]_{\hat{X}_b} && ~ \varsigma_b \imm(\sigma^3_b)  }
\end{equation}
and by Lemma \ref{lem-CommDiagLift} we have again that at one of the solutions related by the discrete symmetries the lifting to $\SL(2,\C)$ with $R_{ab} = \nu_{ab} \exp\left(|\varsigma_a + \varsigma_b| \frac{\pi}{2} L\right)$, with $L$ the normalised generator of $\sl(2,\C)$ asosciated to the bivector $B_{ab}$ in the spinor representation, holds:

\begin{equation}
\label{eq-ftjCommDiagSL2C}
\xymatrixcolsep{4pc}\xymatrixrowsep{4pc}
\xymatrix{\ar @{} [dr]  \Phi_a(\sigma^3_a) \ar[d]_{g_{ba}} \ar[rr]^{X_a} && ~\varsigma_a \imm(\sigma^3_a) \ar[d]^{ D_{ba} R_{ab}}   \\
\Phi_b(\sigma^3_b)\ar[rr]_{X_b} && ~\varsigma_b \imm(\sigma^3_b)  }
\end{equation}

To evaluate the action on the critical points we use equation \eqref{eq-LorCritPoint2} and the Regge state condition to obtain

\be
\label{eq-LorDihedralEV}
X_a \, \alpha(\nb_{ab}) = \frac{\parallel \z_{ab} \parallel}{\parallel \z_{ba} \parallel} e^{i \theta_{ab}} D_{ba}R_{ba} X_a \, \alpha(\nb_{ab}).
\ee

This is an eigenvalue equation for $D_{ba} R_{ba}$. Since $D_{ba}$ is the lift of a pure boost and $R_{ba}$ a pure rotation it follows that $e^{- i \theta_{ab}}$ is the eigenvalue of $R_{ab}$ and thus $$e^{- i \theta_{ab}} = \nu_{ab} e^{- i \pi/2}$$ and $\frac{\parallel \z_{ab} \parallel}{\parallel \z_{ba} \parallel}$ is the eigenvalue of $D_{ba}$ associated to the eigenstate state $X_a \, \alpha(\nb_{ab})$. However from the explicit form of the dihedral connection $\eqref{eq-LorDiConn}$ and the discussion in Section \ref{sec-Spinor} we know that this eigenvalue is $e^{-\mu_{\imm} \Theta_{ab}}$. On the other hand, as mentioned above, using the critical point equations the action can be written in terms of these eigenvalues directly:

\be
S_L[\imm] = i \sum_{a<b} p_{ab} \, \log \, \frac{\parallel \z_{ba} \parallel^2}{\parallel \z_{ab} \parallel^2} + 2 k_{ab} \, \theta_{ab},\nn
\ee
and thus
\be
S_L[\imm] = i \sum_{a<b} \gamma 2 k_{ab} \mu_{\imm} \Theta_{ab} + 2 k_{ab} \theta_{ab}.\nn
\ee

Where we have ignored the signs $\nu_{ab}$ which will appear in the total asymptotics. Now $2 \theta_{ab}$ is either $\pi$ or $0$ defining $M = \sum_{thin\, wedges} k_{ab}$ we can see that the integrality conditions of the boundary states force $M$ to be an integer. Thus the sign in front of the second term doesn't actually matter and it merely multiplies the overall asymptotics by a sign. The term $\sum_{a<b} k_{ab} \Theta_{ab}$ is the Regge action $S_R$ of the 4-simplex with areas $k_{ab}$ and we finally have

\be\label{eq-LorActionEvaluated}
S_L[\imm] = i 2\gamma \mu_{\imm} S_R[\imm] + i M \pi. \nn
\ee

For the $\SU(2)$ cases the boosts are all zero and the eigenvalue $\theta_{ab}$ is identified with the self-dual or anti-self-dual part of the dihedral rotation. In the case of a 4-simplex immersed into 3d both perspectives coincide as the dihedral rotations are all $\pi$ or zero.

\subsection{Asymptotic Formula}\label{sec-LorAsymptFormula}

Throughout we write $\nu_{crit} = \prod_{a<b} \nu_{ab}^{2k_{ab}}$ evaluated for a critical point. Summarising and combining these results we can again give the asymptotic behaviour for $Z(\Psi_\lambda, \lambda p_{ab})$. We have assumed that the tetrahedra entering $\Psi$ are non-degenerate:
\begin{itemize}

\item Unless there exists a parameter $\gamma k_{ab} = p_{ab}$ the amplitude is exponentially suppressed:

$$Z(\Psi_\lambda, \lambda p_{ab}) = o(\lambda^N) \,\, \;\mbox{for all}\; N.$$

We will assume $\gamma k_{ab} = p_{ab}$ from here on.

\item If the boundary state $\Psi$ is a coherent boundary state that does not have a Regge geometry and does not define a constant $\su(2)$-valued 2-form on a 4-simplex it is exponentially suppressed:

$$Z(\Psi_\lambda, \lambda p_{ab}) = o(\lambda^N) \,\, \;\mbox{for all}\; N.$$

\item If the boundary state $\Psi$ is a coherent boundary state that does not have a Regge geometry but arises as the boundary of a constant $\su(2)$-valued 2-form on a 4-simplex there is exactly one solution $X$ to the critical and stationary phase equations up to symmetries and we have:

\be
|Z(\Psi_\lambda, \lambda p_{ab})| =  \frac{1}{\lambda^{12}} |N_{| X}| + o(\lambda^{-12}),
\ee

where $N$ is a constant independent of $\lambda$ evaluated below.

\item If the boundary state $\Psi$ is a coherent boundary state that does have the boundary geometry of a geometric 4-simplex immersed in 3d Euclidean space we have exactly one solution $X$ to the critical and stationary phase equations up to symmetries. If it is the associated Regge state we have
\be
Z(\Psi_\lambda, \lambda p_{ab}) =  (-1)^{\chi+\lambda M} \nu^{\lambda} \frac{1}{\lambda^{12}} N_{| X} + o(\lambda^{-12}),
\ee
where $M = \sum_{thin\, wedges} k_{ab}$ as in the preceding section.

\item If the boundary state $\Psi$ is a coherent boundary state that does have the boundary geometry of a geometric 4-simplex embedded in 4d Euclidean space we have exactly two solution $X^\pm$ to the critical and stationary phase equations, associated to the embeddings $\imm$ and $P\imm$ into $\R^4$, up to symmetries. If it is the associated Regge state we have
\bea
Z(\Psi_\lambda, \lambda p_{ab}) &=&  (-1)^{\chi} \frac{1}{\lambda^{12}}\times\nn\\&& \left( \nu_+^{\lambda} N_{| X^+} e^{i \mu_{\imm} \lambda S_{R}(\imm)} + \nu_-^{\lambda} N_{| X^-} e^{i \mu_{P\imm} \lambda S_{R}(P\imm)}\right) \nn\\&&+ o(\lambda^{-12}),
\eea
where $S_R(\imm) = S_R(P\imm)$ is the Euclidean Regge action associated to the embedding $\imm$.

\item If the boundary state $\Psi$ is a coherent boundary state that does have the boundary geometry of a geometric 4-simplex embedded in 4d Lorentzian space we have exactly two solutions $X$, $X'$ to the critical and stationary phase equations, associated to the embeddings $\imm$ and $P\imm$ into $\R^{(3,1)}$, up to symmetries. If it is the associated Regge state we have
\bea
Z(\Psi_\lambda, \lambda p_{ab}) &=&  (-1)^{\chi + \lambda M} \frac{1}{\lambda^{12}}\times\nn\\&&\left( \nu_\imm^{\lambda} N_{| X} e^{i \mu_{\imm} \lambda \gamma S_{R}(\imm)} + \nu_{P\imm}^{\lambda} N_{| X'} e^{i \mu_{P\imm} \lambda \gamma S_{R}(P\imm)}\right) \nn\\&&+ o(\lambda^{-12}),
\eea
where $S_R(\imm) = S_R(P\imm)$ is the Lorentzian Regge action associated to the embedding $\imm$.

\item There are no other cases.
\end{itemize}

The numerical factors $N$ of the stationary phase formula have to be evaluated at the critical points and are given by
\bea
N_{|\mbox{\tiny crit}} &=&   (2\pi)^{22}        \frac{2^4}{    \sqrt{ \det -H_{|\mbox{\tiny crit}} }} \prod_{a<b}   2k_{ab}   c_{ab}  {\Omega_{ab}}_{|\mbox{\tiny crit}}
\nn \\
&=&
 2^{36} \pi^{12}\left( \frac{1+i\gamma}{1-i\gamma}\right)^5 \frac{1}{\sqrt{ \det -H_{|\mbox{\tiny crit}} }}
\prod_{a<b}  k_{ab} {\Omega_{ab}}_{|\mbox{\tiny crit}}.
\eea

The factor $(2\pi)^{22} $ arises from the stationary phase formula. We have a $(6\times 4)$-dimensional integration over $\SLtwoC^6$ and a 20-dimensional integration over $(CP^1)^{10}$.  Asymptotically, we have $d_{\lambda k} \sim 2\lambda k$ which cancels some of the $\lambda$ of the stationary phase formula.  The factor $2^4$ counts the volume of the orbit of the discrete symmetry that is not fixed in the integral. $H_{|\mbox{\tiny crit}}$ is the Hessian matrix of the action \eqref{eq-LorAction} evaluated at the critical points; its explicit form is given in the appendix of \cite{Barrett2010}. The measure $\prod{\Omega_{ab}}_{|\mbox{\tiny crit}}$ is the measure term evaluated at the critical points. The ratio of $\prod\Omega_{ab}$ and $\sqrt{ \det -H }$ is independent of coordinates even though each individually must of course be evaluated with respect to some coordinates. The constant $c_{ab}$ is equal to $\frac1\pi\sqrt{\frac{1+i\gamma}{1-i\gamma}}$ when $p = \gamma k$, which is a pure phase.

\chapter{Conclusions}

In this thesis we connected two themes, the construction of topological quantum field theories from a graphical calculus and the definition of the same through the representation theory of $\SU(2)$, and the geometricity of the representation theory of $\SU(2)$. The latter allowed the leading order behaviour of some of the amplitudes of the former in geometric terms.

To this end, the first chapter gave a detailed sketch of the construction of TQFTs (Definition \ref{defi-TQFT}) from state sums (Definition \ref{defi-statesum}), that is, the sum of local amplitudes associated to states on triangulations. Pachner's theorem (Theorem \ref{Theo-PAmoves}) gave us a way to construct TQFTs as state sums. A graphical calculus (Section \ref{sec-DualNets}) was used as a convenient intermediate step to algebratise the Pachner moves and solve them in terms of graphical equalities. Such a solution was then given through $\SU(2)$ representation theory, leading to the definition of the Ponzano-Regge state sum (Definition \ref{defi-PRStatesum}). A key consequence of the graphical calculus that allowed us to ignore the problem of regularising the model was a dual expression for state sum amplitudes (Theorem \ref{theo-dualevalofPF}) that is manifestly finite for the Ponzano-Regge model and was the basis of the analysis of geometricity in Chapter 3.

In Chapter 2 we reviewed a host of geometricity results for $\SU(2)$, $\Spin(4)$ and $\SL(2,\C)$. The key tool for understanding the geometry of the representations of $\SU(2)$ were the coherent states of Equation \ref{eq-costa-def}. Starting from these and using symplectic reduction we defined coherent triangles, tetrahedra and from these coherent two- and three-manifolds (Definitions \ref{defi-2dReggeState}, \ref{defi-3dCoherentBoundaryStates} and \ref{defi-3dReggeStates}). For $\Spin(4)$ we showed how to interpret the bivectors on $\R^4$ as its generators and how to use Hodge duality to decompose them into two copies of $\SU(2)$. We also showed how to characterise those bivectors corresponding to the faces of a 4-simplex embedded in $\R^4$ (Theorem \ref{Theo-BivReconstruction}). Similarly we showed how bivectors in $\R^{3,1}$ have the structure of a complexified $\su(2)$, and in Section \ref{sec-Spinor} how the spinor equations and geometricity arises as a result.

Chapter 3 set about to combine the first two chapters for the Ponzano-Regge amplitude. Using an exponential form of the dual expression for the state sum \ref{eq-PRAmpExpo} the overlap between a coherent boundary manifold and the Ponzano-Regge amplitude was calculated to first order in the large spin expansion (Theorem \ref{theo-PRAsymptotics}). The key geometricity result enabling this was Lemma \ref{lem-3dGeometricity} which gave the solutions to the stationary and critical point equations that provide the leading order in the expansion in terms of immersions of the geometry underlying the coherent boundary manifold into $\R^3$. We furthermore could evaluate the phase of the asymptotics as a type of Regge action of the geometry in question.

The Euclidean 4-dimensional case was consider in Chapter 4. While the 4-dimensional state sum was not defined in Chapter 1, its 4-simplex weight in terms of the dual evaluation (Theorem \ref{theo-dualevalofPF}) is essentially the same as for the Ponzano-Regge model. Thus we could again calculate the overlap of the state sum weight with coherent boundary states in Theorem \ref{theo-ftjAsymptotics}. Theorem \ref{theo-ClassificationEuc4d} gave a full classification of the possible type of solutions in terms of $\su(2)$ valued bivectors on the 4-simplex for general boundary states and, as a special case, geometric embeddings of the boundary manifold in $\R^4$ for Regge states. In the latter case we again find the Regge action of the embedded geometry. As a corollary we could immediately give the leading order behaviour of the EPRL-FK type amplitudes.

We gave the full leading order behaviour of the Lorentzian EPRL amplitude in Section \ref{sec-LorAsymptFormula}. The boundary states were again the coherent boundary states of Chapter 2. The two key inputs here were the spinor geometricity considerations of Chapter 2 and the full classification of $\SU(2)$ solutions in Chapter 4. In the geometric sectors of the theory we again could derive the Regge action as the phase of the leading order terms.

The work of this thesis covers geometricity for $\SU(2)$ state sum amplitudes and the recent theories based on them, that is, the squaring of the $\SU(2)$ amplitude for the Euclidean EPRL-FK type amplitudes and the embedding of $\SU(2)$ into the unitaries of its complexification in the Lorentzian EPRL amplitude. A key deficit in the analysis is the failure to understand the Hessians appearing in the asymptotics in geometric terms. Nevertheless the results here allowed for example the calculation of the so called ``graviton propagator'' in the Euclidean models \cite{Bianchi:2006uf}.

The two aspects of $\SU(2)$ this thesis are based on, the algebraic properties of its representation theory allowing us to define topological invariants on the one hand and the geometric asymptotics based on 3-dimensional geometries on the other hand concern vastly different structures. As a result, any deformations or generalisations of the theory that attempt to preserve or enhance either the topological character or the geometric one are likely to perturb the other. The EPRL-FK type models use the $\SU(2)$ geometricity results and attempt to reduce the $\SU(2)$ theory to its geometric sector. This was originally motivated by the Plebanski approach to gravity in which a set of constraints reduces an $\su(2)$ field to a geometric one. The result of this restriction is then, of course, no longer triangulation independent. On the other hand, the very general spherical categories of Barrett and Westbury \cite{Barrett1999} give graphical evaluations that are sufficient for constructing topological quantum field theories but have almost no geometric flavour left to them. However, note that in three dimensions the Turaev-Viro model \cite{Turaev1992} retains some geometric flavours while sharpening the topological sensitivity of the theory.

In four dimensions the situation is more limited. The issue of constructing sensitive TQFTs using triangulations is still open, and while proposals for appropriate categories for 4-dimensional theories exist these seem to be too restrictive and lacking in examples. The $\SU(2)$ based boundary states seem too restrictive to capture the full dynamical content of a theory as complex as GR. It is important to note that in the 4-dimensional case presented here the appearance of the Regge action and the calculations based on it do not test the dynamics of the theory. To do so we would need to perform the sum over representations and intertwiners, in other words we would need to understand the asymptotics of larger triangulations. In the same vein it would be very interesting to understand the behaviour of the triangulation dependent models under Pachner moves. This can be considered a sort of coarse graining or renormalisation procedure. 

In both the Euclidean EPRL-FK and the Lorentzian EPRL amplitude the reduction to the geometric subsector is incomplete. Instead a diagonal unconstrained $\SU(2)\, BF$ type sector remains. This has the potential to spoil the dynamics. Going forward it will be important to understand whether it is possible to ignore, factor out or constrain away this sector. Finally it should be noted that conversely the clear appearance of the geometric sector in the $\ftj$ symbol of the 4-dimensional $\SU(2)$ theory suggests the possibility to mimic the self-dual formulation of euclidean general relativity and reduce to a geometric sector using much simpler constraints than those of the EPRL-FK model.

\appendix

\chapter{Appendix}

\section{Discrete Connections} \label{sec-DiscConn}

\begin{defi}[Discrete Connection]\label{defi-DiscConn}
Given an oriented manifold $\bo^n$ with a triangulation $\TT^n$ with simplices $\sigma_a^n$, a discrete connection is an assignment of a group element $h_{ab}$ to every oriented face $\sigma^{n-1}_{ba}$, satisfying $h_{ab} = h_{ba}^{-1}$. $h_{ab}$ can be interpreted as the parallel transport from $\sigma^n_b$ to $\sigma^n_a$ via the face $\sigma^{n-1}_{ba}$.
\end{defi}

In terms of the Poincar\'e dual of the triangulation we can think of it as the parallel transport along the edge from $b$ to $a$.

We can lift the group elements $\hat{h}_{ba}$ of a discrete $\SO(3)$ connection to $\SU(2)$ elements $h_{ab}$ that cover $\hat{h}_{ba}$ thus defining a discrete $\SU(2)$ connection. However in doing so we are of course free to choose $\pm h_{ab}$. Thus there is a set of $\SU(2)$ connections covering $\hat{g}_{ba}$ related by $h'_{ba} = h_{ba} \nu_{ba}$ with $\nu_{ba} = \pm 1$. This can be fixed up to gauge by using a spin structure on $\bo^n$. The spin structures $\bo^n$ are parametrised by the elements $\omega\in H^1(\bo^n,\Ztwo)$, however, not canonically. We wil here give a specific prescription to lift an $\SO(3)$ connection which will depend on a number of choices.

Consider first that to fix the signs $\nu_{ab}$ up to gauge we need to give the signs for every loop in the 1-skeleton of the Poincar\'e dual. The values around this loops are however not unrelated, but can be expressed in terms of a basis of the first homology class of the 1-skeleton with values in $\Ztwo$. This decomposes into two parts, the basis elements contractible on the 2-skeleton and those not. The first correspond to a set of generators of $H^2$ of the 2-skeleton of the dual triangulation, the latter then correspond to a basis of $H^1(\bo^n,\Ztwo)$ which is equal to the first homology on the two skeleton.

We will now use the following prescription:
\begin{itemize}
 \item For every loop $\gamma$ that is the boundary of one of the chosen generators of $H^2$ of the 2-skeleton, let the discrete holonomy along this loop be the $\SU(2)$ group element with $h_{\gamma} = \exp (\theta \nb\cdot L_{\frac12})$, $0 < \theta \leq \pi$.
 \item For every loop $\gamma$ that is one of the chosen basis elements of $H^1(\bo^n,\Ztwo)$, let the discrete holonomy along this loop be the $\SU(2)$ group element with $h_{\gamma} = \omega(\gamma)\exp (\theta \nb\cdot L_{\frac12})$, $0 < \theta \leq \pi$. 
\end{itemize}

This fixes the group elements $h_{ab}$ up to gauge.
\begin{defi}[Covering Lift]\label{defi-CoveringLift}
A discrete $\SU(2)$ connection $h_{ab}$ on a manifold $\bo^n$ with triangulation $\TT^n$ and a spin structure is called a covering lift of a discrete $\SO(3)$ connection if it satisfies the above criteria.
\end{defi}

We then have the following lemma:

\begin{lem}[Spin Lift of Gauge Transformations]\label{lem-CommDiagLift}
Let $\hat{h}_{ab}$ and $\hat{h}'_{ab}$ be two discrete $\SO(3)$ connections on a triangulation $\TT^n$ of $\bo^n$ that are related by a gauge transformation $\hat{h}_a$: $\hat{h}_a \hat{h}_{ab} \hat{h}_b^{-1} = \hat{h}'_{ab}$. Then given a spin structure on $\bo^n$ any covering lifts of $\hat{h}_{ab}$ and $\hat{h'}_{ab}$ to $\SU(2)$ connections $h_{ab}$ and $h'_{ab}$ are related by a covering of the gauge transformation $\hat{h}_a$; $h_a$. That is, $h_a h_{ab} h_b^{-1} = h'_{ab}$.
\end{lem}
\begin{proof}
Note that for any lift $h'_a$, the connection $h'_a h_{ab} {h'}_b^{-1}$ is a covering lift of $\hat{h}'_{ab}$ compatible with the spin structure. This is immediate as the conditions on being a covering lift are gauge invariant. Therefore it differs from  $h'_{ab}$ at most by signs $\epsilon_a$, and the lift $h_a = \epsilon_a h'_a$ defines the required lift of the gauge transformation $\hat{h}_a$ relating $h_{ab}$ and $h'_{ab}$. \end{proof}

The lifting from $\SO(3,1)^+$ to its double cover $\SL(2,\C)$ proceeds among the same lines.

\section{Stationary Phase}\label{sec-StatPhase}

We briefly give a summary of stationary phase techniques used throughout this thesis. Our main reference is \cite{Hormander}.

Take a closed manifold $D$ of dimension $n$ and consider smooth, complex-valued functions $a$ and $S$ on $D$
such that the real part $\Reel (S) \le 0$. We will then asymptotically evaluate the function
 \be\label{eq-XAsymptoticIntegral}
 f(\lambda) = \int_D d x \, a(x) \, e^{\lambda
S(x)}. \ee

We will need the Hessian $H$ of $S$, that is, the $n\times n$  matrix of second derivatives of $S$. First consider the case that the stationary points of $S$ are isolated and thus that the Hessian is non-degenerate. We will then need those stationary points that are in addition critical points. That is, those point $x_c$ such that $\delta_x S(x)|x_c = 0$ and $\Reel (S(x_c)) = 0$.

If $S$ has no critical stationary points the integral \eqref{eq-XAsymptoticIntegral} is exponentially suppressed for large $\lambda$. That means the function $f$ decreases
faster that any power of $\lambda^{-1}$, for all $N \geq 1$: \be \label{eq-XAsymptoticSuppression} f(\lambda) =
o(\lambda^{-N}). \ee

If isolated critical stationary points exist, and $a$ is non vanishing at them, the asymptotics of $f$ is given by a sum over such points over terms of order $\lambda^{-n/2}$. The expansion of $f$ in $\lambda^-1$ is given by \be \label{eq-XAsymptoticSeperated} \sum_{x_c} a(x_c) \left(\frac{2 \pi}{\lambda}\right)^{n/2} \frac 1{\sqrt{ \det (-H|_{x_c})}} \,
e^{\lambda S(x_c)} \left[1+ O(1/\lambda) \right]. \ee

For a rigorous definition of the square root of the determinant in this context see \cite{Hormander}.

At a critical stationary point, the matrix $-H$ has a positive-definite real part,  and the square root of the determinant of this matrix is the unique square root which is continuous on matrices with positive-definite real part, and positive on real ones. For further details see \cite{Hormander}.

If there are stationary critical points where the determinant vanishes more care is needed and the degenerate directions need to be modded out. We call $\mathcal{C}$ the set of stationary critical points $\mathcal{C}:=\{y\in D~|~ \delta S(y)=0,~\Reel S(y)=0 \}$. Now if $\mathcal{C}$ is a disjoint union of closed submanifolds of $D$, $S$ is called a Morse-Bott function in the literature and a Morse function in the special case when all manifolds are 0-dimensional. In the latter case we simply have isolated critical points as discussed above.

In general it is a sum over critical manifolds again where, again assuming $a$ is non-vanishing on the critical manifolds, each critical manifold $\mathcal{C}_{x_0}$ of dimension $p$, labelled by some $x_0$ on the critical manifold, contributes the term\cite{ramacher-2009}
\be
\left(\frac{2 \pi}{\lambda}\right)^{(n-p)/2}
e^{\lambda S(x_0)}
\int_{\mathcal{C}_{x_0}} d\omega_{\mathcal{C}_{x_0}}(y) \frac {a(y)}{\sqrt{ \det (-H^\perp(y))}} \,
\left[1+ O(1/\lambda) \right],
\ee
where $H^\perp(y)$ is the restriction of the Hessian to the normal directions to
$\mathcal{C}_{x_0}$ as defined by some a Riemannian metric on $D$, and $d\omega_{\mathcal{C}_{x_0}}$ is the measure induced on the critical submanifold by the same Riemannian measure on the domain space. This can be extended to the case where $\mathcal{C}$ is a manifold-with-boundary.

\section{Table of Symbols}

\begin{itemize}

\item [{$\la \cdot \ra$}\;:] An evaluation of coloured diagrams.
\item [{$(\cdot,\cdot)$}\;:] The bilinear inner product on $\SU(2)$ irreps. 
\item [{$\la\cdot|\cdot\ra$}\;:] The Hermitian inner product on $\SU(2)$ irreps.
\item [{$\ftj$}\;:] The 15j symbol of $\SU(2)$ recoupling theory.
\item [{$\alpha_j$}\;:] A state in the $\SU(2)$ irrep $j$.
\item [{$\alpha(\nb)_j$}\;:] Coherent $\SU(2)$ state in the direction $\nb$ and the representation $j$.
\item [{$\ad$}\;:] The admissibility condition on a tripplet of $\SU(2)$ irreps.
\item [{$B^{IJ}$}\;:] A bivector.
\item [{$(\bb^+, \bb^-)$}\;:] A bivector written in terms of its self-dual, anti-self-dual decomposition.
\item [{$\ci$}\;:] The circle.
\item [{$c^k$}\;:] A colouring function.
\item [{$\nCob$}\;:] $n$-dimensional cobordisms.
\item [{$\epsilon$}\;:] The fully antisymmetric tensor in any dimension with $\epsilon^{01} = \epsilon^{0123} = \epsilon^{1234} = \epsilon^{123} = 1$.
\item [{$\epsilon_j$}\;:] The symmetrised tensor product of $2j$ 2-dimensional $\epsilon$ tensors.
\item [{$f^k$}\;:] A state sum amplitude, a function from a coloured $k$-simplex to $\C$.
\item [{$g$}\;:] An $\SU(2)$ group element.
\item [{$\hat{g}$}\;:] An $\SO(3)$ group element.
\item [{$\Hom(\cdot,\cdot)$}\;:] The space of linear group homomorphisms between two representation spaces.
\item [{$\Inv$}\;:] Invariant subspace of the tensor product of representations.
\item [{$\iota$}\;:] State in the invariant subspace of the tensor product of representations.
\item [{$\imm$}\;:] Immersion of a surface or a standard simplex into flat $\R^3$, $\R^4$ or $\R^{(1,3)}$.
\item [{$J$}\;:] The anti-linear $\SU(2)$ group homomorphism relating the bilinear and the Hermitian inner products. 
\item [{$L^i_j$}\;:] The standard basis of the Lie algebra $\su(2)$ in the irrep $j$ with $[L^i, L^j] = - \epsilon^{ijk} L^k$.
\item [{$\cbo$}\;:] Cobordism.
\item [{$\pole$}\;:] The vector $(1,0,0,0)$.
\item [{$\nb_{ab}$}\;:] 3-vector associated to the edge or triangle bordering the triangles or tetrahedra $\sigma_a$, $\sigma_b$ defining a coherent boundary state.
\item [{$N$}\;:] Normal vector in 4d.
\item [{$\phi_a$}\;:] Boundary immersion of the simplex $\sigma_a$ into $\R^3$.
\item [{$\Phi_a$}\;:] Extension of the boundary immersion of the simplex to a map into $\R^4$.
\item [{$\sigma^n$}\;:] The $n$-dimensional simplex.
\item [{$\sigma^i_P$}\;:] The Pauli matrices with $L^i_{\frac12} = \frac{i}{2} \sigma^i_P$.
\item [{$\bo$}\;:] Boundary manifolds.
\item [{$\T$}\;:] Triangulation of a manifold.
\item [{$\T_k$}\;:] The $k$-simplices of a triangulation $\T$.
\item [{$\tet$}\;:] The tetrahedral network.
\item [{$\thet$}\;:] The theta network.
\item [{$\Vect$}\;:] Category of Vector spaces.
\item [{$\vb_{ab}$}\;:] Edge vector of an immersed surface in $\R^3$.
\item [{$X_a$}\;:] An $\SU(2)$, $\Spin(4)$ or $\SL(2,\C)$ group element associated to the triangle or tetrahedron $\sigma_a$.
\item [{$\hat{X}_a$}\;:] An $\SO(3)$, $\SO(4)$ or $\SO(1,3)^+$ group element associated to the triangle or tetrahedron $\sigma_a$.
\item [{$\Psi$}\;:] A state in the boundary state space of a particular model.
\item [{$\Z$}\;:] Partition function or amplitude of a particular model.

\end{itemize}
\bibliography{thesis}
\bibliographystyle{hep}

\end{document}

%% file: Macros.tex
\newcommand{\bea}{\begin{eqnarray}}
\newcommand{\eea}{\end{eqnarray}}
\newcommand{\nn}{\nonumber}
\newcommand{\be}{\begin{equation}}
\newcommand{\ee}{\end{equation}}

\theoremstyle{definition}
\newtheorem{defi}{Definition}[section]
\theoremstyle{plain}
\newtheorem{theo}[defi]{Theorem}
\newtheorem{lem}[defi]{Lemma}
\newtheorem{Cor}[defi]{Corollary}

\newcommand{\nb}{{\bf n}}
\newcommand{\mb}{{\bf m}}
\newcommand{\no}{{\bf n}_0}
\newcommand{\vb}{{\bf v}}
\newcommand{\N}{\mathbb{N}}
\newcommand{\Ztwo}{\mathbb{Z}^2}
\newcommand{\Z}{\mathcal{Z}}
\newcommand{\T}{\mathcal{T}}

\newcommand{\R}{\mathbb{R}}
\newcommand{\C}{\mathbb{C}}

\newcommand{\CP}{\mathbb{CP}}

\newcommand{\imm}{\mathfrak{i}}

\newcommand{\id}{\mathbf{1}}
\DeclareMathOperator{\tr}{tr}

\newcommand{\Vect}{V\! ect} 
\newcommand{\nCob}{nCob} 
\newcommand{\bo}{\Sigma} 
\newcommand{\cbo}{M} 
\newcommand{\ZZ}{\mathcal{Z}}
\newcommand{\TT}{\mathcal{T}}

\newcommand{\Inv}{\rm Inv}

\newcommand{\ad}{\rm ad}

\newcommand{\tet}{\rm Tet} 
\newcommand{\thet}{\rm Theta} 
\newcommand{\ci}{\rm Circ} 

\newcommand{\tensor}{\otimes}
\newcommand{\hodge}{\ast}

\newcommand{\ftj}{\mbox{15}j}   
\newcommand{\lalg}[1]{\mathfrak{#1}}  
\newcommand{\SU}{\mathrm{SU}}
\newcommand{\SO}{\mathrm{SO}}
\newcommand{\OO}{\mathrm{O}}
\newcommand{\SL}{\mathrm{SL}}

\newcommand{\Spin}{\mathrm{Spin}}
\newcommand{\Hom}{\mathrm{Hom}}
\newcommand{\Reel}{\mathrm{Re}}
\newcommand{\su}{\lalg{su}}
\renewcommand{\sl}{\lalg{sl}}

\newcommand{\so}{\lalg{so}}

\newcommand{\spin}{\lalg{spin}}
\newcommand{\dd}{\mathrm{d}}

\newcommand{\pole}{\mathcal N}

\newcommand{\SLtwoC}{\SL(2,\mathbb{C})}

\newcommand{\z}{Z}  

\newcommand{\bb}{\mathbf{b}} 
\def\la{\langle}
\def\ra{\rangle}

\def\bra#1{\mathinner{\langle{#1}|}}
\def\ket#1{\mathinner{|{#1}\rangle}}
\def\braket#1#2{\mathinner{\langle{#1}|{#2}\rangle}}

\def\ph#1{{\phantom{#1}}}

%% file: fig-k2.eps_tex

\begingroup
  \makeatletter
  \providecommand\color[2][]{%
    \errmessage{(Inkscape) Color is used for the text in Inkscape, but the package 'color.sty' is not loaded}
    \renewcommand\color[2][]{}%
  }
  \providecommand\transparent[1]{%
    \errmessage{(Inkscape) Transparency is used (non-zero) for the text in Inkscape, but the package 'transparent.sty' is not loaded}
    \renewcommand\transparent[1]{}%
  }
  \providecommand\rotatebox[2]{#2}
  \ifx\svgwidth\undefined
    \setlength{\unitlength}{336.93242187pt}
  \else
    \setlength{\unitlength}{\svgwidth}
  \fi
  \global\let\svgwidth\undefined
  \makeatother
  \begin{picture}(1,0.43925722)%
    \put(0,0){\includegraphics[width=\unitlength]{fig-k2.eps}}%
    \put(0.55599282,0.35483298){\color[rgb]{0,0,0}\makebox(0,0)[lb]{\smash{=}}}%
    \put(0.45554941,0.10244007){\color[rgb]{0,0,0}\makebox(0,0)[lb]{\smash{=}}}%
    \put(0.87620428,0.15291862){\color[rgb]{0,0,0}\makebox(0,0)[lb]{\smash{$-1$}}}%
    \put(0.18514307,0.10451764){\color[rgb]{0,0,0}\makebox(0,0)[lb]{\smash{=}}}%
    \put(0.03489449,0.11926625){\color[rgb]{0,0,0}\makebox(0,0)[lb]{\smash{\svg}}}%
    \put(0.28728744,0.11926625){\color[rgb]{0,0,0}\makebox(0,0)[lb]{\smash{\svg}}}%
    \put(0.55650659,0.11926625){\color[rgb]{0,0,0}\makebox(0,0)[lb]{\smash{\svg}}}%
    \put(0.84255191,0.11926625){\color[rgb]{0,0,0}\makebox(0,0)[lb]{\smash{\svg}}}%
    \put(0.95262867,0.12003694){\color[rgb]{0,0,0}\makebox(0,0)[lb]{\smash{\svg}}}%
  \end{picture}%
\endgroup